\documentclass[12pt]{article}

\pdfoutput=1
\usepackage{commands}
\usepackage{putex} 
\usepackage[all,cmtip]{xy}
\usepackage{fancybox}
\usepackage{subcaption}
\usepackage{graphicx}
\usepackage{cite}
\usepackage{dsfont}
\usepackage{physics}
\usepackage{mciteplus}
\usepackage{comment}
\usepackage{skak}
\usepackage{esint}
\usepackage{amsmath}
 \usepackage[compat=1.1.0]{tikz-feynman}\tikzfeynmanset{warn luatex=false}
\usepackage{multirow}
\usepackage{calc}
\usepackage{mathtools,amssymb}
\usepackage{tikz}
\usepackage{pgfplots}
\pgfplotsset{compat=1.18}
\usetikzlibrary{decorations.markings,decorations.pathmorphing}
\usetikzlibrary{arrows.meta}
\usetikzlibrary{decorations.markings}
\tikzset{
    midarrow/.style={postaction={decorate},decoration={
            markings,
            mark=at position .5 with {\arrow{stealth}}
        }
    }
}

\DeclareFontFamily{OT1}{pzc}{}
\DeclareFontShape{OT1}{pzc}{m}{it}{<-> s * [1.10] pzcmi7t}{}
\DeclareMathAlphabet{\mathpzc}{OT1}{pzc}{m}{it}

\def\({\left(}
\def\){\right)}

\usepackage{tikz-cd}
\makeatletter
\DeclareFontFamily{OMX}{MnSymbolE}{}
\DeclareSymbolFont{MnLargeSymbols}{OMX}{MnSymbolE}{m}{n}
\SetSymbolFont{MnLargeSymbols}{bold}{OMX}{MnSymbolE}{b}{n}
\DeclareFontShape{OMX}{MnSymbolE}{m}{n}{
    <-6>  MnSymbolE5
   <6-7>  MnSymbolE6
   <7-8>  MnSymbolE7
   <8-9>  MnSymbolE8
   <9-10> MnSymbolE9
  <10-12> MnSymbolE10
  <12->   MnSymbolE12
}{}
\DeclareFontShape{OMX}{MnSymbolE}{b}{n}{
    <-6>  MnSymbolE-Bold5
   <6-7>  MnSymbolE-Bold6
   <7-8>  MnSymbolE-Bold7
   <8-9>  MnSymbolE-Bold8
   <9-10> MnSymbolE-Bold9
  <10-12> MnSymbolE-Bold10
  <12->   MnSymbolE-Bold12
}{}

\let\llangle\@undefined
\let\rrangle\@undefined
\DeclareMathDelimiter{\llangle}{\mathopen}%
                     {MnLargeSymbols}{'164}{MnLargeSymbols}{'164}
\DeclareMathDelimiter{\rrangle}{\mathclose}%
                     {MnLargeSymbols}{'171}{MnLargeSymbols}{'171}
\makeatother

\usepackage{latexsym}
\usepackage{stmaryrd}

\usepackage{cite}
\usepackage{framed}
\definecolor{shadecolor}{rgb}{0.95,0.95,0.97}
\definecolor{refkey}{rgb}{0.5,0.5,0}
\definecolor{labelkey}{rgb}{0.5,0.5,0}
\definecolor{citekey}{rgb}{0.5,0.5,0}
\definecolor{darkgreen}{rgb}{0,0.5,0}
\definecolor{darkblue}{cmyk}{0.9,0.9,0,0}
\definecolor{darkred}{rgb}{0.6,0,0.3}
\colorlet{mydarkblue}{blue!50!black}
\colorlet{myred}{red!65!black}
\usepackage{ifpdf}
\ifpdf
\usepackage{graphicx} 
\usepackage[setpagesize=false,pagebackref=false, linktocpage, bookmarksopen=true, colorlinks=true, linkcolor=blue,citecolor=blue,urlcolor=blue]{hyperref}
\usepackage{hyperref}

\else
\usepackage{graphicx}

\fi
\newcommand{\id}{{\bf 1}}

\def\XXint#1#2#3{{\setbox0=\hbox{$#1{#2#3}{\int}$}
		\vcenter{\hbox{$#2#3$}}\kern-.5\wd0}}

\newcommand{\beq}{\begin{equation}}
\newcommand{\eeq}{\end{equation}}
\def\nullify#1{}
\makeatletter
\def\section{\@startsection {section}{1}{\z@}{-3.5ex plus -1ex minus 
		-.2ex}{2.3ex plus .2ex}{\large\bf}}
\makeatother
\makeatletter
\def\subsection{\@startsection {subsection}{1}{\z@}{-3.5ex plus -1ex minus 
		-.2ex}{2.3ex plus .2ex}{\normalsize\bf}}
\makeatother
\begin{document}
	
	\preprint{DESY-25-035}
	
	\institution{DESY}{Deutsches Elektronen-Synchroton DESY, Notkestrasse 85, 22607 Hamburg, Germany}

    \institution{Perimeter}{Perimeter Institute for Theoretical Physics, Waterloo, ON N2L 2Y5, Canada}
	
	\title{Renormalization Group flow in Schur quantization}

	\authors{Federico Ambrosino\worksat{\DESY} and Davide Gaiotto\worksat{\Perimeter}}

		\abstract
	{
    We develop a general formalism to describe the Renormalization Group Flow of Schur indices and fusion algebras of BPS line defects in four-dimensional ${\cal N}=2$ Supersymmetric Quantum Field Theories. The formalism includes and extends known results about the Seiberg-Witten description of these structures. Another application of the formalism is to describe the spectrum of BPS partices of ${\cal N}=2$ gauge theories with matter in terms of the spectrum of pure ${\cal N}=2$ gauge theories. Applications to the theory of quantum groups and to the quantization of cluster varieties are also discussed.
}
	\date{\today}

	\maketitle
	
	\tableofcontents
	\newpage	
 %Please don't use \be for \beta

\section{Introduction}
Four-dimensional $\CN=2$ Supersymmetric Quantum Field Theories are associated to a rich collection of physical and mathematical results \cite{Seiberg:1994rs,Seiberg:1994aj,Witten:1994cg,Argyres:1995jj,Donagi:1995cf,Nekrasov:2002qd,Gaiotto:2009we,Alday:2009aq}, many of which were developed to understand the Seiberg-Witten solution of $SU(2)$ supersymmetric Yang-Mills theory and its generalizations. The Seiberg-Witten solution provides a description of the space of vacua of the gauge theory, as well as a large amount of information about the low-energy effective field theory in each of these vacua and about the RG flow itself, in the form of a complete spectrum of supersymmetry-preserving (BPS) massive particles. 

From the very beginning, the formalism included sequential RG flows passing through one or more intermediate (mesoscopic) effective description, due to hierarchical spectra of massive particles. A classical example was the low effective description of pure $SU(2)$ gauge theory as an $U(1)$ gauge theory coupled to a light hypermultiplet, which provided an interpretation of confinement as a monopole condensation \cite{Seiberg:1994rs}. This feature was soon employed to define novel SQFTs as intermediate steps of known RG flows \cite{Argyres:1995jj}. 
  
Another notion which was present from the very beginning is ``wall-crossing'': discrete features of the RG flow such as the BPS spectrum are locally constant as we vary the choice of vacua or couplings of the theory, but may jump across special ``walls'' in the space of parameters. The mathematical work of Kontsevich and Soibelman \cite{Kontsevich:2008fj}, which arose in the context of wall-crossing in String Theory, provided a conjectural universal description of such wall-crossing phenomena and motivated much work to understand the physical and mathematical aspects of the formula \cite{Cecotti:2009uf,Gaiotto:2009hg,Dimofte:2009tm,Gaiotto:2010be,Gaiotto:2011tf,Gaiotto:2010okc,Manschot:2010qz,Alim:2011ae}. 

One of the tools developed in this context is the fusion algebra $\CA$ of half-BPS loop operators \cite{Kapustin:2007wm,Gaiotto:2009hg,Gaiotto:2010be,Drukker:2009id,Alday:2009fs,Dimofte:2011py,Braverman:2016wma}. This is an algebra over $\bZ[\fq, \fq^{-1}]$ which inherits additional properties from its physical definition. For example, $\CA$ is equipped with an automorphism $\rho$ mapping a loop operator to its (right) dual
as well as a ``Schur index'' pairing
\begin{equation}
    I_{a,b}(\fq) = I_{1,\rho(a)b}(\fq) = I_{1,b \rho^{-1}(a)}(\fq) \in \bC((\fq))
\end{equation}
which is expected to converge for $|\fq|<1$. From now on, we specialize $\fq$ to be a real number in the range $-1<\fq<1$.

Recently, a conjectural positivity constraint 
\begin{equation}
    I_{a,a} > 0 \qquad \qquad a \neq 0 \qquad \qquad -1<\fq<1
\end{equation}
was employed \cite{Gaiotto:2024osr} to introduce a representation-theoretic framework called Schur quantization. The output of Schur quantization 
is an Hilbert space $\cH$ which carries an action of $\CA$ as unbounded normal operators $W_a$. The Hilbert space includes a {\it spherical vector}, a normalizable vector $|1\rangle\in\cH$ satisfying 
\begin{equation}
    |a\rangle \equiv W_a|1\rangle=W_{\rho(a)}^\dagger |1\rangle
\end{equation}
such that the expectation values
\begin{equation}
    \mathrm{Tr}(a):=\langle 1|a|1\rangle
\end{equation}
are finite for all $a\in\CA$ and coincide with Schur indices:  
\begin{equation}
I_{a,b}=\langle a|b\rangle = \Tr \rho(a) b \, .
\end{equation}
The vectors $|a\rangle$ give a natural domain for the action of $\CA$:
\begin{equation}
    W_a \,W^\dagger_{\rho(c)} |b\rangle = |a b c\rangle \, .
\end{equation}

Concrete representations for the output of Schur quantization exist when explicit results for $\CA$ and $I_{ab}$ are available. A general class of examples are $\CN=2$ gauge theories, defined by a UV Lagrangian whose form is determined by a choice of gauge group $G$ and symplectic matter representation\footnote{Subject to a discrete anomaly cancellation condition which is automatically satisfied if $M$ is of cotangent type when restricted to all simple factors of $G$.} $M$. We will denote such a theory as $T[G,M]$ when necessary. 

The algebra $\CA[G,M]$ for $T[G,M]$ is linearly generated by ``'t Hooft-Wilson'' loop operators $D_{m,e}$, labelled by a magnetic weight $m$ and an electric weight $e$ for $G$ defined up to the Weyl group action on the pair.
The algebra structure can be computed via localization \cite{Gomis:2011pf}, though that may require advanced algebraic-geometric tools \cite{Braverman:2016wma,Braverman:2022zei}. It is known mathematically as the K-theoretic Coulomb branch algebra. The Schur index is also computed via localization by a relatively simple integral formula \cite{Gadde:2011ik}. 

It is natural to ask how Schur quantization interacts with RG flow. We expect an RG flow from a theory $T_1$ to a theory $T_2$ to induce an algebra map 
\begin{equation}
    \RG: \CA_1 \to \CA_2 \, ,
\end{equation}
with multi-step flows $T_1 \to T_2 \to T_3$ giving compatible maps:
\begin{equation}
    \RG_{31} = \RG_{32} \circ \RG_{21}\, .
\end{equation}
There are two obstructions to lifting the $\RG$ maps to Schur quantization:
\begin{itemize}
    \item The RG flow and dualization maps do not commute:
    \begin{equation}
        \rho_2 \circ \RG \neq \RG \circ \rho_1 \, .
    \end{equation}
    \item The Schur indices do not match under the naive RG flow:
    \begin{equation}
        I^{(2)}_{\RG(a),\RG(b)} \neq I^{(1)}_{a,b} \, .
    \end{equation}
\end{itemize}

For the special case of $T_2$ being an $U(1)^r$ SYM gauge theory and the RG flow being the Seiberg-Witten solution of the theory $T_1$, it is known that both obstructions \cite{Gaiotto:2010be,Cordova:2015nma,Cordova:2016uwk} are removed by the same object: the quantum spectrum generator $S$, also known as (the inverse of) the generating function of Donaldson-Thomas invariants \cite{Kontsevich2014}. 

We will generalize this result and associate to any RG flow as above a quantum spectrum generator $S$ defined as an invertible element in a formal completion of $\CA_2((\fq))$ such that 
\begin{itemize}
    \item Conjugation by $S$ intertwines the $\rho_1$ and $\rho_2$ dualization maps:
    \begin{equation}
        \RG(a)\cdot S = S \cdot \rho_2^{-1}(\RG(\rho_1(a))) \, .
    \end{equation}
    \item Right multiplication by $S$ intertwines the Schur indices:
    \begin{equation}
        I^{(2)}_{\RG(a)\cdot S,\RG(b)\cdot S} = I^{(1)}_{a,b} \, .
    \end{equation}
\end{itemize}
Furthermore, if we have multi-step flow $T_1 \to T_2 \to T_3$, 
\begin{equation}
    S_{31} = \RG_{32}(S_{21})\cdot S_{32}
\end{equation}

Accordingly, we find a relation between the Schur quantization for $T_1$ and $T_2$: the map 
\begin{equation}
    |1\rangle \to |S\rangle \equiv W_S|1\rangle
\end{equation}
extends to an isometry $\cH_1 \to \cH_2$ which intertwines the standard action of $\CA_1$ on $\cH_1$ and the action by $W_{\RG(a)}$ on $\cH_2$. This composes nicely for multi-step flows. We will also formulate a set of conjectures which implies that $\cH_1 \to \cH_2$ is an unitary transformation. 

This formalism has a variety of algebraic and representation-theoretic consequences:
\begin{enumerate}
    \item It simplifies the derivation of spectrum generators. First of all, standard strategies to build $S$ can be simplified by considering multi-step flows. Furthermore, we find an explicit formula for the spectrum generator associated to the flow $T[G,T^*N] \to T[G,0]$, so the conventional spectrum generator for any such gauge theory can be derived from that of a SYM theory. 
    \item It allows access to Schur quantization for non-Lagrangian theories for which the spectrum generator is otherwise available and gives new strategies to derive it. 
    \item Specialized to theories of class $\CS$, it gives tools to quantize Chern-Simons theory with complex gauge group in a manner analogous to the well-known quantization of (higher) Teichm\"uller space \cite{kashaev1997quantizationteichmullerspacesquantum,fock2006clusterxvarietiesamalgamationpoissonlie,Fock_2008,LE_2019,Teschner:2014vca,fock2006modulispaceslocalsystems,fock2009clusterensemblesquantizationdilogarithm}.
    \item Specialized to four-dimensional lifts of $T_\rho[G]$ theories \cite{Gaiotto:2023ezy}, it gives a rich collection of unitary representations of ``complex'' quantum groups and provides acces to their spectral analysis.
\end{enumerate}

\subsection{Seiberg-Witten RG flows, Donaldson-Thomas invariants and tropical labels}
An Abelian SYM $T[U(1)^r,0]$ of rank $r$ is equipped with a charge lattice $\Gamma$ of rank $2r$ and an anti-symmetric, integral Dirac pairing $\langle \bullet, \bullet \rangle$ which we take to be non-degenerate and unimodular. The corresponding fusion algebra $\CA[U(1)^r,0]$ of 't Hooft-Wilson loop operators is the {\it quantum torus} algebra $\CQ_\Gamma$: it has generators $X_\gamma$ for $\gamma \in \Gamma$, relations 
$X_{\gamma} X_{\gamma'} = \fq^{\langle \gamma, \gamma'\rangle} X_{\gamma + \gamma'}$, 
and a canonical automorphism $\rho_\CQ(X_\gamma) = X_{-\gamma}$. The Schur indices are 
\begin{equation}
    I_{X_{\gamma}, X_{\gamma'}} = (\fq^2)_\infty^{2 r}\delta_{\gamma, \gamma'}
\end{equation}
and Schur quantization is implemented on the Hilbert space $\ell^2(\Gamma)$. Concretely, $|X_\gamma\rangle$ can be identified with the function $(\fq^2)_\infty^{r}$ supported at $\gamma$ and 
\begin{equation}
    W_{X_{\gamma}} |X_{\gamma'}\rangle = |X_{\gamma} X_{\gamma'}\rangle = \fq^{\langle\gamma,\gamma'\rangle} |X_{\gamma+ \gamma'}\rangle \, .
\end{equation}

Any known ${\CN=2}$ SQFT $T$ is endowed with a moduli space of ``Coulomb'' vacua, which do not spontaneously break the $SU(2)_R$ symmetry of the theory. As generic Coulomb vacua, we expect an RG flow $T \to T[U(1)^r,0]$ to an Abelian SYM. In particular, we have an RG flow map $\RG: \CA \to \CQ_\Gamma$ from the UV fusion algebra $\CA$ to the appropriate quantum torus algebra $\CQ_\Gamma$:
\begin{equation}
F_a:=\RG(a)=\sum_{\gamma\in\Ga}\Omega_{a}^{\gamma}(\fq)X_{\gamma} \,.
\end{equation}
These Laurent polynomials are generating functions for ``framed BPS degeneracies'' $\Omega_{a}^{\gamma}(\fq)$, which count the low energy ground states of a loop operator $a$ which carry gauge charge $\gamma$. The $\fq$ variable plays the role of a spin fugacity.\footnote{Conjecturally, $\Omega_{a}^{\gamma}(\fq)$ are positive linear combinations of $SU(2)$ characters $[n]_\fq$.}

The choice of Coulomb vacuum which triggers the RG flow provides some extra data which constrains the form of the $F_a$'s and of the spectrum generator $S$. For the purpose of this paper, the data takes the form of a generic real linear functional on $\Gamma$, which splits non-zero charges into a positive cone and and its opposite and thus gives an ordering $\gamma \prec \gamma'$ of charges. The spectrum generator is supported on positive charges:
\begin{equation}
    S = 1 + \sum_{\gamma \succ 0} \Sigma_\gamma(\fq) X_\gamma \, ,
\end{equation}
and conjecturally there is a linear ``tropical basis'' $\ell_\gamma$ in $\CA$ such that 
\begin{equation}
    F_{\ell_\gamma} = X_\gamma + \sum_{\gamma'\succ \gamma}\Omega_{a}^{\gamma'}(\fq)X_{\gamma'}
\end{equation}
The physical origin of this particular conjecture is unclear to us, but it holds in all known examples and we will assume it in this paper. We will also observe other remarkable properties of the tropical basis. For example, $\rho(\ell_\gamma)$ appears to belong to the tropical basis as well. 

These support conditions guarantee that $F_{\ell_\gamma} S-X_\gamma$ is also supported on charges $\gamma' \succ \gamma$. We will employ them to argue that the above-defined isometry from the Schur quantization Hilbert space $\cH$ to $\ell^2(\Gamma)$ is an unitary transformation. 

We should observe that given any collection of Laurent polynomials $F_{\ell_\gamma}$ which form a closed algebra and a formal power series $S$ which satisfies the intertwining conditions
\begin{equation}
    F_{\ell_\gamma} \cdot S = S \cdot \rho_{\CQ}(F_{\rho(\ell_\gamma)}) \, ,
\end{equation}
one may formally attempt to run the Schur quantization procedure to build a representation on $\ell^2(\Gamma)$. The existence of Schur indices, i.e. 
\begin{equation}
    I_{a,b} = \langle F_a S|F_b S\rangle <\infty \, ,
\end{equation}
thought, is far from obvious. Indeed, we will see examples of tentative spectrum generators which fail this condition and thus presumably are not associated to an actual 4d ${\cal N}=2$ SQFT. 

Finally, we can compare the Seiberg-Witten solution of two theories related by RG flow, via a multi-step process $T_1 \to T_2 \to T[U(1)^r,0]$. Then the RG flow map and the framed BPS degeneracies will be related as 
\begin{equation}
    F^{(1)}_a = F^{(2)}_{\RG_{21}(a)}
\end{equation}
If we assume that both algebras admit a tropical basis, we learn that 
\begin{equation}
    \RG_{21}(\ell^{(1)}_\gamma) = \ell^{(2)}_\gamma + \sum_{\gamma'\succ \gamma}\Omega_{\ell^{(1)}_\gamma}^{\gamma'}(21;\fq)\ell^{(2)}_{\gamma'}
\end{equation} 
If we ``forget'' about the last step of the RG flow and only preserve the tropical bases and ordering $\ell^{(2)}_\gamma \prec \ell^{(2)}_{\gamma'}$ 
iff $\gamma \prec \gamma'$, we see that RG flows $T_1 \to T_2$ must also be associated to a tropical basis for $\CA_2$, with support conditions both for $\RG$ and for $S$ and novel partial BPS degeneracies $\Omega_{a}^{\gamma'}(21;\fq)$. This will provide an argument for $\cH_1 \to \cH_2$ also being an unitary transformation. 

\subsection{Schur quantization and complex cluster quantization}
In the $\fq \to \pm 1$ limit, the fusion algebra $\CA$ becomes commutative and is identified with the algebra of holomorphic functions on a complex symplectic manifold $\CM$, the space of vacua for a circle compactification of the 4d SQFT. 
In the same limit, the full $*$-double $\fA$ can be identified with a Poisson algebra combining both holomorphic and anti-holomorphic functions on $\CM$, using the imaginary part of the complex symplectic form bracket. Accordingly, Schur quantization gives a quantization of $\CM$ as a real phase space\footnote{Both $\fq \to 1$ and $\fq \to -1$ limit to the Poisson algebra of function on two versions of $\CM$, see \cite{Gaiotto:2010be}.}, say with $\fq = \pm e^{-\hbar}$.  

An RG flow $T_1 \to T_2$ gives a Poisson map $\CM_2 \to \CM_1$: the low energy description $T_2$ only covers a patch of the full space of vacua $\CM_1$. In particular, the Seiberg-Witten RG flow gives a $(\bC^*)^{2r}$ log-Darboux coordinate patch on the space of vacua of the UV theory. The RG flow description of Schur quantization can thus be interpreted as a procedure where we first quantize the coordinate patch to $\ell^2(\Gamma)$ and then express quantized (anti)holomorphic functions on $\CM$ as unbounded normal operators acting on an appropriate domain in $\ell^2(\Gamma)$. 

As one moves on the space of 4d vacua for a theory $T$, the resulting RG flows will give a collection of coordinate patches which cover the whole $\CM$. Usually, there is no guarantee that a ``patch-by-patch'' quantization of a phase space should be possible or consistent, with compatible unitary lifts of the classical coordinate transformations. Here, the unitary equivalences with the underlying UV definition guarantee the procedure success. In particular, the inner products 
\begin{equation}
    I_{a,b} = \langle F_a S|F_b S\rangle \, ,
\end{equation}
must be well defined and coincide in all RG coordinate charts. 

In all known examples, the coordinate patches defined by the Seiberg-Witten RG flows equip $\CM$ with the structure of a cluster variety \cite{fomin2001clusteralgebrasifoundations,Fomin_2003,fomin2006clusteralgebrasivcoefficients}: each chart is related to nearby charts by wall-crossing transformations which take the form of ``cluster transformations'' labelled by a finite collection of charges $\gamma_i$. Turning on $\fq$, cluster transformations act on all the RG flow ingredients:
\begin{itemize}
    \item $F_a \to E_\fq(X_{\gamma_i})^{-1} F_a E_\fq(X_{\gamma_i})$.
    \item $S \to E_\fq(X_{\gamma_i})^{-1} S E_\fq(X_{-\gamma_i})$.
    \item $|a \rangle \to E_\fq(X_{\gamma_i})^{-1} E_\fq(X_{\gamma_i})^\dagger$.
\end{itemize}
The collection of charges $\gamma_i$ and the tropical labeling of $\CA$ also 
mutates in a specific manner we review below.\footnote{A confusing aspect of the relation between cluster technology and SQFTs is that certain cluster transformations may not correspond to allowed wall-crossing transformations in the theory. We will investigate in examples if the full Schur quantization structure may detect these discrepancies.} Accordingly, Schur quantization becomes a form of ``cluster'' quantization, analogous to the better known quantization of the positive part of $\CM$. 

\begin{figure}
\centering
\begin{tikzpicture}[->, thick, main node/.style={circle, draw, minimum size=10mm}]
    
    % Nodes
    \node[main node] (1) {1};
    \node[main node] (2) [right=of 1] {2};
    \node[main node] (3) [right=of 2] {3};
    \node[main node] (4) [right=of 3] {4};

    % Arrows (Alternating directions)
    \draw[->] (1) -- (2);
    \draw[<-] (2) -- (3);
    \draw[->] (3) -- (4);

\end{tikzpicture}
\caption{The BPS quiver for the $[A_1,A_4]$ Argyres-Douglas theory. It describes a collection of charges $\gamma_1, \cdots \gamma_4$ with non-trivial pairings $\langle \gamma_1, \gamma_2\rangle = \langle \gamma_3, \gamma_2\rangle = \langle \gamma_3, \gamma_4 \rangle =1$.}\label{fig:A4}
\end{figure}
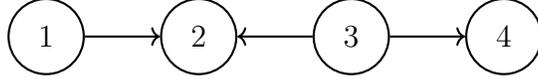

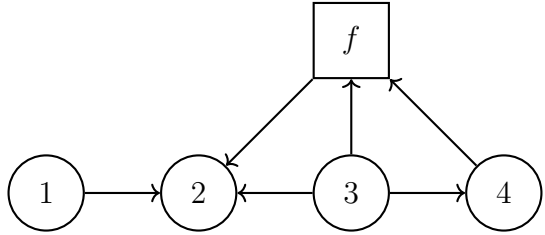
\begin{figure}
\centering
\begin{tikzpicture}[->, thick, 
    main node/.style={circle, draw, minimum size=10mm}, 
    framing node/.style={rectangle, draw, minimum size=10mm}
]

    % Nodes
    \node[main node] (1) {1};
    \node[main node] (2) [right=of 1] {2};
    \node[main node] (3) [right=of 2] {3};
    \node[main node] (4) [right=of 3] {4};
    \node[framing node] (f) [above=of 3] {$f$};

    % Arrows (Alternating directions)
    \draw[->] (1) -- (2);
    \draw[<-] (2) -- (3);
    \draw[->] (3) -- (4);
    \draw[->] (4) -- (f); % Framing node connection
    \draw[->] (3) -- (f);
    \draw[<-] (2) -- (f);
\end{tikzpicture}
\caption{The $A_4$ BPS quiver for the $[A_1,A_4]$ AD theory with an extra framing node. It describes a tropical charge $\gamma$ with non-trivial pairings $\langle \gamma_2, \gamma\rangle = \langle \gamma, \gamma_3\rangle = \langle \gamma, \gamma_4 \rangle =1$.}\label{fig:A4f}
\end{figure}

An important visualization tool for the collection of $\gamma_i$ charges in $\Gamma$ is the ``BPS quiver'', with nodes labelled by $\gamma_i$ and 
\begin{equation}
|i \to j| \equiv \mathrm{max}(0,\langle \gamma_i, \gamma_j\rangle)
\end{equation}
arrows from the $i$-th to the $j$-th nodes. See Figure \ref{fig:A4}. A tropical charge $\gamma$ can be represented by an extra ``framing'' node with label $\gamma$. See Figure \ref{fig:A4f}. Cluster transformations `mutate'' the BPS quiver at one node by a simple rule. 

\subsection{Class $\CS$ Schur quantization and complex Chern-Simons theory}
Class $S$ theories are defined by the data of an ADE Lie algebra $\fg$ and 
a decorated Riemann surface $C$ \cite{Witten:1997sc,Gaiotto:2009we}. They have the property that the associated algebra $\CA$ coincides with the ``skein algebra'' $\mathrm{Sk}(\fg, C)$
and the manifold $\CM$ of vacua coincides with a ``character variety'' of $\fg$ local systems on $C$ with appropriate singularities. The Schur indices have an interpretation in terms of an $\fq$-deformation of 2d YM theory \cite{Gadde:2011ik}. 

These properties, reinforced by an appropriate duality chain, leads to the identification between Schur quantization and the quantization of ``complex'' Chern-Simons theory on $C$ \cite{Gaiotto:2024osr}. 

There is also a well-developed geometric approach \cite{Gaiotto:2010be,Longhi:2015ivt,Neitzke:2020jik,Neitzke:2021gxr} to the study of RG flows, $F_a$'s and quantum spectrum generators, etc. We expect the theory of multi-step RG flows we set up to help simplify this geometric approach. 

\subsection{Computational strategies}
The Schur quantization and RG flow machinery we discuss in this paper is most useful if the data of $F_{\ell_\gamma}$ and $S$ can be explicitly derived or at least proven to exist in a mathematically rigorous manner. Multiple strategies have been developed in the past for this purpose:
\begin{enumerate}
    \item The original work by Kontsevich and Soibelman on Cohomological Hall Algebras (CoHA) takes as an input a BPS quiver equipped with a ``superpotential'' $W$. It outputs a graded algebra whose equivariant character is expected to coincide with $S^{-1}$. The addition of a framing node conjecturally allows the computation of $S^{-1} F_{\ell_\gamma}$ \cite{Kontsevich:2010px,Cirafici:2017wlw,Gaiotto:2024fso}. We are not aware of a physical prescription for $W$, unless the SQFT has an embedding into String Theory, but the choice of a ``generic'' $W$ seems consistent with available data. The superpotential makes computations difficult. A recently conjectured alternative construction of the relevant CoHAs tentatively allows for a fully automated computation of $S^{-1} F_{\ell_\gamma}$ for any BPS quiver (See \cite{Davison:2025ydh} for a further wrinkle in the story). We will verify that ``unphysical'' BPS quivers fail to give convergent Schur indices. 
    \item The theory of ``scattering diagrams''  \cite{Kontsevich2014,MR4729624} appears to offer an alternative combinatorial route to the computation of the $F_{\ell_\gamma}$'s from a BPS quiver, from which one can reconstruct $\rho$ and then $S$.  
    \item Our conjectural partial RG flow $T[G,T^*N] \to T[G,0]$ allows for the computation of $S$ for all gauge theories for which $G$ for which the quantum spectrum generator of $T[G,0]$ is known. The $\rho$ map is known from the gauge theory description and one can then recover the $F_a$'s from that information. Partial RG flows allow one to ``transport'' this information to a larger class of theories. 
    \item Above-mentioned geometric computational tools (``spectral networks'') are available for theories of class $S$.
\end{enumerate}
As the whole structure is very constrained, it is often possible to reconstruct it from partial information. We will review a variety of computational strategies in examples. 

\subsection{Structure of the Paper}
This paper is organized as follows. Section \ref{Sec:alg} is devoted to the illustration of the Schur quantization and its application to Coulomb branch IR description of $4d \cN =2$ theories. In Section \ref{sec:rank2} we focus on the minimal example of Schur quantization applied to 4d $\cN=2$ theories with rank 2 lattice. In Section \ref{sec:addingfla} and \ref{sec:broadconj}, building on the example of $SU(2)$ $\cN=2^*$ and the partial RG flow technique, we illustrate a powerful conjecture  how to compute the spectrum generator of theories with flavor. One of the outcome is the computation of the Spectrum generator for $SU(2)$ $N_f = 4$, which was not known by other means. 
In Section \ref{sec:su3} we test our technology further on $SU(3)$ gauge theories. In Section \ref{sec:Uq} we discuss some applications to the theory of quantum groups. In Section \ref{sec:cut} we briefly discuss how to define RG flows geometrically in theories of class $S$ labelled by $\sl_2$. In Section \ref{sec:teichquant} we discuss complex quantization of $\sl_2$ character varieties 
in a manner parallel to the conventional quantization of Teichm\"uller space. Section \ref{sec:phys} reviews some physical background for our constructions. Appendix \ref{spectralproblem} discusses in detail the spectral problem relating UV and IR Schur quantization in $SU(2)$ gauge theories. 

\section{Schur Quantization}\label{Sec:alg}

The typical algebraic structure we encounter in Schur quantization (and in sphere quantization for 3d theories \cite{Gaiotto:2023hda,Gaiotto:2023kuc}) is a quadruple $(\CA, \rho, \cH,|1\rangle)$, where\footnote{The condition $(c)$ can be better stated in the language of $*$-algebras by defining a $*$-algebra double $\fA \equiv \CA \otimes \CA^{\rm op}$ with $*$-structure $(a \otimes b)^* = \rho(b) \otimes \rho^{-1}(a)$. Then $\cH$ is an unitary representation of $\fA$. The condition $(d)$ is that $(a \otimes 1)|1\rangle = (1 \otimes a)|1\rangle$.} 
\begin{enumerate}
    \item $\CA$ is an algebra over the complex numbers.
    \item $\rho$ is an anti-linear automorphism of $\CA$.
    \item $\cH$ is an Hilbert space equipped with a representation of $\CA$ by unbounded operators such that $[a,b^\dagger]=0$ for all $a,b \in \CA$.\footnote{In principle we should distinguish elements $a$ of $\CA$ from the operators $W_a$ which represent them. To lighten the notation, we will denote both with the same symbol.}
    \item $|1\rangle$ is a cyclic vector in the domain of $\CA$ which satisfies a ``spherical'' condition:
    \begin{equation}
        |a\rangle \equiv a|1\rangle = \rho(a)^\dagger |1\rangle \, .
    \end{equation}
    Here $|a\rangle$ are normalizable states also in the domain of $\CA$.
\end{enumerate}
We will refer to this structure as a $\rho$-spherical unitary representation of $\CA$.

Observe that the vectors of the form $|a\rangle$ give a dense common domain for the action of $\CA$:
\begin{equation}
    a \rho(c)^\dagger |b\rangle = a b \rho(c)^\dagger |1\rangle = |abc\rangle 
\end{equation}
Furthermore,
\begin{equation}
    \Tr \, a \equiv \langle 1|a|1\rangle  
\end{equation}
is a $\rho^2$-twisted trace on $\CA$:
\begin{equation}
    \langle a|b\rangle = \Tr \, \rho(a) b = \Tr \, b \rho^{-1}(a) \,.
\end{equation}
In several physical contexts, $\CA$, $\rho$ and $\Tr$ are given as protected quantities in some SQFT and are computable via localization. Then $\cH$ can be abstractly recovered as the closure of $\CA$ with respect to the inner product $\Tr \rho(a) b$.

Suppose now that we are given such a quadruple $(\CA_\IR, \rho_\IR, \cH_\IR,|1\rangle_\IR)$ and an embedding 
\begin{equation}
    \RG:\CA_\UV \to \CA_\IR
\end{equation}
of a second algebra $\CA_\UV$ into $\CA_\IR$, as well as a choice of $\rho_\UV$ which does {\it not} agree with $\rho_\IR$, i.e. 
\begin{equation}
    \RG \neq \rho_\IR^{-1} \circ \RG \circ \rho_\UV
\end{equation}
Then $\CA_\UV$ acts on $\cH$, but $|1\rangle_\IR$ is not spherical. 

In order to define a spherical vector, we may seek a formal linear combination $S$ of elements in $\CA_\IR$ such that 
\begin{equation}
    \RG(a) \cdot S = S \cdot \rho_\IR^{-1}(\RG(\rho_\UV(a)))
\end{equation}
If we can find such an element, then 
\begin{equation}
    S |1\rangle_\IR = |S\rangle_\IR
\end{equation}
formally satisfies the spherical condition for the $\CA_\UV$ action:
\begin{equation}
    \RG(a) |S\rangle_\IR \equiv S \, \RG(\rho_\UV(a))^\dagger |1\rangle_\IR =  \RG(\rho_\UV(a))^\dagger |S\rangle_\IR
\end{equation}
If the $\RG(a) |S\rangle_\IR$ vectors are in addition normalizable and dense in $\cH_\IR$, then we can define a quadruple $(\CA_\UV, \rho_\UV, \cH_\IR,|S\rangle_\IR)$.

We will apply these constructions to the situation where $\CA$ is the fusion algebra of supersymmetric loop operators in a four-dimensional ${\cal N}=2$ SQFT. The relevant algebras are defined over $\bZ[\fq, \fq^{-1}]$ and we represent them after specializing $\fq$ to be a real number in the interval $-1<\fq<1$. 

A typical example of $\CA_\IR$ is the quantum torus algebra $\CQ_\Gamma$ associated to a lattice $\Gamma$ equipped with an anti-symmetric unital integral pairing $\langle \bullet, \bullet \rangle$: it has linear generators $X_\gamma$ with $\gamma \in \Gamma$ and relations 
\begin{equation}
X_{\gamma} X_{\gamma'} = \fq^{\langle \gamma, \gamma'\rangle} X_{\gamma + \gamma'}    
\end{equation}
This algebra has an automorphism $\rho_{\CQ}(X_\gamma) = X_{-\gamma}$ and a trace\footnote{The overall normalization is chosen for future convenience.} 
\begin{equation}
    \Tr_Q X_\gamma = (\fq^2)_\infty^{\mathrm{rk}\, \Gamma} \delta_{\gamma,0} \, .
\end{equation} 
We can build a quadruple $(\CQ_\Gamma, \rho_Q, \ell^2(\Gamma), |0\rangle)$ with an action 
\begin{equation}
X_{\gamma}|\gamma'\rangle=\fq^{\langle \gamma, \gamma'\rangle}|\gamma+\gamma'\rangle \, .
\end{equation}
Indeed, 
\begin{equation}
X^\dagger_{\gamma}|\gamma'\rangle=\fq^{\langle \gamma, \gamma'\rangle}|-\gamma+\gamma'\rangle \, .
\end{equation}
We denoted as $|\gamma\rangle$ a basis of $\ell^2(\Gamma)$, normalized as 
\begin{equation}
    \langle \gamma|\gamma'\rangle = (\fq^2)_\infty^{\mathrm{rk}\, \Gamma} \delta_{\gamma,\gamma'}
\end{equation}

As discussed in the Introduction, there are well understood physical and mathematical constructions of algebra embeddings $\CA \to \CQ_\Gamma$
\begin{equation}
    \RG(a) = F_a
\end{equation}
and of a ``quantum spectrum generator'' $S$ such that 
\begin{equation} \label{eq:inter}
    F_a \cdot S = S \cdot \rho_Q^{}(F_{\rho(a)}) \, ,
\end{equation}
for a specific automorphism $\rho$. This is precisely the data required to identify a formal spherical vector $|1\rangle \equiv S |0\rangle$ in $\ell^2(\Gamma)$ as discussed above. 

The conjectural IR formulae for Schur indices \cite{Cordova:2015nma,Cordova:2016uwk} claim that 
\begin{itemize}
    \item The inner products 
    \begin{equation}
        I_{a,b} = \langle a|b\rangle 
    \end{equation} 
    for $|a\rangle \equiv F_a S |0\rangle$ are convergent power series in $\fq$.
    \item $I_{a,b}$ computes specific protected quantities (Schur indices)  in the underlying SQFT. 
\end{itemize}
In particular, these conjectures imply that $|1\rangle \in \ell^2(\Gamma)$ is an actual spherical vector, but do not imply that it is cyclic. 

There is an additional mathematical result whose physical meaning is somewhat unclear: the existence of a {\it tropical basis} for $\CA$. This is a linear basis $\ell_\gamma$ of $\CA$ in bijection with elements in $\Gamma$, such that \begin{equation}\label{Fa-tropexp}
F_{\ell_\gamma}=X_{\gamma}+\sum_{\delta\in\Gamma_+}\Omega_{\ell_\gamma}^{\gamma+\delta}(\fq) X_{\gamma+\delta},
\end{equation}
where $\Gamma_+$ is a positive cone in $\Gamma$ and 
\begin{equation}
    S = 1 + \sum_{\delta \in \Gamma_+} \Sigma_\delta(\fq) X_\delta
\end{equation}
with $\Sigma_\delta(\fq) \in \bZ[[\fq]]$.\footnote{In actual examples, $\Sigma_\delta(\fq)$ resum to rational functions of $\fq$ with poles at rational points on the unit circle only. Furthermore, $\fq \to \fq^{-1}$ maps $S$ to $S^{-1}$. It would be nice to understand this structure in greater detail.} We will denote interchangeably $\gamma'-\gamma \in \Gamma_+$ as $\gamma' \succ \gamma$. The latter notation generalizes better. 

The vectors
\begin{equation}
    |\ell_\gamma \rangle = |\gamma\rangle + \sum_{\delta \in \Gamma_+} \Omega_{\ell_\gamma}^{\gamma+\delta}(\fq) |\gamma+\delta\rangle
\end{equation}
has a triangular form and span $\ell^2(\Gamma)$. We conclude that
$(\CA, \rho, \ell^2(\Gamma),S|0\rangle)$ is a quadruple in the above sense.

We typically have many distinct choices of physically-motivated $\RG$ maps, e.g. different cluster coordinate charts of a quantum cluster variety. As the $I_{a,b}$ should not depend on the choice of $\RG$ map, we learn that the corresponding representations of $\CA$ must be unitarily equivalent, with matching spherical vectors. They must also agree with ``Schur quantization'', which defines $\cH$ directly in terms of the abstract Schur indices $I_{a,b}$ \cite{Gaiotto:2024osr}. 

\subsection{On the domains of definition}
\label{Schwartzspace}

One should note that the operators $W_a$ implementing the $a$ action on $\cH$ are unbounded, in general. This means that they can be defined only on a dense subset of $\cH$. This is a standard feature that needs to be treated with some care. We'd like to point that there is a canonical domain of definition, which can be regarded as an analog of the Schwarz space in our context. To this aim, let us introduce a family of semi-norms
\begin{equation}
\nu_a(\psi):=\big\lVert W_a|\psi\rangle \big\rVert_{\cH}^{},\qquad a\in A.
\end{equation}
Note that the vectors $|b\rangle=W_b|1\rangle$ 
in a spherical  unitary representation have finite semi-norms 
$\nu_a$, as we assumed that $\langle 1|W_a^{\dagger}W_a|1\rangle$ is finite in a spherical representation.
We can therefore introduce the generalised Schwartz space $\CS$ as the completion of the 
dense subspace spanned by the vectors $|b\rangle\in\cH$ with respect to the 
topology defined by the family of semi-norms $\nu_a$. The space $\CS$ is, essentially by 
definition, a common domain for all $W_a$ and $\wt W_a \equiv W_{\rho(a)}^\dagger$.

\subsection{An extended tropical conjecture}
A consequence of the intertwining relations (\ref{eq:inter}) is that \begin{equation}\label{Fa-atropexp}
F_{\rho(\ell_\gamma)}=X_{-\gamma}+\sum_{\delta\in\Gamma_+}\Omega_{\rho(\ell_\gamma)}^{-\gamma-\delta}(\fq) X_{-\gamma-\delta},
\end{equation}
gives a tropical basis with respect to the cone $-\Gamma_+$. In examples, we find that the two tropical bases coincide, so that $\rho(\ell_\gamma) = \ell_{\sigma(\gamma)}$ for some ``tropical spectrum generator'' map $\sigma: \Gamma \to \Gamma$. This relation allows one to read off $\rho$ in situations where only the tropical basis is given. 

Combining the two restrictions, we learn that 
\begin{equation}\label{eq:tropicaldomain}
F_{\ell_\gamma}=X_{\gamma}+\left[\sum_{ - \sigma^{-1}(\gamma)\succ \delta \succ \gamma}\Omega_{\ell_\gamma}^{\delta}(\fq) X_{\delta}\right] + X_{- \sigma^{-1}(\gamma)},
\end{equation}
is supported on a finite collection of charges bounded by $\gamma$ and $- \sigma^{-1}(\gamma)$.

This support condition implies that such a ``doubly-tropical'' basis $F_{\ell_\gamma}$ is only ambiguous up to shifts by $F_{\ell_{\gamma'}}$ such that both $\gamma'$ and $- \sigma^{-1}(\gamma')$ sit in the $\gamma + \Gamma_+ \cap - \sigma^{-1}(\gamma) - \Gamma_+$ intersection. We will encounter both examples of $F_{\ell_\gamma}$ where the definition is unambiguous and ambiguous examples, such as  situations where $0$ belongs to the intersection and thus $X_0$ can be freely added. 

A final experimental observation which we will discuss at length in later Sections is that there is a canonical and physically reasonable choice of a basis $\ell_i$ in $\CA$ such that for all physical definitions of RG map $F$ we can find a labelling $i(\gamma)$ such that $\ell_\gamma = \ell_{i(\gamma)}$ is a doubly-tropical basis. This observation strongly suggests the existence of a physical definition of the $\ell_i$. It would be interesting to find it. 

\subsection{Restoring flavour.}
A slightly more general situation involves a lattice extension  
\begin{equation}
    0\to \Gamma_f \to \Gamma \to \Gamma_g \to 0
\end{equation}
and the pairing $\langle \bullet, \bullet \rangle$ is defined on $\Gamma_g$ and pulled back to $\Gamma$. 

The quantum torus algebra $Q_\Gamma$ now has central elements $X_{\gamma_f}$ for $\gamma_f \in \Gamma_f$. The IR formulae for the Schur index employ an auxiliary vector space which we will denote as $\ell^2(\Gamma_g)$, with a small subtlety: the 
basis vectors $|\gamma\rangle$ are labelled by elements of $\Gamma$ with identification 
\begin{equation}
    |\gamma + \gamma_f\rangle  = \mu_{\gamma_f} |\gamma\rangle
\end{equation}
where $\mu$ is an unitary character for $\Gamma_f$. The action of $\CQ_\Gamma$ is as before and the inner products/Schur indices $I_{a,b}(\fq,\mu)$ are now functions of $\fq$ and $\mu$.  

The output of the construction is thus a family of $\rho$-spherical representations of $\CA$ parameterized by the unitary character $\mu$.

\subsection{Partial RG flow}
Another situation where our construction can be made explicit involves a chain of algebra morphisms
\begin{equation}
    \CA_\UV \to \CA_\IR \to \CQ_\Gamma \, .
\end{equation}
We denote the image of $\CA_\UV$ elements in $\CQ_\Gamma$
as $F^\UV_a$ and the image of $\CA_\IR$ elements as $F^\IR_a$, 
so that 
\begin{equation}
    F^\UV_a = F^\IR_{\RG(a)}
\end{equation}
Suppose that we have quantum spectrum generators $S_\UV$ and $S_\IR$ such that 
\begin{align}
    F^\UV_a \cdot S_\UV &= S_\UV \cdot \rho_Q(F^\UV_{\rho_\UV(a)}) \cr
    F^\IR_a \cdot S_\IR &= S_\IR \cdot \rho_Q(F^\IR_{\rho_\IR(a)})
\end{align}
which we combine into 
\begin{equation}
    F^\IR_{\RG(a)} \cdot S_\UV S_\IR^{-1} = S_\UV S_\IR^{-1} \cdot F^\IR_{\rho_\IR^{-1}(\RG(\rho_\UV(a)))} 
\end{equation}
We can then find a candidate $S$ for the $\CA_\UV \to \CA_\IR$ map by solving the equation
\begin{equation}
    F^\IR_S = S_\UV S_\IR^{-1} \, ,
\end{equation}

Conversely, the relation 
\begin{equation}
    S_\UV = F^\IR_S S_\IR 
\end{equation}
can be sometimes used to compute the spectrum generator $S_\UV$ of a complicated system by first computing $S_\IR$ and $F^\IR$ for a simpler subsystem and then finding $S$. 

This setup arises in specific physical situations we describe below. It leads to a generalization of the IR formulae for Schur indices and to a broad collection of interesting algebraic identities. 

As discussed in the Introduction, $\RG$ and $S$ inherit some tropical properties 
from this multi-step flow, where we define $\ell^\IR_\gamma \succ \ell^\IR_{\gamma'}$ if $\gamma \succ \gamma'$ and then abstract the tropical bases of $\CA_\UV$ and $\CA_\IR$ from the $\Gamma$ labelling.

\subsection{Cluster Structure}
\label{introcluster}
A general phenomenon frequently occurs: each $\CA \to \CQ_\Gamma$ ``quantum chart'' can undergo mutations $\mu_{\gamma_i}$ (and $\mu^{-1}_{-\gamma_i}$) for some specific finite collection of charges $\gamma_i$, which have the property that the condition $\gamma \succ \gamma'$ can be refined in all tropical formulae to $\gamma-\gamma'$ being a non-negative linear combination of the $\gamma_i$'s.

The resulting quantum charts have the same property with new set of charges 
\begin{equation}
    \mu^t_i:\qquad \gamma_i \to - \gamma_i \qquad \qquad \gamma_j \to \gamma_j + \mathrm{max}(\langle \gamma_j, \gamma_i\rangle,0) \gamma_i
\end{equation}
Such situations have a close relationship to the theory of cluster algebras, cluster varieties and their quantization.  The relation between Schur quantization and IR formulae for the Schur index for such theories can be encoded as a general strategy to quantize cluster varieties as complex phase spaces. We will see an importance instance of this in section when we discuss the Quantization of Teichm\"uller space in Section \ref{sec:teichquant}.

In a cluster algebra perspective, one focuses on the $x_i \equiv X_{\gamma_i(u)}$
generators and the integers $\langle \gamma_i(u), \gamma_j(u)\rangle$. Define 
\begin{equation}
    |i \to j| = \mathrm{max}(\langle \gamma_i, \gamma_j\rangle,0) 
\end{equation}
which can be conveniently interpreted as the number of arrows in an auxiliary ``BPS quiver'' \cite{ Cecotti:2011rv,Alim:2011kw,Alim:2011ae,Lee:2012naa,Gabella:2017hpz,derksen2008quiverspotentialsrepresentationsi,Cirafici:2013bha,Cecotti:2014zga,Cecotti:2010fi}.

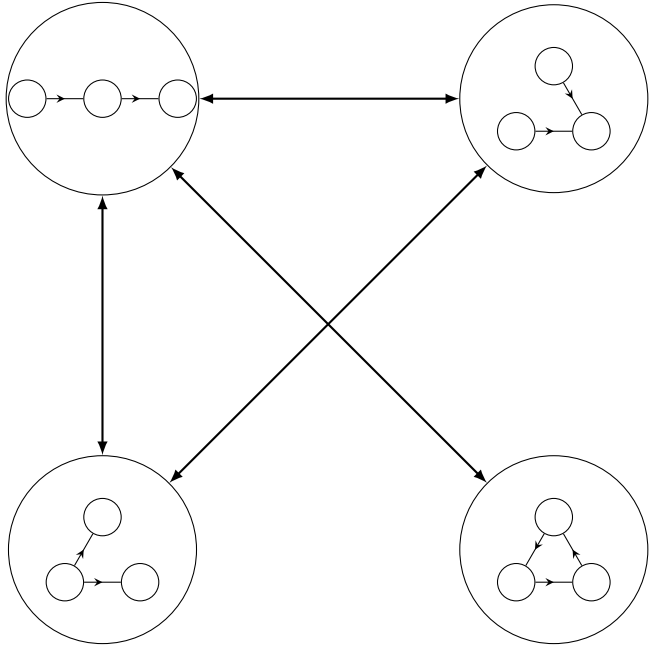
\begin{figure}
\centering
\begin{tikzpicture}[scale=1.5, every node/.style={scale=1}]
% Styles for small quivers and arrows
\tikzstyle{smallquiver}=[circle, draw, minimum size=2.5cm, inner sep=0pt]
\tikzstyle{vertex}=[circle, draw, inner sep=5pt, fill=white, minimum size=.5mm, font=\tiny]

% Define small quivers as nodes
\node[smallquiver] (q1) at (0,0) {
    \begin{tikzpicture}
        \node[vertex] (a0) at (0,0) {};
        \node[vertex] (a1) at (1,0) {};
        \node[vertex] (a2) at (2,0) {};
        \draw[midarrow] (a0) -- (a1);
        \draw[midarrow] (a1) -- (a2);
    \end{tikzpicture}
};
\node[smallquiver] (q2) at (4,0) {
    \begin{tikzpicture}
        \node[vertex] (b0) at (0,0) {};
        \node[vertex] (b1) at (1,0) {};
        \node[vertex] (b2) at (0.5,0.866) {};
        \draw[midarrow] (b0) -- (b1);
        \draw[midarrow] (b2) -- (b1);
    \end{tikzpicture}
};
\node[smallquiver] (q3) at (0,-4) {
    \begin{tikzpicture}
        \node[vertex] (c0) at (0,0) {};
        \node[vertex] (c1) at (1,0) {};
        \node[vertex] (c2) at (0.5,0.866) {};
        \draw[midarrow] (c0) -- (c1);
        \draw[midarrow] (c0) -- (c2);
    \end{tikzpicture}
};
\node[smallquiver] (q4) at (4,-4) {
    \begin{tikzpicture}
        \node[vertex] (d0) at (0,0) {};
        \node[vertex] (d1) at (1,0) {};
        \node[vertex] (d2) at (0.5,0.866) {};
        \draw[midarrow] (d0) -- (d1);
        \draw[midarrow] (d1) -- (d2);
        \draw[midarrow] (d2) -- (d0);
    \end{tikzpicture}
};

% Large quiver arrows with bidirectional style
\tikzstyle{largequiverarrow}=[<->, >=latex, thick]

% Draw two-way arrows between specified quiver-nodes
\draw[largequiverarrow] (q1) -- (q2) node[midway, above] {};
\draw[largequiverarrow] (q1) -- (q3) node[midway, left] {};
\draw[largequiverarrow] (q1) -- (q4) node[midway] {};
\draw[largequiverarrow] (q2) -- (q3) node[midway] {};

\end{tikzpicture}
\caption{The mutation orbit of the $A_3$ BPS quiver. Figure from \cite{Gaiotto:2024fso}.}\label{fig:A3orbit}
\end{figure}

The actual embedding of the charges in an overall lattice $\Gamma$ is forgotten. This allows one to disregard the difference between $\mu_{\gamma_i(u)}$ and $\mu_{-\gamma_i(u)}^{-1}$, which are unified into the notion of mutation of the BPS quiver:
\begin{enumerate}
    \item Add arrows between nodes connected to the $i$-th node:
    \begin{equation}
        |j\to k| \to |j\to k|+|j\to i||i\to k| \,.
    \end{equation}
    \item Remove pairs of edges with opposite orientation until $|j \to k||k \to j|=0$.
    \item Flip the edges connected to the $i$-th node: $|i \to j|\leftrightarrow |j \to i|$.
\end{enumerate}
Some BPS quivers generate finite orbits under mutations, see e.g. Figure \ref{fig:A3orbit}, but most have infinite orbits. Sometimes, the $\gamma_i$ do not generate $\Gamma$ and we need to keep track of some extra generators. These are usually denoted as ``frozen'' nodes. We will denote them as dashed in figures. 

The spectrum generator and functions $F_a$ are not included in the cluster data, but there are multiple strategies to recover them which we review in the next Section. 

For now, we describe a simple strategy which requires a certain amount of guesswork \cite{keller2011clustertheoryquantumdilogarithm}. One can simply restore the $\Gamma$ lattice (as the span of the $\gamma_i$) and apply mutations (but not inverse mutations) while keeping track of the charges. If we can find a sequence of mutations $\mu_{\gamma(k)}$ after which $\gamma_i \to - \gamma_i$, then we can postulate
\begin{equation}
    S=E_\fq(X_{\gamma(1)}) E_\fq(X_{\gamma(2)}) \cdots 
\end{equation}
so that the sequence of mutations transforms it into 
\begin{equation}
    S=E_\fq(X_{-\gamma(1)}) E_\fq(X_{-\gamma(2)}) \cdots  \, .
\end{equation}
We can then recover the $F_a$ from $S$. This is facilitated by knowing the action of mutation on tropical labels. In particular, knowing both the ``smallest'' and the ``largest'' monomial in $F_{\ell_\gamma}$ leaves a finite collection of unknown coefficients in each.

Another important expected property is that the coefficients in $F_{\ell_\gamma}$
are positive integral linear combinations of $\fq$-integers 
\begin{equation}
    [n]_\fq = \fq^{n-1} + \fq^{n-3}+ \cdots + \fq^{1-n} \, .
\end{equation}

In many physically relevant examples, the BPS quiver is available but 
it does not allow for a definition of $S$ as a finite sequence of mutations. Nevertheless, it is believe that the data of the BPS quiver determines uniquely the whole structure of $S$ and $F$'s. We will come back to this in later Sections.

\section{A minimal sequence of examples: rank 2 lattice}
\label{sec:rank2}
We now discuss a few examples based on the rank $2$ lattice $\Gamma = \bZ^2$, with inner product
\begin{equation}
    \langle (a,b),(c,d) \rangle = a d - b c
\end{equation}

\subsection{The minimal quantum torus algebra}
The first example is the quantum torus algebra $\CQ_\Gamma$ itself, the algebra associated to pure $U(1)$ gauge theory. A minimal presentation has generators $u$, $u^{-1}$, $v$, $v^{-1}$ and relation 
\begin{equation}
    u v = \fq^2 v u \, .
\end{equation}
The 
\begin{equation}
    X_{a,b}\equiv \fq^{-ab} u^a v^b
\end{equation}
linear basis makes manifest the $SL(2,\bZ)$ automorphisms inherited from $\Gamma$. Also, $\rho_Q(X_{a,b}) = X_{-a,-b}$.

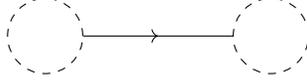
\begin{figure}
    \centering
    \begin{tikzpicture}
        % Nodes
        \node (A) at (0,0) [circle, draw, dashed, minimum size=1cm]{};
        \node (B) at (3,0) [circle, draw, dashed, minimum size=1cm]{};

        % Arrow
         \draw[postaction={decorate},
              decoration={markings, mark=at position 0.5 with {\arrow{>}}}] 
              (A) -- (B);
    \end{tikzpicture}
    \caption{A BPS quiver diagram for a pure $U(1)$ gauge theory has no nodes, as the RG flow is trivial, $S=1$ and there are no mutations available. We thus include frozen nodes (dashed) with generators $(1,0)$ and $(0,1)$.}
    \label{fig:quiver1}
\end{figure}

A standard way to represent $\CQ_\Gamma$ is to give $v$ as a multiplication operator and $u$ as a shift operator. E.g. we can act on $L^2(S^1) \otimes \ell^2(\bZ)$ by
\begin{align}
    v\cdot \psi_m(\zeta) &= \fq^{-m} \zeta \psi_m(\zeta) \cr
    u\cdot \psi_m(\zeta) &= \psi_{m-1}(\fq \zeta) \, .
\end{align}
so that $[u,v^\dagger]=0$, etc. This is the representation which naturally arises in localization calculations of Schur indices. 

It is convenient to Fourier-transform along $S^1$ to recover the $\rho_Q$-spherical representation on $\ell^2(\bZ^2)$. Introducing orthogonal basis elements $|a,b\rangle$ supported on $(a,b)$ with norm $(\fq^2)_\infty^2$,
we have 
\begin{align}
    u |a,b\rangle &= \fq^b |a+1,b\rangle \cr
    v |a,b\rangle &= \fq^{-a} |a,b+1\rangle
\end{align}
Essentially by definition, 
\begin{equation}
    |a,b\rangle = X_{a,b} |0,0\rangle = X_{-a,-b}^\dagger |0,0\rangle
\end{equation}
and $|0,0 \rangle$ is the spherical vector. 

\subsubsection{The $\fq$-Weyl algebra}
The next example is (one of the possible definition of) the $\fq$-deformed Weyl algebra $\CW$, the algebra associated to the theory SQED$_1$, aka $T[U(1),T^*\bC]$. It has generators $u_\pm$, $v$ and $v^{-1}$, with 
\begin{align}
    u_+ v &= \fq^2 v u_+ \cr
    u_- v &= \fq^{-2} v u_- \cr
    u_+ u_- &= 1 + \fq v \cr
    u_- u_+ &= 1+ \fq^{-1} v
\end{align}
The analogy with the Weyl algebra arises from the relation 
\begin{equation}
    \fq^{-1} u_+ u_- - \fq u_- u_+ = \fq^{-1}- \fq \, .
\end{equation}
The algebra has a linear basis consisting of 
\begin{equation}
    D_{a,b} \equiv \fq^{- ab} u_+^a v^b \qquad \qquad D_{-a,b} \equiv \fq^{ab} u_-^a v^b \, ,
\end{equation}
with $a \geq 0$, and Serre automorphism 
\begin{equation}
    \rho(D_{a,b}) = D_{-a,-a-b} \qquad \qquad \rho(D_{-a,b}) = D_{a,-b} \, ,
\end{equation}
with $a \geq 0$.

Localization calculations naturally lead to multiplication-and-shift representations on $L^2(S^1) \otimes \ell^2(\bZ)$. Either
\begin{align}
    v\cdot \psi_m(\zeta) &= \fq^{-m} \zeta \psi_m(\zeta) \cr
    u_+\cdot \psi_m(\zeta) &= \psi_{m-1}(\fq \zeta) \, .\cr
    u_-\cdot \psi_m(\zeta) &= (1 + \fq^{-m-1} \zeta)\psi_{m+1}(\fq^{-1} \zeta) \, ,
\end{align}
or 
\begin{align}
    v\cdot \psi_m(\zeta) &= \fq^{-m} \zeta \psi_m(\zeta) \cr
    u_+\cdot \psi_m(\zeta) &= (1 + \fq^{-m+1} \zeta)\psi_{m-1}(\fq \zeta) \, .\cr
    u_-\cdot \psi_m(\zeta) &= \psi_{m+1}(\fq^{-1} \zeta) \, ,
\end{align}
These two representations are related by an unitary map
\begin{equation} \label{eq:unitary}
    \psi_m(\zeta) \to \left[\prod_{k=0}^\infty \frac{1 + \fq^{2 k + m + 1}\zeta^{-1}}{1 + \fq^{2 k + m + 1}\zeta }\right]\zeta^{-m} \psi_m(\zeta) \, .
\end{equation}

These representations have a spherical vector. In the first representation, it can be taken to be the wavefunction 
\begin{equation}
    \psi[1]_m(\zeta) \equiv \frac{(\fq^2)_\infty \delta_{m,0}}{\prod_{k=0}^\infty (1 + \fq^{2k+1} \zeta)} \, ,
\end{equation}
We can then compute
\begin{align}
    \psi[D_{a,b}]_m(\zeta) &= \frac{(\fq^2)_\infty \zeta^b \delta_{m,a}}{\prod_{k=0}^\infty (1 + \fq^{2k+1+a} \zeta)} \cr
    \psi[D_{-a,b}]_m(\zeta) &= \frac{(\fq^2)_\infty \zeta^b \delta_{m,-a}}{\prod_{k=0}^\infty (1 + \fq^{2k+1+a} \zeta)} \, ,
\end{align}
with $a \geq 0$. These are examples of vectors in $\CU(\CW)$. 
In the second representation, 
\begin{equation}
    \psi[1]_m(\zeta) \equiv \frac{(\fq^2)_\infty \delta_{m,0}}{\prod_{k=0}^\infty (1 + \fq^{2k+1} \zeta^{-1})} \, .
\end{equation}

As mentioned, these representations are already essentially presented as special cases of the general formalism. In the first case, we write 
\begin{equation}
    F_v = X_{0,1} \qquad \qquad F_{u_+} = X_{1,0} \qquad \qquad F_{u_-} = X_{-1,0}+ X_{-1,1}
\end{equation}
and 
\begin{equation}
    S = E_\fq(X_{0,1})
\end{equation}
with the``bosonic'' quantum dilogarithm defined as 
\begin{equation}
    E_\fq(x) \equiv (-\fq x;\fq^2)^{-1}_\infty = \frac{1}{\prod_{n=0}^\infty (1+\fq^{2n+1} x)} = \sum_{n\geq 0} \frac{(-\fq)^n}{(\fq^2)_n} x^n
\end{equation}

In the second we write 
\begin{equation}
    F_v = X_{0,1} \qquad \qquad F_{u_+} = X_{1,1} +  X_{1,0}\qquad \qquad F_{u_-} = X_{-1,0}
\end{equation}
and 
\begin{equation}
    S = E_\fq(X_{0,-1})
\end{equation}

\begin{figure}[h]
    \centering
    \begin{tikzpicture}
        % First quiver
        \node (A1) at (0,0) [circle, draw, dashed, minimum size=1cm] {};
        \node (B1) at (3,0) [circle, draw, minimum size=1cm] {};
        \draw[postaction={decorate},
              decoration={markings, mark=at position 0.5 with {\arrow{>}}}] 
              (A1) -- (B1);

        % Second quiver (shifted right)
        \node (A2) at (6,0) [circle, draw, dashed, minimum size=1cm] {};
        \node (B2) at (9,0) [circle, draw, minimum size=1cm] {};
        \draw[postaction={decorate},
              decoration={markings, mark=at position 0.5 with {\arrow{>}}}] 
              (B2) -- (A2);
    \end{tikzpicture}
    \caption{In this example we have two possible RG flows, exchanged by mutations. A BPS quiver for the first RG flow (left) has a normal node for the mutable charge $\gamma_1 = (0,1)$ and a frozen node to keep track of the second generator $(1,0)$ in $\Gamma$. An mutation at $\gamma_1$ produces the second quiver, associated to the mutable charge $(0,-1)$. This is a way to derive $S$ in this example.}
    \label{fig:quiver2}
\end{figure}
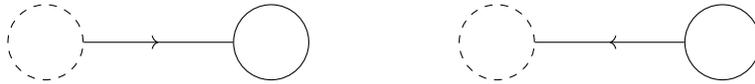

\subsubsection{The Pentagon algebra}
We are ready for the third example, associated to the $[A_1,A_2]$ Argyres-Douglas theory. We will denote it as the Pentagon algebra $\CP$. This algebra has five generators, which we denote as $L_i$ with $L_i = L_{i+5}$ to make a $\bZ_5$ symmetry manifest: 
\begin{align}
    L_{i+1} L_i &= \fq^2 L_i L_{i+1} \cr
    L_{i+1} L_{i-1} &= 1+ \fq L_i \cr
    L_{i-1} L_{i+1} &= 1+ \fq^{-1} L_i
\end{align}
The full linear basis for $\CP$ is given by 
\begin{equation}
    D_{i;a,b} \equiv \fq^{ab} L_i^a L_{i+1}^b
\end{equation}
with an obvious $\bZ_5$ action. Finally, $\rho(L_i) = L_{i+2}$.

In this example, localization formulae for Schur indices are not available and we do not have a ready-made $\rho$-spherical representation for $\CP$. We can instead apply our formalism to the five known cluster parameterizations $\CP \to \CQ_{\bZ^2}$, intertwined by the $\bZ_5$ action. Later on, we will recover this from a partial RG flow of the theory $T[SU(2),T^* \bC^2]$. We have 
\begin{align}
    F_{L_0} &= X_{1,0} \cr
    F_{L_1} &= X_{0,-1} \cr
    F_{L_2} &= X_{-1,-1} + X_{-1,0}\cr
    F_{L_3} &= X_{-1,0} + X_{-1,1} + X_{0,1} \cr
    F_{L_4} &= X_{0,1} + X_{1,1} 
\end{align}
We already ordered the monomials in a manner which anticipates the tropical labeling appropriate for this parameterization: $\gamma_1 = (1,0) \succ 0$ and $\gamma_2 = (0,1) \succ 0$. Note that in this example we do not need to specify a full choice of positive cone $\Gamma_+$: terms in a sum will always have charges which differ by a non-negative integral combination of ``mutable'' charges such as $\gamma_1$ and $\gamma_2$. This is a hallmark of systems with a cluster description. 

\begin{figure}[h]
    \centering
    \begin{tikzpicture}
        % Nodes
        \node (A) at (0,0) [circle, draw, minimum size=1cm]{};
        \node (B) at (3,0) [circle, draw, minimum size=1cm]{};

        % Arrow
         \draw[postaction={decorate},
              decoration={markings, mark=at position 0.5 with {\arrow{>}}}] 
              (A) -- (B);
    \end{tikzpicture}
    \caption{The BPS quiver diagram now has two standard nodes of charges $\gamma_1=(1,0)$ and $\gamma_2 = (0,1)$. A mutation at the first node gives charges $(-1,0)$ and $(0,1)$, which can be mutated further at the second node to get $(-1,0)$ and $(0,-1)$. This is one way to compute $S$. If we start with a mutation at the second node we get $(1,1)$ and $(0,-1)$, then $(-1,-1)$ and $(1,0)$ and then $(0,-1)$ and $(-1,0)$, explaining the alternative form for $S$.}
    \label{fig:quiver3}
\end{figure}
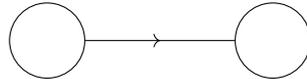

We read off the tropical labels: 
\begin{align}
    L_0 &= \ell_{1,0} = \rho(\ell_{-1,0}) \cr
    L_1 &= \ell_{0,-1} = \rho(\ell_{0,1}) \cr
    L_2 &= \ell_{-1,-1} =\rho(\ell_{1,0})\cr
    L_3 &= \ell_{-1,0} = \rho(\ell_{0,-1}) \cr
    L_4 &= \ell_{0,1} = \rho(\ell_{-1,-1}) 
\end{align}
It is not hard to verify that the $D_{i;a,b}$ form the tropical basis, decomposing $\bZ^2$ into five conical sectors. 

The spectrum generator takes a simple form:
\begin{equation}
    S = E_\fq(X_{\gamma_1})E_\fq(X_{\gamma_2}) 
\end{equation}
Famously, the spectrum generator can also be written as 
\begin{equation}
    S = E_\fq(X_{\gamma_2})E_\fq(X_{\gamma_1 + \gamma_2})E_\fq(X_{\gamma_1}) 
\end{equation}
thanks to the pentagon identity for quantum dilogarithms.

We can verify the intertwining relations $L_i S = S \rho_{\CQ}(L_{i+2})$ step-by-step, using the basic properties of the quantum dilogarithm: 
\begin{align}
    E_\fq(X_\gamma) X_{\gamma'} &= \sum_{k=0}^{\langle \gamma, \gamma'\rangle} {\langle \gamma, \gamma'\rangle \choose k}_\fq X_{\gamma'+k \gamma}E_\fq(X_\gamma) \qquad \langle \gamma, \gamma'\rangle \geq 0 \cr
     X_{\gamma'} E_\fq(X_\gamma)&= \sum_{k=0}^{\langle \gamma', \gamma\rangle} {\langle \gamma', \gamma\rangle \choose k}_\fq E_\fq(X_\gamma) X_{\gamma'+k \gamma} \qquad \langle \gamma, \gamma'\rangle \leq 0 
\end{align}
We leave this as an exercise for the reader.

We are now in position to describe the $\rho$-spherical representation of $\CP$ on $\ell^2(\bZ^2)$. The spherical vector is  
\begin{equation}
      |1\rangle \equiv S|0\rangle = (\fq^2)_\infty \sum_{n,m\geq 0} \frac{(-1)^{n+m} \fq^{n+m+n m}}{(\fq^2)_n(\fq^2)_m} |n,m\rangle 
\end{equation}
It has finite norm in the desired range $-1<\fq<1$:
\begin{equation}
    I(\fq) \equiv \langle 1|1\rangle = (\fq^2)^2_\infty \sum_{n,m\geq 0} \frac{\fq^{2n+2m+2n m}}{(\fq^2)^2_n(\fq^2)^2_m}
\end{equation}
The same is true of the other states $D_{i;a,b}|1\rangle$, but we should stress that this statement is not immediately obvious. It is clearly true for 
\begin{equation}
      |D_{0;a,b}\rangle = (\fq^2)_\infty \sum_{n,m\geq 0} \frac{(-1)^{n+m} \fq^{n+m+n m}}{(\fq^2)_n(\fq^2)_m} \fq^{a m + b n}|a+n,-b+m\rangle 
\end{equation}
but it would not be the case for a generic $X_{a,b} |1\rangle$ if $a>0$ or $b<0$. 

Normalizability of all the $|a\rangle$ states can be reduced to finiteness of the expectation value
\begin{equation}
    |\langle 1|a|1\rangle|<\infty
\end{equation}
which is invariant under the action of $\rho$ thanks to the 
intertwining property. Hence finiteness of $\langle 1|D_{0;a,b}\rangle$ is thus sufficient to prove finiteness of all inner products $\langle a|b\rangle$.

Notice that $\rho$ also intertwines the five different choices of cluster map $\CP \to \CQ_{\bZ^2}$. We immediately learn that these give equivalent $\rho$-spherical unitary representations of $\CP$, matched by identifying the $|a\rangle$ vectors in each of them. The unitary equivalence can be reduced to the action of transformations of the form (\ref{eq:unitary}). 

Each of the $D_{i;a,b}$ can be mapped to $D_{0;a,b}$ by some power of $\rho$ and then diagonalized by a Fourier transform, with a spectrum of the form $e^{i\phi} \fq^k$ for integer $k$. 

We are now ready to discuss our first example of multi-step RG flow: $\CP \to \CW \to \CQ_{\bZ^2}$. We can identify 
\begin{align}
    S_\UV &= E_\fq(X_{\gamma_1})E_\fq(X_{\gamma_2}) \cr
    S_{\IR} &= E_\fq(X_{\gamma_2}) \cr
    S &= E_\fq(u_+)
\end{align}
Comparing the $F$'s for the two theories gives the $\RG$ map:
\begin{align}
    \RG(L_0) &= u_+ \cr
    \RG(L_1) &= v^{-1} \cr
    \RG(L_2) &= \fq v^{-1} u_-\cr
    \RG(L_3) &= u_- + v \cr
    \RG(L_4) &= v + \fq v u_+ 
\end{align}
We also get a tropical structure on $\CW$ adapted to this: $v \succ 0$, $u_+ \succ 0$, $v \succ u_-$, etc. For every $D_{a,b}$ in $\CA_\IR$, we can find 
a $D_{i;c,d}$ such that $\RG(D_{i;c,d}) = D_{a,b} + \cdots$ where the ellipsis 
is ``larger'' than $D_{a,b}$.

\subsection[The last rank $2$ example: $\su(2)$ Seiberg-Witten theory]{\boldmath The last rank $2$ example: $\su(2)$ Seiberg-Witten theory}
This example has a small subtlety: there really are two closely related theories, $T[SU(2),0]$ and $T[SO(3),0]$, and two corresponding algebras $\CA_{SU(2)}$ and $\CA_{SO(3)}$. These share a large sub-algebra and are themselves sub-algebras of a larger algebra $\CA$ defined over $\bZ[\fq^{\frac12}, \fq^{-\frac12}]$. We will employ $\CA$ through the Section and restrict to the sub-algebras $\CA_{SU(2)}$ and $\CA_{SO(3)}$ only if strictly necessary, at the price of restricting $\fq>0$. 

A crucial role in the story is played by a special element of the quantum torus algebra, the Wilson line 
\begin{equation}
    F_{w_1} \equiv X_{0,-1}+ X_{-1,1} + X_{0,1}
\end{equation}
which is fixed by $\rho$. Together with 
\begin{equation}
    F_{H_0} \equiv X_{\frac12,0}
\end{equation}
it generates the image of $\CA$ in an extended quantum torus algebra $\CQ_{\frac12\bZ \oplus \bZ}$. In particular, one finds (the image of) a sequence of increasingly complicated operators $H_i$ which satisfy
\begin{equation}
    w_1 H_i = \fq^{\frac12} H_{i-1} + \fq^{-\frac12} H_{i+1} \qquad \qquad H_i w_1 =\fq^{-\frac12} H_{i-1} + \fq^{\frac12} H_{i+1}
\end{equation}
and thus can be obtained from repeated $\fq^{\frac12}$-commutators with $w_1$:
\begin{align}
    \cdots &= \cdots \cr
    F_{H_{-3}} &= X_{\frac12,-3}+(\fq + \fq^{-1})X_{-\frac12,-1}+ X_{-\frac32,1} + X_{-\frac12,1} \cr
    F_{H_{-2}} &= X_{\frac12,-2}+X_{-\frac12,0} \cr
    F_{H_{-1}} &= X_{\frac12,-1} \cr
    F_{H_0} &= X_{\frac12,0} \cr
    F_{H_1} &= X_{-\frac12,1}+ X_{\frac12,1} \cr
    F_{H_2} &= X_{-\frac12,0}+ X_{-\frac32,2}+ (\fq + \fq^{-1}) X_{-\frac12,2}+X_{\frac12,2} \cr
    \cdots &= \cdots 
\end{align}
These are part of a tropical basis with $(1,0) \succ 0$ and $(-1,2) \succ 0$. The first monomial is the ``smallest'', giving the tropical charge. Looking at the ``largest'' monomials, we recognize $\rho(H_a) = H_{a-2}$. The transformation $H_a \to H_{a+1}$ generates a $\bZ$ symmetry of the algebra. 

The $H_a$ themselves satisfy quadratic relationships:
\begin{align}
    H_{i-1} H_i &= \fq H_i H_{i-1} \cr
    H_{i-1} H_{i+1} &= 1 + \fq H_i^2 \cr
    H_{i+1} H_{i-1} &= 1 + \fq^{-1} H_i^2 \cr
    H_{i-2} H_{i+1} &= \fq^{-\frac12} w_1 + \fq H_{i-1} H_i \cr
    H_{i+1} H_{i-2} &= \fq^{\frac12} w_1 + \fq^{-1} H_i H_{i-1} 
\end{align}
etcetera. More generally, a linear basis for the whole algebra consists of $w_a$ polynomials in $w_1$ defined by
\begin{equation}
    w_1 w_a =w_{a+1}+ w_{a-1}
\end{equation}
together with
\begin{equation}
    D_{i;a,b} \equiv \fq^{-\frac{a b}{2}}H_{i-1}^a H_i^b
\end{equation}
It is not hard to verify that the tropical charges of these basis elements cover $\Gamma$ exactly once. 

The spectrum generator is known explicitly:
\begin{equation}
    S = E_\fq(X_{1,0}) E_\fq(X_{-1,2})
\end{equation}
notice that the charges $\gamma_1 = (1,0)$ and $\gamma_2 = (-1,2)$ have pairing $2$. These will be the mutable charges. It is easy to check that the intertwining relations hold for $w_1$ and $H_0$, and thus for all elements in the algebra. 

\begin{figure}[h]
    \centering
    \begin{tikzpicture}
        % Nodes
        \node (A) at (0,0) [circle, draw, minimum size=1cm]{};
        \node (B) at (3,0) [circle, draw, minimum size=1cm]{};

        % Arrow
         \draw[postaction={decorate},
              decoration={markings, 
              mark=at position 0.45 with {\arrow{>}},
              mark=at position 0.55 with {\arrow{>}}}] 
              (A) -- (B);
        \draw[postaction={decorate},
              decoration={markings, 
              mark=at position 0.45 with {\arrow{>}},
              mark=at position 0.55 with {\arrow{>}}}] 
              (A) -- (B);
    \end{tikzpicture}
    \caption{The BPS quiver diagram has two standard nodes of charges $\gamma_1=(1,0)$ and $\gamma_2 = (-1,2)$, with two arrows between them. Either mutation flips the direction of the arrow, leading to an isomorphic quiver. A mutation at the first node followed by the second node flips the charges and explains the form of $S$ in the text.}
    \label{fig:quiver4}
\end{figure}
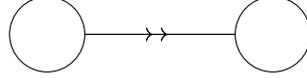

The corresponding state in $\ell^2(\Gamma)$ becomes 
\begin{equation}
    |1\rangle =  (\fq^2)_\infty \sum_{n,m\geq 0} \frac{(-1)^{n+m} \fq^{n+m+2 n m}}{(\fq^2)_n(\fq^2)_m} |n,m\rangle
\end{equation}
The Schur index is thus given by the convergent power series
\begin{equation}
    I(\fq) = (\fq^2)^2_\infty \sum_{n,m\geq 0} \frac{\fq^{2n+2m+4n m}}{(\fq^2)^2_n(\fq^2)^2_m} 
\end{equation}
which experimentally coincides with the UV gauge theory answer 
\begin{equation}
    I(\fq) = 1+\fq^4 + \fq^{12} + \fq^{24}+\fq^{40} + \cdots= \sum_{k=0}^\infty \fq^{2 k (k+1)} 
\end{equation}
Finiteness of $\langle 1|D_{i;a,b}|1\rangle$ is not immediately obvious, but finiteness of $\langle 1|D_{0;a,b}|1\rangle$ is. An argument based on $\rho$-invariance then buys finiteness of $\langle 1|D_{2i;a,b}|1\rangle$. Odd labels require a bit more brute-force work. Also, the presence of factors of $\fq^{\frac12}$ when working with $\CA$ means that we can only work in the  $0<\fq<1$ range until we restrict to 
\begin{itemize}
        \item $\CA_{\mathrm{SU}(2)}$ consisting of all $w_a$ and $D_{i;2a,b}$, with tropical charges of the form $(n,m)$ defining the integral lattice $\Gamma_{\mathrm{SU}(2)}$.
        \item $\CA_{\mathrm{SO}(3)}$ consisting of $w_{2a}$ and $D_{i;a,2b}$, with tropical charges of the form $(n/2,2m)$ defining the integral lattice $\Gamma_{\mathrm{SO}(3)}$.
\end{itemize}
In any case, we obtain a $\rho$-spherical unitary representation of these algebras on $\ell^2(\Gamma_{\mathrm{SU}(2)})$ or $\ell^2(\Gamma_{\mathrm{SO}(3)})$ respectively. 

Differently from the case of $\CP$, where all basis elements could be mapped to a monomial by one of the cluster maps and then diagonalized by a Fourier transform, the $w_1$ operator is special here and its spectrum is more interesting. We will now argue that the spectrum takes the form 
\begin{equation}
\label{spectrumwilsonline}
    \lambda + \lambda^{-1} \qquad\qquad \lambda= e^{i \phi} \fq^k \, .
\end{equation}
for $SU(2)$ gauge theory. For $SO(3)$, $w_2$ has spectrum of the form 
\begin{equation}
\label{spectrumwilsonline2}
    \lambda + 1+ \lambda^{-1} \qquad\qquad \lambda= e^{i \phi} \fq^k \, .
\end{equation}

It is not difficult to find some formal eigenfunctions. Observe that 
\begin{equation}
    F[w_1] - \lambda - \lambda^{-1} = X_{-1,1} + X_{0,1} (1 - \lambda X_{0,-1})(1 - \lambda^{-1} X_{0,-1})\, .
\end{equation}
In the Fourier-transformed presentation with diagonalized $X_{1,0}$ we can write 
a test wave-function
\begin{equation}
\label{eigenfunction}
    \psi_m(\zeta;\lambda) = (-\zeta)^{m-1} \frac{(\fq^{2+m} \lambda^\dagger\zeta;\fq^2)_\infty (\fq^{2+m} (\lambda^\dagger)^{-1} \zeta;\fq^2)_\infty}{(\fq^m \lambda \zeta^{-1};\fq^2)_\infty (\fq^m \lambda^{-1} \zeta^{-1};\fq^2)_\infty}
\end{equation}
Formally, this is annihilated by both $F[w_1]- \lambda - \lambda^{-1}$ and its Hermitean conjugate. We really have two such solution, one supported at integral $m$, the other at half-integral $m$. in the representation over  $\bL^2(\bZ \times S^1)$ $F[w_1]$ acts as the difference operator
 \begin{equation}\label{eq:intham}
F[w_1]\, g_{n}(\zeta) =g_{n+1}(\fq \zeta ) +  g_{n-1}(\fq^{-1} \zeta) + \fq^{-n} \zeta g_n(\zeta)\comma
\end{equation}
To prove that this distributional wave function solves the spectral problem  we note that \eqref{eq:intham} coincides  with the difference operator associated to the trace of the Lax Operator of (discrete) Quantum Liouville theory.  In Appendix \ref{spectralproblem}, we exploit this  connection to the Liouville spectral problem to provide a proof for the spectrum \eqref{spectrumwilsonline} and for the formal solution \eqref{eigenfunction}.
Strictly speaking, one should test the distributional eigenfunction $\psi$ against the $|a\rangle$ vectors. This is in general challenging, but it follows from the basic test against the spherical vector: 
\begin{equation}
    \langle \psi(\fq^n e^{i \theta}) |1\rangle = \cdots \delta_{n,0} (\fq^2 e^{2 i \theta};\fq^2)_\infty (\fq^2 e^{-2 i \theta};\fq^2)_\infty \comma
\end{equation}
as the intertwining relations for the 't Hooft operators allow one to compute more general $\langle \psi(\fq^n e^{i \theta}) |a\rangle$ inner products.
Conversely one can express: 
\begin{equation}
   |1\rangle = \frac{1}{4 \pi} \int_0^{2 \pi} d \theta \,  (\fq^2;\fq^2)_\infty (e^{2 i \theta};\fq^2)_\infty(e^{-2 i \theta};\fq^2)_\infty \psi_m(\zeta;e^{i \theta}) \period
\end{equation}
Covariance of the $\psi$ kernel under mutations is a remarkable identity which can be reduced to the unitary pentagon identity. See \cite{Dimofte:2013lba} for a discussion and a geometric construction for many useful kernels associated to full or partial RG flows in class $S$ theories labelled by $\sl_2$. The reference defines 3d ${\cal N}=2$ theories whose superconformal indices \cite{Imamura:2011su} give the desired kernels \cite{Dimofte:2011py}.\footnote{Ellipsoid partition functions \cite{Hama:2011ea} give kernels for Teichm\"uller quantization \cite{Dimofte:2011ju}. }

\subsubsection{Weak coupling factorization}
In later examples, we will need a remarkable alternative factorization of $S$:
\begin{equation}
   E_\fq(X_{-1,2})E_\fq(X_{-1,4}) \cdots E^{-1}_\fq(-\fq^{-1} X_{0,2})E^{-1}_\fq(-\fq X_{0,2})\cdots E_\fq(X_{1,4})E_\fq(X_{1,2})E_\fq(X_{1,0})
\end{equation}
which employs two semi-infinite sequences of quantum dilogs sandwiching an extra contribution
\begin{equation}
    (X_{0,2};\fq^2)_\infty  (\fq^2 X_{0,2};\fq^2)_\infty 
\end{equation}
which is formally identical to (twice!) the integration measure in the UV definition of the Schur index. This pattern persists for other $SU(2)$ gauge theories. We will come back to this in later Sections.

This factorized expression for $S$ is rather ``dangerous'', in the sense that it involves heavy concellations between terms with arbitrarily negative powers of $\fq$. The same factorization for $S^{-1}$ is safer, as it only involve positive terms and no cancellations. 

\subsubsection{Some extra formulae}
In later Sections, we will need to express charges as linear combinations of $\gamma_1$ and $\gamma_2$:
\begin{align}
    \cdots &= \cdots \cr
    F_{H_{-3}} &= X_{- \gamma_1-\frac32 \gamma_2}+(\fq + \fq^{-1})X_{- \gamma_1-\frac12 \gamma_2}+ X_{- \gamma_1+\frac12 \gamma_2} + X_{+\frac12 \gamma_2} \cr
    F_{H_{-2}} &= X_{-\frac12 \gamma_1-\gamma_2}+X_{-\frac12 \gamma_1} \cr
    F_{H_{-1}} &= X_{-\frac12 \gamma_2} \cr
    F_{H_0} &= X_{\frac12 \gamma_1} \cr
    F_{H_1} &= X_{\frac12 \gamma_2}+ X_{\gamma_1+\frac12 \gamma_2} \cr
    F_{H_2} &= X_{-\frac12 \gamma_1}+ X_{-\frac12 \gamma_1+\gamma_2}+ (\fq + \fq^{-1}) X_{\frac12 \gamma_1+\gamma_2}+X_{\frac32 \gamma_1+\gamma_2} \cr
    \cdots &= \cdots \cr
    F_{w_1} &= X_{-\frac12 \gamma_1-\frac12\gamma_2}+ X_{-\frac12 \gamma_1+\frac12\gamma_2} + X_{\frac12 \gamma_1+\frac12\gamma_2}
\end{align}
Notice that all terms in a given sum have charges which differ by integral multiples of $\gamma_1$ and $\gamma_2$. This is also the hallmark of a cluster RG map.

\subsection{Fake examples}
We should caution the reader that not all cluster seeds/BPS quivers are expected to occur in a physical context, and our general construction is thus not expected to apply to generic seeds. 

For example, consider $S=E_\fq(\gamma_1)E_\fq(\gamma_2)$ with $\langle \gamma_1, \gamma_2\rangle = 3$. This can be analyzed in the same manner as our previous examples, generating a sequence of candidate $F_{L_i}$ from mutations of $F_{L_0}=X_{\gamma_1/3}$. But we find experimentally that $F_{L_i} F_{L_j} |1\rangle$ is not normalizable for $|i-j|$ sufficiently large.
\section{\boldmath Adding Flavor: the $\cN=2^* \, SU(2)$ example}
\label{sec:addingfla}
We now consider a rich example based on a $\Gamma = \bZ^3$ lattice, associated to ${\cal N}=2^*$ $SU(2)$ gauge theory, aka $T[SU(2), T^* \bC^3]$.  The inner product on the lattice is 
\begin{equation}
    \langle (c_1,c_2,c_3),(d_1,d_2,d_3)\rangle= 2c_1 d_2 + 2c_2 d_3 + 2c_3 d_1 - 2c_2 d_1 - 2c_3 d_2 -2c_1 d_3
\end{equation}
The element $\mu = X_{\frac12,\frac12,\frac12}$ is central.

As for the pure $SU(2)$ case, it is convenient to enlarge the associated algebra at the pice of having half-integral powers of $\fq$ at intermediate steps. We now have three analogues of the Wilson line operator:
\begin{align}
    F_{D_{1,0}} &= X_{0,-\frac12,-\frac12}+ X_{0,-\frac12,\frac12} + X_{0,\frac12,\frac12} \cr
    F_{D_{0,1}} &= X_{-\frac12,0,-\frac12}+ X_{\frac12,0,-\frac12} + X_{\frac12,0,\frac12}\cr
    F_{D_{1,1}} &= X_{-\frac12,-\frac12,0}+ X_{-\frac12,\frac12,0} + X_{\frac12,\frac12,0}
\end{align}
Here the linear basis for the algebra $\CA$ consists of $D_{a,b}=D_{-a,-b}$ with integer $a$ and $b$, all invariant under $\rho$, with an overall $SL(2,\bZ)$ symmetry. It is a special case of Double-Affine Hecke Algebra.

The product of two operators $D_{p,q}$ and $D_{p',q'}$ is simple if $|p q' - p' q|=1$:
\begin{equation}
   D_{p,q} D_{p',q'} = \fq^{-\frac12} D_{p+p',q+q'}+ \fq^{\frac12} D_{p-p',q-q'} \qquad \qquad p q' - p' q = 1\, ,
\end{equation}
but becomes more complicated as $p q' - p' q$ increases. Each of the $D_{p,q}$ with $p$ and $q$ coprime behaves as a fundamental Wilson line, i.e. 
\begin{equation}
    D_{p,q} D_{ap,aq} = D_{ap,aq}D_{p,q} = D_{(a+1)p,(a+1)q}+D_{(a-1)p,(a-1)q}
\end{equation}
and $D_{ap,aq}$ from a sub-algebra isomorphic to the representation ring of $SU(2)$, which can be identified with Wilson lines in a specific Lagrangian description of the theory. 

We can compute more $F$'s from these relations. For example, from $F_{D_{1,0}}F_{D_{0,1}}$ we compute
\begin{equation}
    F_{D_{1,-1}}= X_{-\frac12,-\frac12,-1}+ X_{\frac12,-\frac12,-1} + (\fq^{-1}+ \fq) X_{\frac12,-\frac12,0}+ X_{\frac12,-\frac12,1} + X_{\frac12,\frac12,1}
\end{equation}
and cyclic rotations
\begin{equation}
    F_{D_{1,2}}= X_{-1,-\frac12,-\frac12}+ X_{-1,\frac12,-\frac12} + (\fq^{-1}+ \fq) X_{0,\frac12,-\frac12}+ X_{1,\frac12,-\frac12} + X_{1,\frac12,\frac12}
\end{equation}
and
\begin{equation}
    F_{D_{2,1}}= X_{-\frac12,-1,-\frac12}+ X_{-\frac12,-1,\frac12} + (\fq^{-1}+ \fq) X_{-\frac12,0,\frac12}+ X_{-\frac12,1,\frac12} + X_{\frac12,1,\frac12}
\end{equation}
We can recover $F_{D_{p,q}}$ in this manner. 

The three mutations available in this chamber involve charges $\gamma_1 = (1,0,0)$, $\gamma_2 = (0,1,0)$ and $\gamma_3 = (0,0,1)$.

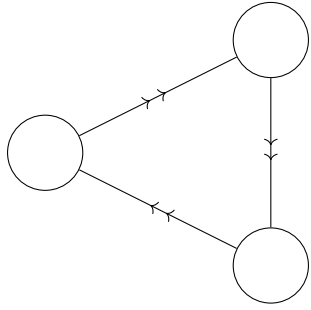
\begin{figure}[h]
    \centering
    \begin{tikzpicture}
        % Nodes arranged in a triangle
        \node (A) at (0,0) [circle, draw, minimum size=1cm] {};
        \node (B) at (3,1.5) [circle, draw, minimum size=1cm] {};
        \node (C) at (3,-1.5) [circle, draw, minimum size=1cm] {};

        % Arrows with double arrowheads in the middle
        \draw[postaction={decorate},
              decoration={markings, 
              mark=at position 0.45 with {\arrow{>}},
              mark=at position 0.55 with {\arrow{>}}}] 
              (A) -- (B);
              
        \draw[postaction={decorate},
              decoration={markings, 
              mark=at position 0.45 with {\arrow{>}},
              mark=at position 0.55 with {\arrow{>}}}] 
              (B) -- (C);
              
        \draw[postaction={decorate},
              decoration={markings, 
              mark=at position 0.45 with {\arrow{>}},
              mark=at position 0.55 with {\arrow{>}}}] 
              (C) -- (A);

    \end{tikzpicture}
    \caption{The BPS quiver in this example. Mutation at any node gives an isomorphic quiver. Note the pure $SU(2)$ sub-quiver, associated to the RG flow described below.}
    \label{fig:cyclic_quiver}
\end{figure}

For example the mutation by $X_{0,0,1}$ gives new degeneracies: 
\begin{align}
    F'_{D_{1,0}} &= X_{0,-\frac12,-\frac12} + X_{0,\frac12,\frac32}+X_{0,\frac12,\frac12}  \cr
    F'_{D_{0,1}} &=  X_{-\frac12,0,\frac12}+ X_{-\frac12,0,-\frac12}+ X_{\frac12,0,-\frac12} \cr
   F'_{D_{1,-1}}&= X_{-\frac12,-\frac12,-1}+ X_{\frac12,-\frac12,-1} + X_{\frac12,\frac12,1}
\end{align}
which can be related to the original $F$ by a combination of an automorphism of $\CA$ and a charge redefinition $(c_1,c_2,c_3) \to (c_2,c_1,2 c_2-c_3)$, so that the next available set of mutations would involve charges $(1,0,0)$, $(0,1,2)$, $(0,0,-1)$. 

This is an example where the spectrum generator is not expected to be expressible as a sequence of mutations! Instead, a ``weak coupling'' infinite product is available in the literature. Observe that the intertwining relation with the $F$'s only determines $S$ up to a function of $\mu$. This ambiguity can be fixed in two ways: we can compare it to the UV localization formulae for the Schur index or we can modify the theory to enlarge the algebra of loop operators without changing the BPS spectrum. We will explore all of these strategies in the following. 

\subsection{Intertwining recursion}
First, we can explore the intertwining relations for the available $F$'s. 
Starting from  
\begin{equation}
    S =  \sum_{p=0}^\infty\sum_{q=0}^\infty\sum_{r=0}^\infty \Sigma_{p,q,r}(\fq) X_{p,q,r}
\end{equation}
with $\Sigma_{0,0,0}(\fq)=1$ and $\Sigma_{p,q,r}(\fq)=0$ outside that summation range, we have 
\begin{align}
     \Sigma_{p,q,r}(\fq) &=\frac{\fq^{2p-r-q+1}}{\fq^{q-r}-\fq^{r-q}}(\Sigma_{p,q-1,r}(\fq)-\Sigma_{p,q,r-1}(\fq))+\Sigma_{p,q-1,r-1}(\fq) 
\end{align}
and cyclic rotations. Curiously, the equation is symmetric in $q$ and $r$. Also, if $q=r$ we should replace it with $\Sigma_{p,q-1,q}(\fq)=\Sigma_{p,q,q-1}(\fq)$, which is also compatible with $S_3$ invariance. As expected, $\Sigma_{p,p,p}(\fq)$ is not fixed by the equations. 

We thus find
\begin{align}
    \Sigma_{1,0,0}(\fq)&=\Sigma_{0,1,0}(\fq)=\Sigma_{0,0,1}(\fq) = -\frac{\fq}{1-\fq^2} \cr
    \Sigma_{1,1,0}(\fq)&=\Sigma_{0,1,1}(\fq)=\Sigma_{1,0,1}(\fq) = \frac{\fq^4}{(1-\fq^2)^2} \cr
    \Sigma_{2,0,0}(\fq)&=\Sigma_{0,2,0}(\fq)=\Sigma_{0,0,2}(\fq) = \frac{\fq^2}{(1-\fq^2)(1-\fq^4)}\cr
    \Sigma_{3,0,0}(\fq)&=\Sigma_{0,3,0}(\fq)=\Sigma_{0,0,3}(\fq) = -\frac{\fq^3}{(1-\fq^2)(1-\fq^4)(1-\fq^6)}\cr
    \Sigma_{2,1,0}(\fq)&=\cdots = -\frac{\fq^7}{(1-\fq^2)^2(1-\fq^4)}\cr
    \Sigma_{4,0,0}(\fq)&=\Sigma_{0,4,0}(\fq)=\Sigma_{0,0,4}(\fq) = -\frac{\fq^4}{(1-\fq^2)(1-\fq^4)(1-\fq^6)(1-\fq^8)}\cr
    \Sigma_{3,1,0}(\fq)&=\cdots = \frac{\fq^{10}}{(1-\fq^2)^2(1-\fq^4)(1-\fq^6)}\cr
    \Sigma_{2,2,0}(\fq)&=\cdots = \frac{\fq^{12}}{(1-\fq^2)^2(1-\fq^4)^2}\cr
    \Sigma_{2,1,1}(\fq)&=\cdots = -\frac{\fq}{1-\fq^2} \left(\Sigma_{1,1,1}(\fq) +\frac{\fq^7}{(1-\fq^2)^2(1-\fq^4)} \right)+\frac{\fq^4}{(1-\fq^2)^2}
\end{align}
As expected, $\Sigma_{p,p,p}(\fq)$ are undetermined. We can potentially compare $||S\rangle|^2$ and the UV formula for the Schur index:
\begin{equation}
    I(\fq;\mu) = \oint \frac{d\zeta}{4 \pi i \zeta} \frac{( \fq^2;\fq^2)_\infty^2( \zeta^2;\fq^2)_\infty(\fq^2 \zeta^2;\fq^2)_\infty(\zeta^{-2};\fq^2)_\infty(\fq^2 \zeta^{-2};\fq^2)_\infty}{(\fq \mu;\fq^2)_\infty(\fq \mu^{-1};\fq^2)_\infty ( \fq \mu \zeta^2;\fq^2)_\infty(\fq \mu^{-1} \zeta^2;\fq^2)_\infty(\fq \mu \zeta^{-2};\fq^2)_\infty(\fq \mu^{-1} \zeta^{-2};\fq^2)_\infty} \, ,
\end{equation}
to fix them recursively, by making some reasonable assumptions about the power of $\fq$. This is un-illuminating, so we will move to a more conceptual approach. We note a curious observation: a simple choice such as $\Sigma_{p,p,p}=0$ for $p>0$ leads to a candidate spherical vector which does not even appear to be normalizable. Finding the correct $\mu$ dependence is thus necessary in order to complete the quantization program. 

\subsection{Improved intertwining recursion}
A physical way to address the problem is to weakly gauge the Cartan subgroup of the flavour symmetry group in the physical theory, resulting in an $U(1) \times SU(2)$ gauge theory with matter in $T^* \bC^3$. Effectively, this enlarges the charge lattice to $\bZ^4$ and makes the pairing non-degenerate, without changing the BPS spectrum of the theory and thus $S$ itself. 

A general feature of gauge theories is that the algebra of loop operators contains sub-algebras generated by the 't Hooft-Wilson operators charged under a single factor of the gauge group. Here that means the extended $\CA$ will contain both the $SU(2)$ $\CN=2^*$ algebra and the algebra of SQED${}_3$ (aka $T[U(1),T^*\bC^3]$) as sub-algebras. Wilson lines charged under one factor of the gauge group commute with 't Hooft-Wilson lines in other factors, but 't Hooft lines for different factors do not typically commute. 

We build a candidate IR charge lattice by adding an extra charge $\gamma_m$ which pairs with one of the elementary ones. If we denote the charges which appear in mutations as 
\begin{align}
    \gamma_1 &= (0;1,0,0) \cr
    \gamma_2 &= (0;0,1,0) \cr
    \gamma_3 &= (0;0,0,1)
\end{align}
we can add an extra ``frozen'' charge
\begin{equation}
    \gamma_m = (1;0,0,0) 
\end{equation}
with $\langle \gamma_m, \gamma_i\rangle=\delta_{i1}$. This definition naively breaks the $\bZ_3$ symmetry of the problem, but that can be restored by a charge re-definition. We will see this choice is compatible with the identification of $D_{1,0}$ as the Wilson line for the $SU(2)$ gauge group.  

We conjecture that $F_{u_+} = X_{1;0,0,0}$ is a valid (and tropical) choice of $F$, associated to an ``Abelian 't Hooft loop'' operator. It obviously behave well under the three possible mutations: two leave it unchanged and the third maps it to $X_{1;0,0,0}+ X_{1;1,0,0}$. Rather than testing the behaviour under further mutations, we can look at the interplay with known $F$'s: 
\begin{align}
    F_{u_+} \cdot F_{D_{1,0}} &= X_{1;0,-\frac12,-\frac12}+ X_{1;0,-\frac12,\frac12} + X_{1;0,\frac12,\frac12}  = F_{D_{1,0}}\cdot F_{u_+} \cr
    F_{u_+} \cdot F_{D_{0,1}} &= \fq^{-\frac12} X_{1;-\frac12,0,-\frac12}+ \fq^{\frac12}(X_{1;\frac12,0,-\frac12} + X_{1;\frac12,0,\frac12})\cr
    F_{u_+} \cdot F_{D_{1,1}} &= \fq^{-\frac12} (X_{1;-\frac12,-\frac12,0}+ X_{1;-\frac12,\frac12,0}) + \fq^{\frac12} X_{1;\frac12,\frac12,0}
\end{align}
The new $F_{u_+}$ commutes with $F_{D_{1,0}}$. We identify the latter with the $w_1$ Wilson line of the $SU(2)$ gauge theory, which commutes with the $U(1)$ 't Hooft operators. The remaining summands on the last two lines provide new candidate $F$'s for 't Hooft operators of mixed charges. It can be easily verified that they behave well under all three mutations, supporting our conjecture. Decomposing further products we can generate many more candidate $F$'s and provide further support for our conjecture.

With some patience we could identify each of these $F$'s with 't Hooft-Wilson operators $D_{m_1,e_1;m_2,e_2}$ for the extended theory. We leave this as an exercise for the enthusiastic reader. 

Identifying $F$'s with opposite tropical charge is a bit more challenging. 
We can try to build a candidate tropical $F$ of the form 
\begin{equation}
    X_{-1;0,0,0}+ \cdots
\end{equation}
In order to survive all three mutations, it must include an extra term 
\begin{equation}
    X_{-1;0,0,0}+X_{-1;1,0,0}+ \cdots
\end{equation}
Good behavior under the $(0,0,1)$ mutation then requires some extra terms
\begin{equation}
    X_{-1;0,0,0}+X_{-1;1,0,0}+(\fq + \fq^{-1}) X_{-1;1,0,1}+X_{-1;1,0,2}+ \cdots
\end{equation}
The last term requires an extra correction to survive a $(0,1,0)$ mutation:
\begin{equation}
    X_{-1;0,0,0}+X_{-1;1,0,0}+(\fq + \fq^{-1}) X_{-1;1,0,1}+X_{-1;1,0,2}+(\fq + \fq^{-1})X_{-1;1,1,2}+X_{-1;1,2,2}+ \cdots \, .
\end{equation}
Then the $(1,0,0)$ mutation requires 
\begin{align}
    X_{-1;0,0,0}&+X_{-1;1,0,0}+(\fq + \fq^{-1}) X_{-1;1,0,1}+X_{-1;1,0,2}+(\fq + \fq^{-1})X_{-1;1,1,2}+ \cr
    &+ X_{-1;1,2,2}+X_{-1;2,2,2} \, .
\end{align}
at which point we are fortunately done! We record the effect of the three mutations: 
\begin{align}
    (1,0,0): X_{-1;0,0,0}&+(\fq + \fq^{-1}) X_{-1;1,0,1}+(\fq + \fq^{-1}) X_{-1;2,0,1}+X_{-1;1,0,2}+ \cr 
    &+[3]_\fq X_{-1;2,0,2}+[3]_\fq X_{-1;3,0,2}+X_{-1;4,0,2}+(\fq + \fq^{-1})X_{-1;1,1,2}+ \cr &+(\fq + \fq^{-1})X_{-1;2,1,2}
    + X_{-1;1,2,2}\cr
    (0,1,0): X_{-1;0,0,0}&+X_{-1;1,0,0}+(\fq + \fq^{-1})X_{-1;1,1,0}+X_{-1;1,2,0}+ \cr &+(\fq + \fq^{-1}) X_{-1;1,0,1}+X_{-1;1,0,2}+X_{-1;2,2,2} \cr
    (0,0,1): X_{-1;0,0,0}&+X_{-1;1,0,0}+(\fq + \fq^{-1})X_{-1;1,1,2}+ \cr
    &+ X_{-1;1,2,2}+(\fq + \fq^{-1})X_{-1;1,2,2}+X_{-1;1,2,3}+X_{-1;2,2,2} \, . 
\end{align}
As an extra test of our conjecture, note that 
\begin{equation}
    F_{u_+} \cdot (X_{-1;0,0,0}+X_{-1;1,0,0}+[2]_\fq X_{-1;1,0,1}+X_{-1;1,0,2}+[2]_\fq X_{-1;1,1,2}+ X_{-1;1,2,2}+X_{-1;2,2,2})
\end{equation}
contains $1$, $\mu^2$ and 
\begin{equation}
    X_{0;1,0,0}+[2]_\fq X_{0;1,0,1}+X_{0;1,0,2}+[2]_\fq X_{0;1,1,2}+ X_{0;1,2,2} = \mu F_{D_{2,0}-1} \, .
\end{equation}
This candidate $F$ is nice but not quite what we need: it behaves as the image of a negative charge 't Hooft loop in SQED${}_2$ rather than SQED${}_3$. Indeed, if we write $D_{1,0}=\zeta + \zeta^{-1}$, we find that the product with $F_{u_+}$ factors as 
\begin{equation}
    (1 + \fq \mu \zeta)(1+\fq \mu \zeta^{-1})
\end{equation}
Instead, we want the image of
\begin{equation}
    u_+ u_- = (1 + \fq \mu \zeta)(1+\fq \mu)(1+\mu \zeta^{-1})
\end{equation}
as the $U(1)$ gauge fields couple to three hypermultiplets. 

We can fix the problem by adding the $(1+\fq \mu)$ factor by hand, setting 
\begin{align}
   F_{u_-} =  X_{-1;0,0,0}&+X_{-1;1,0,0}+(\fq + \fq^{-1}) X_{-1;1,0,1}+X_{-1;1,0,2}+X_{-1;1,1,1}+X_{-1;2,1,1}+\cr
    &+(\fq + \fq^{-1}) X_{-1;2,1,2}+(\fq + \fq^{-1})X_{-1;1,1,2}+  X_{-1;1,2,2}+X_{-1;2,2,2}+\cr
    &+(\fq + \fq^{-1})X_{-1;2,2,3}+ X_{-1;2,3,3}+X_{-1;3,3,3} \, .
\end{align}
We are now ready to impose an intertwining relation 
\begin{align}
    X_{1;0,0,0}\cdot S = S &\cdot ( X_{1;0,0,0}+ X_{1;1,0,0}+(\fq + \fq^{-1})X_{1;1,1,0}+X_{1;1,1,1}+ X_{1;2,1,1}+ \cr &+(\fq + \fq^{-1})X_{1;2,2,1}+ X_{1;1,2,0}+(\fq + \fq^{-1}) X_{1;1,2,1}+X_{1;1,2,2}+\cr &+ X_{1;2,2,2}+
    X_{1;2,3,1}+(\fq + \fq^{-1}) X_{1;2,3,2}+X_{1;2,3,3}+ X_{1;3,3,3} )\, .
\end{align}
This equation reproduces $\Sigma_{1,0,0}$ as well as $\Sigma_{1,1,0}$. It also fixes $\Sigma_{1,1,1} = \frac{2 \fq^3 - 3 \fq^5}{(1-\fq^2)^3}$. More importantly, the norm of the spherical vector now matches the Schur index of the theory. 

If we had employed the naive $F$ without the extra $(1+\fq \mu)$, the final answer for $S$ would have missed an $E_\fq(\mu)$ factor and failed to reproduce the correct Schur index.

\subsection{Partial RG flow}
There is a natural RG flow relating this example to the pure $SU(2)$ example. We can identify 
\begin{equation}
    \RG(D_{1,0}) = w_1
\end{equation}
the Wilson lines in a natural way, as well as the action of the $(0,1,0)$ and $(0,0,1)$ mutations. Then we can identify
\begin{align}
    F^\IR(H_{-2}) &= X_{0,-\frac12,-1}+X_{0,-\frac12,0} \cr
    F^\IR(H_{-1}) &= X_{0,0,-\frac12} \cr
    F^\IR(H_{0}) &= X_{0,\frac12,0} \cr
    F^\IR(H_{1}) &= X_{0,0,\frac12} +X_{0,1,\frac12} 
\end{align}
so that
\begin{align}
    \RG(D_{0,1}) &= \mu^{-\frac12} H_0 + \mu^{\frac12} H_{-2} \cr
     \RG(D_{1,1}) &= \mu^{-\frac12} H_1 + \mu^{\frac12} H_{-1} 
\end{align}

The spectrum generator for the partial RG flow will thus satisfy
\begin{align}
    w_1 \cdot S &= S \cdot w_1 \cr
    (\mu^{-\frac12} H_0 + \mu^{\frac12} H_{-2}) \cdot S &= S \cdot (\mu^{\frac12} H_{2} + \mu^{-\frac12} H_{0}) \cr
    (\mu^{-\frac12} H_1 + \mu^{\frac12} H_{-1}) \cdot S &= S \cdot (\mu^{\frac12} H_{3} + \mu^{-\frac12} H_{1})  \, .
\end{align}
The first equation suggests that $S$ should actually be built from $w_1$ and $\mu$ only. As $F^\IR(S)$ should contain $X_{1,0,0}$, $S$ should contain $\ell_{1,0,0} = \mu w_2$, or more precisely, 
\begin{equation}
    S = 1 -\frac{\fq}{1-\fq^2}w_2 + \cdots
\end{equation}

A ``weak-coupling'' expression for $S_\UV$ available in the literature (with a mild correction to include the flavour factor we computed above) is 
\begin{equation}
   E_\fq(X_{0,0,1}) \cdots \left[E_\fq(X_{1,0,0})E_\fq(X_{1,1,1})E_\fq(X_{1,2,2})E^{-1}_\fq(-\fq^{-1} X_{0,1,1})E^{-1}_\fq(-\fq X_{0,1,1})\right]\cdots E_\fq(X_{0,1,0})
\end{equation}
It is not difficult to argue that acting with $w_n$ on the left of the first semi-infinite sequence of operators is the same as inserting 
\begin{equation}
    \chi_n(X_{0,1,1}) \equiv X_{0,-\frac{n}{2},-\frac{n}{2}} + X_{0,1-\frac{n}{2},1-\frac{n}{2}}+ \cdots X_{0,\frac{n}{2},\frac{n}{2}}
\end{equation}
in the middle of the expression. Accordingly, if we decompose 
\begin{equation}
E_\fq(\mu \zeta^{-1})E_\fq(\mu)E_\fq(\mu \zeta) = f_3(\mu,\chi_n(\zeta))
\end{equation}
we deduce that 
\begin{equation}
    S_\UV = \left[\sum_{m,n} c_{n,m} \mu^m F^\IR(w_n) \right] S_\IR
\end{equation}
with $S_\UV = E_\fq(\gamma_2)E_\fq(\gamma_3)$, and in particular 
\begin{equation}
    S = f_3(\mu,w_n)
\end{equation}
With some work, one can successfuly compare this solution to the recursive solution presented above. 

A comparison with UV formulae for the Schur index is suprisingly simple, and can be recast as a direct derivation of $S_\UV$. We have
\begin{equation}
    \langle 1;\UV|(F^\UV_{D_{1,0}})^n|1;\UV\rangle = \langle 1;\IR|(F^\IR_{w_1})^n \left[\sum_{m,n} c_{n,m} \mu^m F^\IR(w_n) \right]^2|1;\IR\rangle
\end{equation}
and thus it inserts the factor 
\begin{equation}
E_\fq(\mu \zeta^{-1})E_\fq(\mu)E_\fq(\mu \zeta) 
\end{equation}
in the UV index contour integral formula, which precisely converts the pure $SU(2)$ calculation to $SU(2)$ $N=2^*$. 

\section{The Spectrum Generator for RG flows triggered by large masses.}
\label{sec:broadconj}
The example in the last Section suggests a broad generalization to any gauge theory $T[G,T^*N]$ with cotangent matter. The basic idea is that there is a natural RG flow $T[G,T^*N] \to T[G,0]$ triggered by giving large masses to all of the matter fields. 

We conjecture that this RG flow will send Wilson loops to Wilson loops for the same representation. As $\rho$ acts on Wilson lines by conjugating the representation, the spectrum generator $S$ for this flow should commute with Wilson loop operators. Accordingly, we also conjecture that $S$ is a function of Wilson loop operators. 

More precisely, consider the matter contribution to the UV formulae for Schur indices
\begin{equation}
    \prod_{(w,w_f) \in T^*N} E_\fq(\mu^{w_f} \zeta^{w}) = \prod_{(w,w_f) \in N} E_\fq(\mu^{w_f} \zeta^{w})E_\fq(\mu^{-w_f} \zeta^{-w})
\end{equation}
where $w$ are the weights of the $G$ Cartan torus action on $T^*N$ and $w_f$ are the weights of a flavor group action, say a $U(1)$ for each irreducible representation in $N$.

We can take the contribution of $N$ and organize it into characters for $G$:
\begin{equation}
    \prod_{(w,w_f) \in N} E_\fq(\mu^{w_f} \zeta^{w}) = f_N(\chi_R(\zeta),\mu)
\end{equation}
We conjecture 
\begin{equation}
    S = f_N(w_R,\mu)
\end{equation}

It is easy to see that such $S$ reproduces the correct Schur index with Wilson loop insertions only:
\begin{equation}
    \langle S|w_a|S\rangle = \langle 1| w_a \prod_{(w,w_f) \in N} E_\fq(\mu^{w_f} \zeta^{w})E_\fq(\mu^{-w_f} \zeta^{-w})|1\rangle
\end{equation}
As Wilson loops contribute to the UV Schur index formulae by inserting characters in the contour integral, the product of $E_\fq$'s will simply reproduce the matter contribution to the UV index. 

Ideally, we should also produce candidate RG flow images of 't Hooft-Wilson lines $D^N_{m,e}$ for the theory with matter, as linear combinations of the $D_{m',e'}$ loop operators of the pure gauge theory. 

The $D^N_{m,e}$ loop operators are described in the UV by ``Abelianized'' formulae \cite{Bullimore:2015lsa}. Essentially, one maps them to an extension of the algebra of loop operators for $T[H,T^*N]$, where $H$ is the Cartan torus of $G$, which includes certain rational functions of Abelian Wilson generators $v$. A natural strategy is then to map the extended algebra for $T[H,T^*N]$
to the extended algebra for $T[H,0]$ by the RG flow map associated to the spectrum generator 
\begin{equation}
    \prod_{(w,w_f) \in N} E_\fq(\mu^{w_f} v^{w}) 
\end{equation}
We expect this map to descend to a map between $D_{m,e}$'s, compatible with our conjectural $S$. 

If we have available any Coulomb RG flow for $T[G,0]$, with some 
spectrum generator $S_{\IR}$ and images $F^\IR_{w_R}$ for the Wilson loops, we can then predict a spectrum generator for $T[G,T^*N]$:
\begin{equation}
    S_\UV = f_N(F^{IR}_{w_R},\mu) S_\IR \,.
\end{equation}
with $F^\UV_{w_R} = F^\IR_{w_R}$ and appropriate images for more general $D_{m,e}$. 

Furthermore, suppose that $N$ is the direct sum of irreps $N_i$, associated to flavour parameters $\mu_i$. Observe that the tropical labels of Wilson lines map the weight lattice of $G$ into $\Gamma$.  Then each factor of $f_{N_i}( \mu_i F^{IR}_{w_R})$ adds to the support of $S_\UV$ 
a collection of charges generated from the images of roots, which are already in $S_\IR$, and of the lowest weight $w_i$ in $N_i$. This is also the tropical charge of $w_{N_i}$. 

Accordingly, if we assume a cluster structure for both UV and IR theories, it appears that each $N_i$ adds a new simple charge given by the image of $w_i$ plus the flavour charge of $\mu_i$. In the language of BPS quivers, each $N_i$ adds a new node labelled by the tropical charge of $w_N$. This was clearly the case in the previous Section: $\gamma_1$ is the tropical charge of $w_2$ in the IR theory. This principle has already been observed in the literature, with a focus on simple gauge groups \cite{Cecotti:2012sf,Cecotti:2012gh,Cecotti:2012se}.

We can broaden the conjecture further to a situation where we gauge the $G$ flavour symmetry of any theory $T$, which may not be a gauge theory:
\begin{itemize}
    \item Take the spectrum generator $S_T$ as a function of the $G$ flavour fugacities $\zeta$.
    \item Expand it into characters $\chi_R(\zeta)$.
    \item Replace  $\chi_R(\zeta)$ with Wilson loop operators $w_R$ for a pure $G$ gauge theory to obtain a candidate $S$.
    \item Define $S_\UV = F^\IR(S) S_\IR$ as before. 
\end{itemize}

\subsection[$SU(2)$ gauge theory with flavor]{\boldmath$SU(2)$ gauge theory with flavour}
There are four more examples of well-defined gauge theories with gauge group $SU(2)$, coupled with a number $N_f$ of fundamental hypermultiplets which goes from $1$ to $4$. In other words, $T[SU(2),T^* \bC^{2 N_f}]$. 

The BPS quivers for these theories are known in an uniform manner: the pair of nodes from $SU(2)$, labeled by charges $\gamma_1$ and $\gamma_2$ with $\langle \gamma_1, \gamma_2 \rangle = 2$, are augmented by $N_f$ extra nodes $\gamma_i$, $i>2$, such that $\langle \gamma_2, \gamma_i \rangle = \langle \gamma_i, \gamma_1 \rangle =1$ and all other pairings vanishing. As expected from the general conjecture, $\gamma_i$ are precisely $N_f$ copies of the tropical charge for $w_1$ in pure $SU(2)$ gauge theory. 

\begin{figure}[h]
    \centering
    \begin{tikzpicture}
        % Nodes
        \node (A1) at (-3,0) [circle, draw, minimum size=1cm] {}; % Leftmost copy
        \node (A) at (1,0) [circle, draw, minimum size=1cm] {}; % Main leftmost node
        \node (B) at (3,1.5) [circle, draw, minimum size=1cm] {};
        \node (C) at (3,-1.5) [circle, draw, minimum size=1cm] {};

        % Ellipsis between A1 and A
        \node at (-.5,0) {\Large $\cdots$};

        % Arrows from A1 (same as from A)
        \draw[postaction={decorate},
              decoration={markings, 
              mark=at position 0.5 with {\arrow{>}}}] 
              (A1) -- (B);
              
        \draw[postaction={decorate},
              decoration={markings, 
              mark=at position 0.5 with {\arrow{>}}}] 
              (A1) -- (C);

        % Arrows from A (same as from A1)
        \draw[postaction={decorate},
              decoration={markings, 
              mark=at position 0.5 with {\arrow{>}}}] 
              (A) -- (B);
              
        % Vertical arrow with double arrowheads in the middle
        \draw[postaction={decorate},
              decoration={markings, 
              mark=at position 0.45 with {\arrow{>}},
              mark=at position 0.55 with {\arrow{>}}}] 
              (B) -- (C);
              
        \draw[postaction={decorate},
              decoration={markings, 
              mark=at position 0.5 with {\arrow{>}}}] 
              (C) -- (A);
              
    \end{tikzpicture}
    \caption{The BPS quiver for $SU(2)$ with $N_f$ flavours. In a partial RG perspective, the rightmost nodes are inherited from pure $SU(2)$ gauge theory, with charges $\gamma_1$ and $\gamma_2$, and the collection of nodes on the left represent the $N_f$ extra matter representations, with charges $\gamma_3, \cdots \gamma_{2+N_f}$.}
    \label{fig:quiver_collection}
\end{figure}
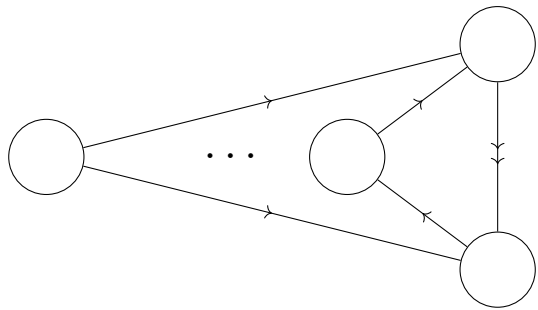

These BPS quivers have loops, which makes a direct calculation of $S$ complicated. For $N_f=1,2,3$ there is a simple strategy to produce a candidate $S$: a sequence of mutations leads to quivers with no loops, for which $S$ is known to be an ordered product of $E_\fq$ factors for each node of the quiver. We can then mutate back to the original BPS quiver and compare with our conjectural prescription for the quantum spectrum generator of gauge theories.  

For $SU(2)$ $N_f=4$ no such simplification is possible, but we can still compare  our conjecture to known solutions for the spectrum generator and $F$'s. 

\subsubsection{$N_f=1$}
A mutation at the $\gamma_3$ node gives a BPS quiver with nodes $\gamma_1$, $\gamma_2+\gamma_3$ and $-\gamma_3$. The spectrum generator in this chamber is known to be 
\begin{equation}
    S' = E_{\fq}(X_{\gamma_1})E_{\fq}(X_{-\gamma_3})E_{\fq}(X_{\gamma_2+\gamma_3}) 
\end{equation}
The reader may observe the well-known RG flow to the $[A_1,A_2]$ theory, whose spectrum generator can e.g. be recognized in the last two factors.

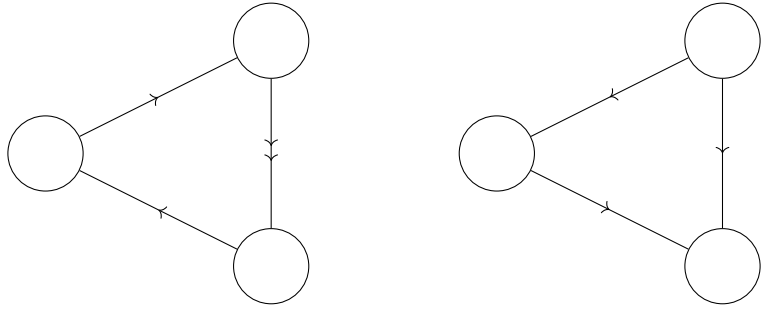
\begin{figure}[h]
    \centering
    \begin{tikzpicture}
        % First Quiver (original)
        \node (A1) at (0,0) [circle, draw, minimum size=1cm] {};
        \node (B1) at (3,1.5) [circle, draw, minimum size=1cm] {};
        \node (C1) at (3,-1.5) [circle, draw, minimum size=1cm] {};

        % Diagonal arrow (single arrowhead in the middle)
        \draw[postaction={decorate},
              decoration={markings, 
              mark=at position 0.5 with {\arrow{>}}}] 
              (A1) -- (B1);
              
        % Vertical arrow with double arrowheads in the middle
        \draw[postaction={decorate},
              decoration={markings, 
              mark=at position 0.45 with {\arrow{>}},
              mark=at position 0.55 with {\arrow{>}}}] 
              (B1) -- (C1);
              
        % Diagonal arrow (single arrowhead in the middle)
        \draw[postaction={decorate},
              decoration={markings, 
              mark=at position 0.5 with {\arrow{>}}}] 
              (C1) -- (A1);

        % Second Quiver (flipped diagonal arrows, single arrowhead on vertical)
        \node (A2) at (6,0) [circle, draw, minimum size=1cm] {};
        \node (B2) at (9,1.5) [circle, draw, minimum size=1cm] {};
        \node (C2) at (9,-1.5) [circle, draw, minimum size=1cm] {};

        % Reversed diagonal arrows
        \draw[postaction={decorate},
              decoration={markings, 
              mark=at position 0.5 with {\arrow{>}}}] 
              (B2) -- (A2);
              
        \draw[postaction={decorate},
              decoration={markings, 
              mark=at position 0.5 with {\arrow{>}}}] 
              (A2) -- (C2);

        % Vertical arrow (now single arrowhead in the middle)
        \draw[postaction={decorate},
              decoration={markings, 
              mark=at position 0.5 with {\arrow{>}}}] 
              (B2) -- (C2);

    \end{tikzpicture}
    \caption{Left: the BPS quiver for the $N_f=1$ theory computed from partial RG flow. Right: a mutation at the leftmost node ($\gamma_3$) gives a simpler quiver, with node labels $\gamma_1$, $-\gamma_3$, $\gamma_2 + \gamma_3$ used in the main text.}
    \label{fig:quiver_side_by_side}
\end{figure}

We can record some $F$'s in this chamber:
\begin{align}
    F'_{D_{1,2}} &= X_{\gamma_2 + \gamma_3}+ X_{\gamma_1+\gamma_2 + \gamma_3}+ X_{\gamma_2}+ (\fq + \fq^{-1})X_{\gamma_1 + \gamma_2}+ X_{2 \gamma_1 + \gamma_2} \cr
    F'_{D_{1,1}} &= X_{-\gamma_3}+ X_{\gamma_1-\gamma_3} \cr
    F'_{D_{1,0}} &= X_{\gamma_1} \cr
    F'_{D_{1,-1}} &= X_{- \gamma_2 - \gamma_3}\cr
    F'_{D_{1,-2}} &= X_{- \gamma_2} + X_{\gamma_3}\cr
    F'_{D_{1,-3}} &= X_{-\gamma_1 - 2 \gamma_2 - \gamma_3}+ (\fq + \fq^{-1}) X_{-\gamma_1 - \gamma_2}+ X_{-\gamma_1 - \gamma_2 - \gamma_3} + X_{\gamma_3 - \gamma_1} + X_{-\gamma_1}\cr
    F'_{D_{0,1}} &= X_{-\frac12 \gamma_1-\frac12 \gamma_2}+X_{-\frac12 \gamma_1+\frac12 \gamma_2+\gamma_3}+X_{-\frac12 \gamma_1+\frac12 \gamma_2}+X_{\frac12 \gamma_1+\frac12 \gamma_2}
\end{align}
Here we denoted as $D_{0,1}$ the Wilson line. 

We can reorder the factors in $S'$ to  
\begin{equation}
    S'= E_{\fq}(X_{\gamma_1})E_{\fq}(X_{\gamma_2 +\gamma_3})E_{\fq}(X_{\gamma_2})E_{\fq}(X_{-\gamma_3}) 
\end{equation}
and then (inverse) mutate to the desired spectrum generator:
\begin{equation}
    S_\UV = E_{\fq}(X_{\gamma_3}) E_{\fq}(X_{\gamma_1})E_{\fq}(X_{\gamma_2 +\gamma_3})E_{\fq}(X_{\gamma_2}) 
\end{equation}
A comparison with pure $SU(2)$ gauge theory is facilitated by a a further re-ordering:
\begin{equation}
    S_\UV = E_{\fq}(X_{\gamma_3}) E_{\fq}(X_{\gamma_2 +\gamma_3})E_{\fq}(X_{\gamma_1+ \gamma_2 + \gamma_3})S_\IR
\end{equation}
with the usual 
\begin{equation}
    S_\IR = E_\fq(X_{\gamma_1})E_\fq(X_{\gamma_2})
\end{equation}
Comparing with pure $SU(2)$ gauge theory, that can be written as 
\begin{equation}
    S_\UV = E_{\fq}(\mu X_{-\frac12 \gamma_1-\frac12 \gamma_2}) E_{\fq}(\mu X_{-\frac12 \gamma_1+\frac12 \gamma_2})E_{\fq}(\mu X_{\frac12 \gamma_1+\frac12 \gamma_2})S_\IR
\end{equation}
with $\mu =X_{\frac12 \gamma_1+\frac12 \gamma_2+ \gamma_3}$.

The three arguments in the $E_\fq$ factors are the same as the summands in $\mu F^\IR_{w_1}$, as 
\begin{equation}
    F^\IR_{w_1} = X_{-\frac12 \gamma_1-\frac12 \gamma_2}+X_{-\frac12 \gamma_1+\frac12 \gamma_2}+X_{\frac12 \gamma_1+\frac12 \gamma_2}
\end{equation}
Remarkably, it appears that the whole product can be written in terms of $w_a$'s, and thus must satisfy:
\begin{equation}
    E_{\fq}(\mu X_{-\frac12 \gamma_1-\frac12 \gamma_2}) E_{\fq}(\mu X_{-\frac12 \gamma_1+\frac12 \gamma_2})E_{\fq}(\mu X_{\frac12 \gamma_1+\frac12 \gamma_2}) = f_2(\mu,w_a) 
\end{equation}
with 
\begin{equation}
    E_{\fq}(\mu \zeta^{-1}) E_{\fq}(\mu \zeta) = f_2(\mu,\chi_a(\zeta)) \, .
\end{equation} 
This verifies our conjecture about the spectrum generator of gauge theory.

We can record some $F$'s in this chamber:
\begin{align}
    F_{D_{1,2}} &=  X_{\gamma_2}+ (\fq + \fq^{-1})X_{\gamma_1 + \gamma_2}+ X_{2 \gamma_1 + \gamma_2}+X_{\gamma_1+\gamma_2 + \gamma_3}+ X_{2 \gamma_1 + \gamma_2+\gamma_3} \cr
    F_{D_{1,1}} &= X_{-\gamma_3}+ X_{\gamma_1-\gamma_3}+ X_{\gamma_1} \cr
    F_{D_{1,0}} &= X_{\gamma_1}+X_{\gamma_1+\gamma_3} \cr
    F_{D_{1,-1}} &= X_{- \gamma_2 - \gamma_3}+X_{- \gamma_2}\cr
    F_{D_{1,-2}} &= X_{- \gamma_2} +X_{- \gamma_2+\gamma_3} + X_{\gamma_3}\cr
    F_{D_{1,-3}} &= X_{-\gamma_1 - 2 \gamma_2 - \gamma_3}+X_{-\gamma_1 - 2 \gamma_2}+ (\fq + \fq^{-1}) X_{-\gamma_1 - \gamma_2}+ X_{-\gamma_1 - \gamma_2 - \gamma_3} + X_{-\gamma_1} \cr
    F_{D_{0,1}} &= X_{-\frac12 \gamma_1-\frac12 \gamma_2}+X_{-\frac12 \gamma_1+\frac12 \gamma_2}+X_{\frac12 \gamma_1+\frac12 \gamma_2}
\end{align}
and thus 
\begin{align}
    \RG(D_{1,2}) &=  H_1^2+\fq^{-\frac12} \mu H_0 H_1 \cr
    \RG(D_{1,1})  &= \mu^{-1} \fq^{-\frac12} H_0 H_1+ H_0^2 \cr
    \RG(D_{1,0})  &= H_0^2+ \fq^{-\frac12} \mu H_{-1} H_0 \cr
    \RG(D_{1,-1})  &= \fq^{-\frac12} \mu^{-1} H_{-1} H_0+H_{-1}^2\cr
    \RG(D_{1,-2})  &= H_{-1}^2 +\fq^{-\frac12} \mu H_{-2} H_{-1}\cr
    \RG(D_{1,-3}) &= \fq^{-\frac12} \mu^{-1} H_{-2} H_{-1}+H_{-2}^2 \cr
    F_{D_{0,1}} &= w_1
\end{align}
The enthusiastic reader is invited to compare these relations with the Abelianized expressions, reviewed for example in \cite{Gaiotto:2024osr}. For example, the right hand side of $\RG(D_{1,0})$ maps to 
\begin{align}
    &(u_+ + u_-)^2 +  \mu ( u_+  v^{-1} + u_- v )(u_+ + u_-) = \cr &= (1+\fq^{-1} \mu v^{-1}) u_+^2 + \frac{\fq+ \fq^{-1}+ \mu(v+v^{-1})}{(\fq v - \fq^{-1} v^{-1})(\fq^{-1} v - \fq v^{-1})} + (1+\fq^{-1} \mu v) u_-^2
\end{align}
This matches precisely the Abelianized expression $(3.130)$ for $D_{1,0}$, after replacing $(1+\fq^{-1} \mu v^{-1}) u_+^2 \to u_+$ and $(1+\fq^{-1} \mu v^{-1}) u_+^2 \to u_-$, which up to naming conventions are the RG map one would encounter in the Abelian version $T[U(1),T^*\bC^2]$ of the theory.

\subsubsection{$N_f=2$}
A mutation of the canonical quiver at the $\gamma_3$ and $\gamma_4$ nodes gives a quiver with nodes $\gamma_1$, $\gamma_2+\gamma_3+\gamma_4$, $-\gamma_3$ and $-\gamma_4$. The spectrum generator in this chamber is known to be 
\begin{equation}
   S' =  E_{\fq}(X_{\gamma_1})E_{\fq}(X_{-\gamma_3})E_{\fq}(X_{-\gamma_4})E_{\fq}(X_{\gamma_2+\gamma_3+\gamma_4}) 
\end{equation}

\begin{figure}[h]
    \centering
    \begin{tikzpicture}
        % First Quiver (Left)
        \node (A1) at (0,0) [circle, draw, minimum size=1cm] {};
        \node (B1) at (1.5,1.5) [circle, draw, minimum size=1cm] {};
        \node (C1) at (1.5,-1.5) [circle, draw, minimum size=1cm] {};
        \node (D1) at (3,0) [circle, draw, minimum size=1cm] {}; % D to the right

        % Diagonal arrows (single arrowhead in the middle)
        \draw[postaction={decorate},
              decoration={markings, mark=at position 0.5 with {\arrow{>}}}] 
              (A1) -- (B1);
              
        \draw[postaction={decorate},
              decoration={markings, mark=at position 0.5 with {\arrow{>}}}] 
              (C1) -- (A1);

        \draw[postaction={decorate},
              decoration={markings, mark=at position 0.5 with {\arrow{>}}}] 
              (D1) -- (B1);
              
        \draw[postaction={decorate},
              decoration={markings, mark=at position 0.5 with {\arrow{>}}}] 
              (C1) -- (D1);

        % Vertical arrow (double arrowheads in the middle)
        \draw[postaction={decorate},
              decoration={markings, 
              mark=at position 0.45 with {\arrow{>}},
              mark=at position 0.55 with {\arrow{>}}}] 
              (B1) -- (C1);

        % Second Quiver (Middle)
        \node (A2) at (5,0) [circle, draw, minimum size=1cm] {};
        \node (B2) at (6.5,1.5) [circle, draw, minimum size=1cm] {};
        \node (C2) at (6.5,-1.5) [circle, draw, minimum size=1cm] {};
        \node (D2) at (8,0) [circle, draw, minimum size=1cm] {}; % D to the right

        % Reversed diagonal arrows
        \draw[postaction={decorate},
              decoration={markings, mark=at position 0.5 with {\arrow{>}}}] 
              (B2) -- (A2);
              
        \draw[postaction={decorate},
              decoration={markings, mark=at position 0.5 with {\arrow{>}}}] 
              (A2) -- (C2);

        \draw[postaction={decorate},
              decoration={markings, mark=at position 0.5 with {\arrow{>}}}] 
              (B2) -- (D2);
              
        \draw[postaction={decorate},
              decoration={markings, mark=at position 0.5 with {\arrow{>}}}] 
              (D2) -- (C2);

        % No vertical arrow (B2 → C2 is removed)

        % Third Quiver (Right)
        \node (A3) at (10,0) [circle, draw, minimum size=1cm] {};
        \node (B3) at (11.5,1.5) [circle, draw, minimum size=1cm] {};
        \node (C3) at (11.5,-1.5) [circle, draw, minimum size=1cm] {};
        \node (D3) at (13,0) [circle, draw, minimum size=1cm] {}; % D to the right

        % Same as Middle quiver but flip bottom two arrows
        \draw[postaction={decorate},
              decoration={markings, mark=at position 0.5 with {\arrow{>}}}] 
              (A3) -- (B3);
              
        \draw[postaction={decorate},
              decoration={markings, mark=at position 0.5 with {\arrow{>}}}] 
              (A3) -- (C3);

        \draw[postaction={decorate},
              decoration={markings, mark=at position 0.5 with {\arrow{>}}}] 
              (D3) -- (B3);
              
        \draw[postaction={decorate},
              decoration={markings, mark=at position 0.5 with {\arrow{>}}}] 
              (D3) -- (C3);

    \end{tikzpicture}
    \caption{BPS quivers for $N_f=2$. Left: The quiver associated to the partial RG flow to pure gauge theory. A mutation at the side nodes $\gamma_3$ and $\gamma_4$ gives the middle quiver, which admits a total order and allows for an immediate computation of the spectrum generator $S'$. A mutation at the top node $\gamma_1$ leads to the quiver on the right, which is particularly symmetric}
    \label{fig:quiver_three_modified_rightD}
\end{figure}
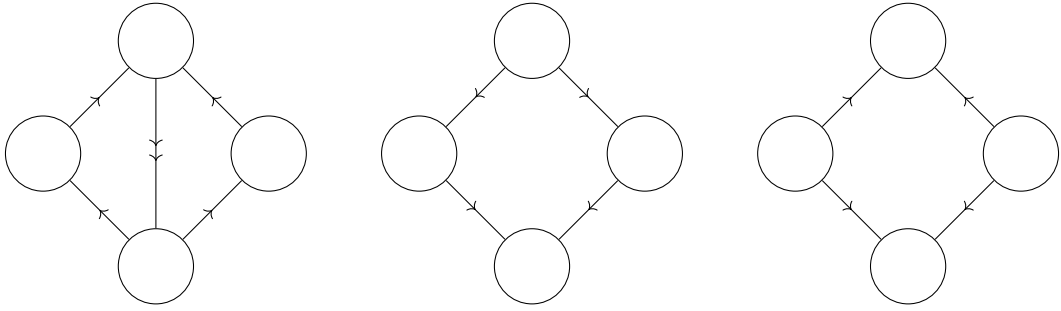

We reorder the factors to 
\begin{equation}
    S' = E_{\fq}(X_{\gamma_1})E_{\fq}(X_{\gamma_2+\gamma_3+\gamma_4}) E_{\fq}(X_{\gamma_2+\gamma_4})E_{\fq}(X_{\gamma_2+\gamma_3})E_{\fq}(X_{\gamma_2}) E_{\fq}(X_{-\gamma_3}) E_{\fq}(X_{-\gamma_4})
\end{equation}
and mutate back to  
\begin{equation}
    S_\UV = E_{\fq}(X_{\gamma_3}) E_{\fq}(X_{\gamma_4})E_{\fq}(X_{\gamma_1})E_{\fq}(X_{\gamma_2+\gamma_3+\gamma_4}) E_{\fq}(X_{\gamma_2+\gamma_4})E_{\fq}(X_{\gamma_2+\gamma_3}) E_{\fq}(X_{\gamma_2}) 
\end{equation}
We can reorganize that to an RG flow form:
\begin{equation}
    S_\UV =  E_{\fq}(X_{\gamma_4}) E_{\fq}(X_{\gamma_2+\gamma_4}) E_{\fq}(X_{\gamma_1 +\gamma_2+\gamma_4})E_{\fq}(X_{\gamma_3}) E_{\fq}(X_{\gamma_2+\gamma_3})E_{\fq}(X_{\gamma_1 + \gamma_2+\gamma_3})S_\IR
\end{equation}
We recognize the factors in front as $f_2(\mu_1,w_a) f_2(\mu_2,w_a)$ with $\mu_1 =X_{\frac12 \gamma_1+\frac12 \gamma_2+ \gamma_3}$ and $\mu_2 =X_{\frac12 \gamma_1+\frac12 \gamma_2+ \gamma_4}$, as expected.

For future use, it is also useful to mutate the original expression at $\gamma_1$ to 
\begin{equation}
   S'' =  E_{\fq}(X_{-\gamma_3})E_{\fq}(X_{-\gamma_4})E_{\fq}(X_{\gamma_2+\gamma_3+\gamma_4}) E_{\fq}(X_{-\gamma_1})
\end{equation}
This form of the spectrum generator makes manifest the full $\mathrm{Spin}(4) = SU(2) \times SU(2)$ symmetry of the system, which is hidden in the other chambers. Indeed, we can write the charges as  
\begin{align}
     S'' &= E_{\fq}(\mu^{-1}_+ X_{-\frac12 \gamma_3- \frac12 \gamma_4 })E_{\fq}(\mu_+ X_{-\frac12 \gamma_3- \frac12 \gamma_4 }) \cdot \cr &\cdot E_{\fq}(\mu_- X_{-\frac12 \gamma_1 +\frac12 \gamma_2+\frac12\gamma_3+\frac12\gamma_4}) E_{\fq}(\mu^{-1}_- X_{-\frac12 \gamma_1 +\frac12 \gamma_2+\frac12\gamma_3+\frac12\gamma_4})
\end{align}
with $\mu_+ = X_{\frac12 \gamma_3- \frac12 \gamma_4 }$ and $\mu_- = X_{\frac12 \gamma_1 +\frac12 \gamma_2+\frac12\gamma_3+\frac12\gamma_4}$
being the flavor parameters for the two $SU(2)$ global symmetries. 

We can transport the Wilson line image predicted by our conjecture 
\begin{equation}
    F^\UV_{D_{0,1}} = X_{-\frac12 \gamma_1-\frac12 \gamma_2}+X_{-\frac12 \gamma_1+\frac12 \gamma_2}+X_{\frac12 \gamma_1+\frac12 \gamma_2}
\end{equation}
to 
\begin{equation}
    F'_{D_{0,1}} = X_{-\frac12 \gamma_1-\frac12 \gamma_2}+X_{-\frac12 \gamma_1+\frac12 \gamma_2}+X_{-\frac12 \gamma_1+\frac12 \gamma_2+ \gamma_3}+X_{-\frac12 \gamma_1+\frac12 \gamma_2+ \gamma_4}+X_{-\frac12 \gamma_1+\frac12 \gamma_2+\gamma_3 + \gamma_4}+X_{\frac12 \gamma_1+\frac12 \gamma_2}
\end{equation}
and then to 
\begin{align}
    F''_{D_{0,1}} &= X_{\frac12 \gamma_1-\frac12 \gamma_2}+(X_{\frac12 \gamma_1+\frac12 \gamma_2+\gamma_3 + \gamma_4}+X_{-\frac12 \gamma_1-\frac12 \gamma_2})+X_{-\frac12 \gamma_1+\frac12 \gamma_2+\gamma_3 + \gamma_4}+ \cr &+(X_{-\frac12 \gamma_1+\frac12 \gamma_2+ \gamma_3}+X_{-\frac12 \gamma_1+\frac12 \gamma_2+ \gamma_4})+X_{-\frac12 \gamma_1+\frac12 \gamma_2}
\end{align}
Here we grouped terms which form doublets for the two $SU(2)$ global symmetries.

A simple guess for $\RG(D_{1,0})$ based on the UV formulae and the $N_f=1$ example is 
\begin{equation}
    \RG(D_{1,0})= H_0^2+ \fq^{-\frac12} (\mu_1+\mu_2) H_{-1} H_0 + \fq^{-1} \mu_1 \mu_2 H_{-2} H_0 \, ,
\end{equation}
though the definition of $D_{1,0}$, even in the UV, could be corrected by subtracting a multiple of $1$ as was done in \cite{Gaiotto:2024osr}. This would replace $\fq^{-1} H_{-2} H_0 \to H_{-1}^2$. 

This gives 
\begin{equation}
    F^\UV_{D_{1,0}}= X_{\gamma_1}+  X_{\gamma_1+\gamma_3}+  X_{\gamma_1+\gamma_4} + X_{\gamma_1+\gamma_3+ \gamma_4} \, ,
\end{equation}
which mutates to a very simple 
\begin{equation}
    F''_{D_{1,0}}= X_{\gamma_1} \, ,
\end{equation}
Another simple monomial $F$ is 
\begin{equation}
    F''_{D_{1,-1}}= X_{\frac12 \gamma_1 + \frac12 \gamma_2} \, ,
\end{equation}
i.e. 
\begin{equation}
    F^\UV_{D_{1,-1}}= X_{\frac12 \gamma_1 + \frac12 \gamma_2}+ X_{\frac32 \gamma_1 + \frac12 \gamma_2}+ X_{\frac32 \gamma_1 + \frac12 \gamma_2+ \gamma_3} + X_{\frac32 \gamma_1 + \frac12 \gamma_2+ \gamma_4}+ X_{\frac32 \gamma_1 + \frac12 \gamma_2+ \gamma_3+ \gamma_4}\, ,
\end{equation}
i.e. 
\begin{equation}
    \RG(D_{1,-1})= H_0 H_1+ (\mu_1+\mu_2) H_0^2 + \mu_1 \mu_2 H_0 H_{-1} \, .
\end{equation}

\subsubsection{$N_f=3$}
For brevity, we can start from the desired expression  
\begin{equation}
    S_\UV = f_2(\mu_1,w_a) f_2(\mu_2,w_a)f_2(\mu_3,w_a)S_\IR
\end{equation}
together with 
\begin{equation}
    F_{w_1} = X_{-\frac12 \gamma_1-\frac12 \gamma_2}+X_{-\frac12 \gamma_1+\frac12 \gamma_2}+X_{\frac12 \gamma_1+\frac12 \gamma_2}
\end{equation}
and borrow some manipulations from the previous example, 
\begin{align}
    S_\UV &=  E_{\fq}(X_{\gamma_5}) E_{\fq}(X_{\gamma_4}) E_{\fq}(X_{\gamma_3})  E_{\fq}(X_{\gamma_2+\gamma_3+\gamma_4+\gamma_5}) E_{\fq}(X_{\gamma_1}) E_{\fq}(X_{\gamma_2+\gamma_4+\gamma_5}) E_{\fq}(X_{\gamma_2+\gamma_3+\gamma_5})    \cdot \cr &\cdot  E_{\fq}(X_{\gamma_2+\gamma_5})  E_{\fq}(X_{\gamma_2+\gamma_3+\gamma_4}) E_{\fq}(X_{\gamma_2+\gamma_4})E_{\fq}(X_{\gamma_2+\gamma_3}) E_{\fq}(X_{\gamma_2}) 
\end{align}
then mutate at the three nodes 
\begin{align}
    S'&= E_{\fq}(X_{\gamma_2+\gamma_3+\gamma_4+\gamma_5}) E_{\fq}(X_{\gamma_1}) E_{\fq}(X_{\gamma_2+\gamma_4+\gamma_5}) E_{\fq}(X_{\gamma_2+\gamma_3+\gamma_5})    E_{\fq}(X_{\gamma_2+\gamma_5})  E_{\fq}(X_{\gamma_2+\gamma_3+\gamma_4}) \cdot \cr &\cdot  E_{\fq}(X_{\gamma_2+\gamma_4})E_{\fq}(X_{\gamma_2+\gamma_3}) E_{\fq}(X_{\gamma_2}) E_{\fq}(X_{-\gamma_3}) E_{\fq}(X_{-\gamma_5}) E_{\fq}(X_{-\gamma_4})  
\end{align}
with 
\begin{align}
    F'_{w_1} &= X_{-\frac12 \gamma_1-\frac12 \gamma_2}+X_{-\frac12 \gamma_1+\frac12 \gamma_2}+X_{-\frac12 \gamma_1+\frac12 \gamma_2+\gamma_3}+X_{-\frac12 \gamma_1+\frac12 \gamma_2+\gamma_4}+X_{-\frac12 \gamma_1+\frac12 \gamma_2+\gamma_3+\gamma_4}+ \cr &+X_{-\frac12 \gamma_1+\frac12 \gamma_2+\gamma_5}+X_{-\frac12 \gamma_1+\frac12 \gamma_2+\gamma_3+\gamma_5}+X_{-\frac12 \gamma_1+\frac12 \gamma_2+\gamma_4+\gamma_5}+X_{-\frac12 \gamma_1+\frac12 \gamma_2+\gamma_3+\gamma_4+\gamma_5}+X_{\frac12 \gamma_1+\frac12 \gamma_2}
\end{align}
reorganize
\begin{align}
    S'&= E_{\fq}(X_{\gamma_2+\gamma_3+\gamma_4+\gamma_5}) E_{\fq}(X_{\gamma_1}) E_{\fq}(X_{\gamma_2+\gamma_4+\gamma_5})E_{\fq}(X_{-\gamma_3})  E_{\fq}(X_{\gamma_2+\gamma_3+\gamma_5}) \cdot \cr &\cdot  E_{\fq}(X_{-\gamma_4}) E_{\fq}(X_{\gamma_2+\gamma_3+\gamma_4})   E_{\fq}(X_{-\gamma_5}) 
\end{align}
and mutate again at $\gamma_2+\gamma_3+\gamma_4+\gamma_5$:
\begin{align}
    S''&= E_{\fq}(X_{\gamma_1}) E_{\fq}(X_{\gamma_2+\gamma_4+\gamma_5})E_{\fq}(X_{-\gamma_3})  E_{\fq}(X_{\gamma_2+\gamma_3+\gamma_5}) \cdot \cr &\cdot  E_{\fq}(X_{-\gamma_4}) E_{\fq}(X_{\gamma_2+\gamma_3+\gamma_4})   E_{\fq}(X_{-\gamma_5}) E_{\fq}(X_{-\gamma_2-\gamma_3-\gamma_4-\gamma_5})  
\end{align}
with 
\begin{align}
    F''_{w_1} &= X_{-\frac12 \gamma_1-\frac12 \gamma_2}+X_{-\frac12 \gamma_1+\frac12 \gamma_2}+(\fq + \fq^{-1})X_{-\frac12 \gamma_1+\frac32 \gamma_2+\gamma_3+\gamma_4+\gamma_5}+X_{-\frac12 \gamma_1+\frac52 \gamma_2+2\gamma_3+2\gamma_4+2\gamma_5}+\cr &+X_{-\frac12 \gamma_1+\frac12 \gamma_2+\gamma_3}+X_{-\frac12 \gamma_1+\frac32 \gamma_2+2\gamma_3+\gamma_4+\gamma_5}+X_{-\frac12 \gamma_1+\frac12 \gamma_2+\gamma_4}+X_{-\frac12 \gamma_1+\frac32 \gamma_2+\gamma_3+2\gamma_4+\gamma_5}+\cr &+X_{-\frac12 \gamma_1+\frac12 \gamma_2+\gamma_3+\gamma_4}+X_{-\frac12 \gamma_1+\frac12 \gamma_2+\gamma_5}+X_{-\frac12 \gamma_1+\frac32 \gamma_2+\gamma_3 + \gamma_4 + 2\gamma_5}+X_{-\frac12 \gamma_1+\frac12 \gamma_2+\gamma_3+\gamma_5}+\cr &+X_{-\frac12 \gamma_1+\frac12 \gamma_2+\gamma_4+\gamma_5}+X_{\frac12 \gamma_1+\frac12 \gamma_2}+X_{\frac12 \gamma_1+\frac32 \gamma_2+\gamma_3+\gamma_4+\gamma_5}
\end{align}
reorganized finally to 
\begin{align}
    S''&=   E_{\fq}(X_{\gamma_1})E_{\fq}(X_{-\gamma_2-\gamma_3-\gamma_4-\gamma_5})E_{\fq}(X_{\gamma_2+\gamma_4+\gamma_5}) E_{\fq}(X_{\gamma_2+\gamma_3+\gamma_5}) E_{\fq}(X_{\gamma_2+\gamma_3+\gamma_4})    
\end{align}
and mutated to a standard form:
\begin{align}
    S'''&= E_{\fq}(X_{-\gamma_2-\gamma_3-\gamma_4-\gamma_5})E_{\fq}(X_{\gamma_2+\gamma_4+\gamma_5}) E_{\fq}(X_{\gamma_2+\gamma_3+\gamma_5}) E_{\fq}(X_{\gamma_2+\gamma_3+\gamma_4}) E_{\fq}(X_{-\gamma_1})
\end{align}
with a quadruplet of commuting factors on the right which makes manifest the $SU(4) = \mathrm{Spin}(6)$ symmetry of the system. 

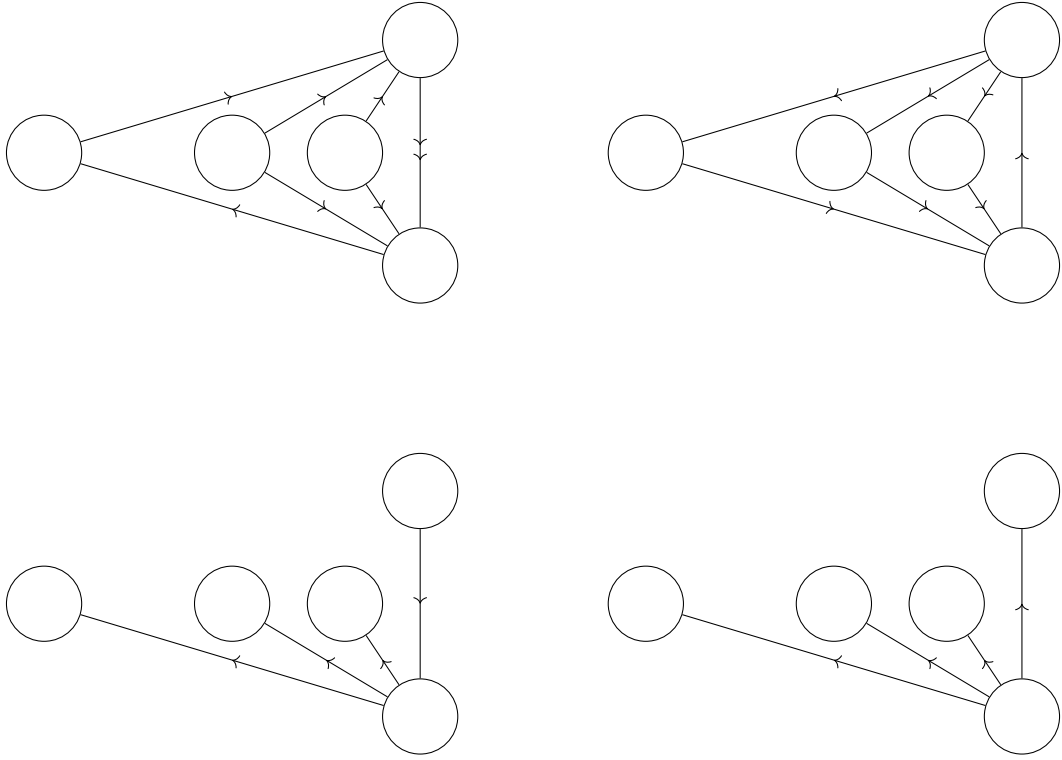
\begin{figure}[h]
    \centering
    \begin{tikzpicture}
        % First Quiver (Top-Left)
        \node (A1) at (0,0) [circle, draw, minimum size=1cm] {};
        \node (B1) at (5,1.5) [circle, draw, minimum size=1cm] {};
        \node (C1) at (5,-1.5) [circle, draw, minimum size=1cm] {};
        \node (D1) at (2.5,0) [circle, draw, minimum size=1cm] {};
        \node (E1) at (4,0) [circle, draw, minimum size=1cm] {};

        % Arrows for First Quiver
        \draw[postaction={decorate},
              decoration={markings, mark=at position 0.5 with {\arrow{>}}}] (A1) -- (B1);
        \draw[postaction={decorate},
              decoration={markings, mark=at position 0.5 with {\arrow{>}}}] (C1) -- (A1);
        \draw[postaction={decorate},
              decoration={markings, mark=at position 0.5 with {\arrow{>}}}] (D1) -- (B1);
        \draw[postaction={decorate},
              decoration={markings, mark=at position 0.5 with {\arrow{>}}}] (D1) -- (C1);
        \draw[postaction={decorate},
              decoration={markings, mark=at position 0.5 with {\arrow{>}}}] (E1) -- (B1);
        \draw[postaction={decorate},
              decoration={markings, mark=at position 0.5 with {\arrow{>}}}] (E1) -- (C1);

        % Vertical arrow (double arrowheads)
        \draw[postaction={decorate},
              decoration={markings, mark=at position 0.45 with {\arrow{>}}, mark=at position 0.55 with {\arrow{>}}}] 
              (B1) -- (C1);

        % Second Quiver (Top-Right)
        \node (A2) at (8,0) [circle, draw, minimum size=1cm] {};
        \node (B2) at (13,1.5) [circle, draw, minimum size=1cm] {};
        \node (C2) at (13,-1.5) [circle, draw, minimum size=1cm] {};
        \node (D2) at (10.5,0) [circle, draw, minimum size=1cm] {};
        \node (E2) at (12,0) [circle, draw, minimum size=1cm] {};

        % Arrows for Second Quiver (flipped)
        \draw[postaction={decorate},
              decoration={markings, mark=at position 0.5 with {\arrow{>}}}] (B2) -- (A2);
        \draw[postaction={decorate},
              decoration={markings, mark=at position 0.5 with {\arrow{>}}}] (A2) -- (C2);
        \draw[postaction={decorate},
              decoration={markings, mark=at position 0.5 with {\arrow{>}}}] (B2) -- (D2);
        \draw[postaction={decorate},
              decoration={markings, mark=at position 0.5 with {\arrow{>}}}] (D2) -- (C2);
        \draw[postaction={decorate},
              decoration={markings, mark=at position 0.5 with {\arrow{>}}}] (B2) -- (E2);
        \draw[postaction={decorate},
              decoration={markings, mark=at position 0.5 with {\arrow{>}}}] (E2) -- (C2);

        % Vertical arrow (single arrowhead)
        \draw[postaction={decorate},
              decoration={markings, mark=at position 0.5 with {\arrow{>}}}] 
              (C2) -- (B2);

        % Third Quiver (Bottom-Left)
        \node (A3) at (0,-6) [circle, draw, minimum size=1cm] {};
        \node (B3) at (5,-4.5) [circle, draw, minimum size=1cm] {};
        \node (C3) at (5,-7.5) [circle, draw, minimum size=1cm] {};
        \node (D3) at (2.5,-6) [circle, draw, minimum size=1cm] {};
        \node (E3) at (4,-6) [circle, draw, minimum size=1cm] {};

        % Arrows for Third Quiver (removed top three diagonal arrows, flipped others)
        \draw[postaction={decorate},
              decoration={markings, mark=at position 0.5 with {\arrow{>}}}] (C3) -- (A3);
        \draw[postaction={decorate},
              decoration={markings, mark=at position 0.5 with {\arrow{>}}}] (C3) -- (D3);
        \draw[postaction={decorate},
              decoration={markings, mark=at position 0.5 with {\arrow{>}}}] (C3) -- (E3);

        % Vertical arrow (single arrowhead)
        \draw[postaction={decorate},
              decoration={markings, mark=at position 0.5 with {\arrow{>}}}] 
              (B3) -- (C3);

        % Fourth Quiver (Bottom-Right)
        \node (A4) at (8,-6) [circle, draw, minimum size=1cm] {};
        \node (B4) at (13,-4.5) [circle, draw, minimum size=1cm] {};
        \node (C4) at (13,-7.5) [circle, draw, minimum size=1cm] {};
        \node (D4) at (10.5,-6) [circle, draw, minimum size=1cm] {};
        \node (E4) at (12,-6) [circle, draw, minimum size=1cm] {};

        % Arrows for Fourth Quiver (flipped vertical arrow)
        \draw[postaction={decorate},
              decoration={markings, mark=at position 0.5 with {\arrow{>}}}] (C4) -- (A4);
        \draw[postaction={decorate},
              decoration={markings, mark=at position 0.5 with {\arrow{>}}}] (C4) -- (D4);
        \draw[postaction={decorate},
              decoration={markings, mark=at position 0.5 with {\arrow{>}}}] (C4) -- (E4);

        % Flipped vertical arrow
        \draw[postaction={decorate},
              decoration={markings, mark=at position 0.5 with {\arrow{>}}}] 
              (C4) -- (B4);

    \end{tikzpicture}
    \caption{Top Left: the BPS quiver compatibl with the RG flow from $N_f=3$ to pure $SU(2)$. Top Right: mutation at the three matter nodes gives this quiver. Bottom Left: Mutation at the bottom node simplifies the quiver drastically. Bottom Right: A final mutation at the top node makes a large symmetry manifest.}
    \label{fig:quiver_2x2}
\end{figure}

The final expression for the Wilson line is somewhat dreadful but $SU(4)$-covariant
\begin{align}
    F'''_{w_1} &= X_{\frac12 \gamma_1-\frac12 \gamma_2}+(X_{-\frac12 \gamma_1-\frac12 \gamma_2}+X_{\frac12 \gamma_1+\frac12 \gamma_2+\gamma_3+\gamma_4}+X_{\frac12 \gamma_1+\frac12 \gamma_2+\gamma_3+\gamma_5}+X_{\frac12 \gamma_1+\frac12 \gamma_2+\gamma_4+\gamma_5})+\cr &+(X_{-\frac12 \gamma_1+\frac12 \gamma_2+\gamma_3+\gamma_4}+X_{-\frac12 \gamma_1+\frac12 \gamma_2+\gamma_3+\gamma_5}+X_{-\frac12 \gamma_1+\frac12 \gamma_2+\gamma_4+\gamma_5}+\cr &+X_{\frac12 \gamma_1+\frac32 \gamma_2+\gamma_3+2\gamma_4+\gamma_5}+X_{\frac12 \gamma_1+\frac32 \gamma_2+2\gamma_3+\gamma_4+\gamma_5}+X_{\frac12 \gamma_1+\frac32 \gamma_2+\gamma_3+\gamma_4+2\gamma_5})+\cr &+(X_{-\frac12 \gamma_1+\frac32 \gamma_2+2\gamma_3+\gamma_4+\gamma_5}+X_{-\frac12 \gamma_1+\frac32 \gamma_2+\gamma_3+2\gamma_4+\gamma_5}+X_{-\frac12 \gamma_1+\frac32 \gamma_2+\gamma_3 + \gamma_4 + 2\gamma_5}+X_{\frac12 \gamma_1+\frac52 \gamma_2+2\gamma_3+2\gamma_4+2\gamma_5})+\cr &+(X_{-\frac12 \gamma_1+\frac12 \gamma_2+\gamma_3}+X_{-\frac12 \gamma_1+\frac12 \gamma_2+\gamma_4}+X_{-\frac12 \gamma_1+\frac12 \gamma_2+\gamma_5}+X_{\frac12 \gamma_1+\frac32 \gamma_2+\gamma_3+\gamma_4+\gamma_5})+\cr &+X_{-\frac12 \gamma_1+\frac52 \gamma_2+2\gamma_3+2\gamma_4+2\gamma_5}+(\fq + \fq^{-1})X_{-\frac12 \gamma_1+\frac32 \gamma_2+\gamma_3+\gamma_4+\gamma_5}+X_{-\frac12 \gamma_1+\frac12 \gamma_2}
\end{align}

Conversely, starting from 
\begin{equation}
    F'''_{D_{1,0}} = X_{\gamma_1}
\end{equation}
we get to 
\begin{align}
    F^\UV_{D_{1,0}} &= X_{\gamma_1}+X_{\gamma_1+\gamma_3}+X_{\gamma_1+\gamma_4}+X_{\gamma_1+\gamma_5}+X_{\gamma_1+\gamma_3+\gamma_4}+ \cr &+X_{\gamma_1+\gamma_3+\gamma_5}+X_{\gamma_1+\gamma_4+\gamma_5}+X_{\gamma_1+\gamma_3+\gamma_4+\gamma_5}+X_{\gamma_1+ \gamma_2 + \gamma_3 + \gamma_4 + \gamma_5}
\end{align}
i.e. 
\begin{equation}
    \RG(D_{1,0}) = H_0^2+(\mu_1 + \mu_2 + \mu_3) \fq^{-\frac12} H_{-1} H_0 + ( \mu_1 \mu_2 + \mu_2 \mu_3 + \mu_1 \mu_3)H_{-1}^2+ \mu_1 \mu_2 \mu_3 H_{-2} H_{-1} 
\end{equation}

\subsubsection{$N_f=4$}
We conjecture the expression 
\begin{equation}
    S_\UV = f_2(\mu_1,w_a) f_2(\mu_2,w_a)f_2(\mu_3,w_a)f_2(\mu_4,w_a)S_\IR
\end{equation}
for the spectrum generator here, i.e.
\begin{align}
    S_\UV &= E_{\fq}(X_{\gamma_4}) E_{\fq}(X_{\gamma_2+\gamma_4}) E_{\fq}(X_{\gamma_1 +\gamma_2+\gamma_4})E_{\fq}(X_{\gamma_3}) E_{\fq}(X_{\gamma_2+\gamma_3})E_{\fq}(X_{\gamma_1 + \gamma_2+\gamma_3}) \cdot \cr &\cdot E_{\fq}(X_{\gamma_5}) E_{\fq}(X_{\gamma_2+\gamma_5}) E_{\fq}(X_{\gamma_1 +\gamma_2+\gamma_5})E_{\fq}(X_{\gamma_6}) E_{\fq}(X_{\gamma_2+\gamma_6})E_{\fq}(X_{\gamma_1 + \gamma_2+\gamma_6})\cdot \cr &\cdot E_\fq(X_{\gamma_1})E_\fq(X_{\gamma_2})
\end{align}
We are not aware of a finite chamber which makes the $SO(8)$ flavour symmetry of the theory manifest. A nice option is to mutate at $\gamma_3$, $\gamma_4$ and $\gamma_2 + \gamma_3+\gamma_4$:
\begin{align}
    S' &=  E_{\fq}(X_{\gamma_2+\gamma_4}) E_{\fq}(X_{\gamma_1 +\gamma_2+\gamma_4}) E_{\fq}(X_{\gamma_2+\gamma_3})E_{\fq}(X_{\gamma_1 + \gamma_2+\gamma_3}) \cdot \cr &\cdot E_{\fq}(X_{\gamma_5}) E_{\fq}(X_{\gamma_2+\gamma_5}) E_{\fq}(X_{\gamma_1 +\gamma_2+\gamma_5})E_{\fq}(X_{\gamma_6}) E_{\fq}(X_{\gamma_2+\gamma_6})E_{\fq}(X_{\gamma_1 + \gamma_2+\gamma_6})\cdot \cr &\cdot E_\fq(X_{\gamma_1})E_\fq(X_{\gamma_2})E_{\fq}(X_{-\gamma_4})E_{\fq}(X_{-\gamma_3})E_{\fq}(X_{-\gamma_2 - \gamma_3-\gamma_4})
\end{align}
The simple charges are now $\gamma_1$, $-\gamma_2 - \gamma_3-\gamma_4$, $\gamma_2+\gamma_3$, $\gamma_2 + \gamma_4$, $\gamma_5$, $\gamma_6$. 
This is an extremely symmetric collection: the BPS quiver is an octahedron. If $S'$ is determined by the BPS quiver, it should have the same symmetry, though it is not manifest. In particular, this includes cyclic permutations of the three pairs of consecutive charges. 

\begin{figure}[h]
    \centering
    \begin{tikzpicture}
        % ======================
        % Left Octahedral Quiver
        % ======================
        % Outer triangle (upward)
        \node (A1) at (0,-4) [circle, draw, minimum size=1cm] {};
        \node (B1) at (-3,1) [circle, draw, minimum size=1cm] {};
        \node (C1) at (3,1) [circle, draw, minimum size=1cm] {};

        % Inner triangle (downward)
        \node (D1) at (0,0) [circle, draw, minimum size=1cm] {};
        \node (E1) at (-.75,-1.5) [circle, draw, minimum size=1cm] {};
        \node (F1) at (.75,-1.5) [circle, draw, minimum size=1cm] {};

        % Outer triangle cyclic arrows
        \draw[postaction={decorate},
              decoration={markings, mark=at position 0.5 with {\arrow{>}}}] 
              (C1) -- (A1);
        \draw[postaction={decorate},
              decoration={markings, mark=at position 0.5 with {\arrow{>}}}] 
              (A1) -- (B1);

        % Inner triangle (clockwise)
        \draw[postaction={decorate},
              decoration={markings, mark=at position 0.5 with {\arrow{>}}}] 
              (D1) -- (E1);
        \draw[postaction={decorate},
              decoration={markings, mark=at position 0.5 with {\arrow{>}}}] 
              (F1) -- (D1);

        % Connections between inner and outer triangles
        \draw[postaction={decorate},
              decoration={markings, mark=at position 0.5 with {\arrow{>}}}] 
              (E1) -- (A1);
        \draw[postaction={decorate},
              decoration={markings, mark=at position 0.5 with {\arrow{>}}}] 
              (B1) -- (D1);
        \draw[postaction={decorate},
              decoration={markings, mark=at position 0.5 with {\arrow{>}}}] 
              (D1) -- (C1);
        \draw[postaction={decorate},
              decoration={markings, mark=at position 0.5 with {\arrow{>}}}] 
              (A1) -- (F1);

        % ======================
        % Right Octahedral Quiver (Shifted)
        % ======================
        % Outer triangle (upward)
        \node (A2) at (8,-4) [circle, draw, minimum size=1cm] {};
        \node (B2) at (5,1) [circle, draw, minimum size=1cm] {};
        \node (C2) at (11,1) [circle, draw, minimum size=1cm] {};

        % Inner triangle (downward)
        \node (D2) at (8,0) [circle, draw, minimum size=1cm] {};
        \node (E2) at (7.25,-1.5) [circle, draw, minimum size=1cm] {};
        \node (F2) at (8.75,-1.5) [circle, draw, minimum size=1cm] {};

        % Outer triangle cyclic arrows
        \draw[postaction={decorate},
              decoration={markings, mark=at position 0.5 with {\arrow{>}}}] 
              (A2) -- (C2);
        \draw[postaction={decorate},
              decoration={markings, mark=at position 0.5 with {\arrow{>}}}] 
              (C2) -- (B2);
        \draw[postaction={decorate},
              decoration={markings, mark=at position 0.5 with {\arrow{>}}}] 
              (B2) -- (A2);

        % Inner triangle (clockwise)
        \draw[postaction={decorate},
              decoration={markings, mark=at position 0.5 with {\arrow{>}}}] 
              (D2) -- (E2);
        \draw[postaction={decorate},
              decoration={markings, mark=at position 0.5 with {\arrow{>}}}] 
              (E2) -- (F2);
        \draw[postaction={decorate},
              decoration={markings, mark=at position 0.5 with {\arrow{>}}}] 
              (F2) -- (D2);

        % Connections between inner and outer triangles
        \draw[postaction={decorate},
              decoration={markings, mark=at position 0.5 with {\arrow{>}}}] 
              (A2) -- (E2);
        \draw[postaction={decorate},
              decoration={markings, mark=at position 0.5 with {\arrow{>}}}] 
              (B2) -- (D2);
        \draw[postaction={decorate},
              decoration={markings, mark=at position 0.5 with {\arrow{>}}}] 
              (C2) -- (F2);
        \draw[postaction={decorate},
              decoration={markings, mark=at position 0.5 with {\arrow{>}}}] 
              (E2) -- (B2);
        \draw[postaction={decorate},
              decoration={markings, mark=at position 0.5 with {\arrow{>}}}] 
              (D2) -- (C2);
        \draw[postaction={decorate},
              decoration={markings, mark=at position 0.5 with {\arrow{>}}}] 
              (F2) -- (A2);

    \end{tikzpicture}
    \caption{Mutating two matter nodes in the BPS quiver associated to the RG flow from pure $N_f=4$ to pure $SU(2)$ produces the Left quiver. Mutation at the bottom node produces the octahedral form of the quiver.}
    \label{fig:double_octahedral_quiver}
\end{figure}
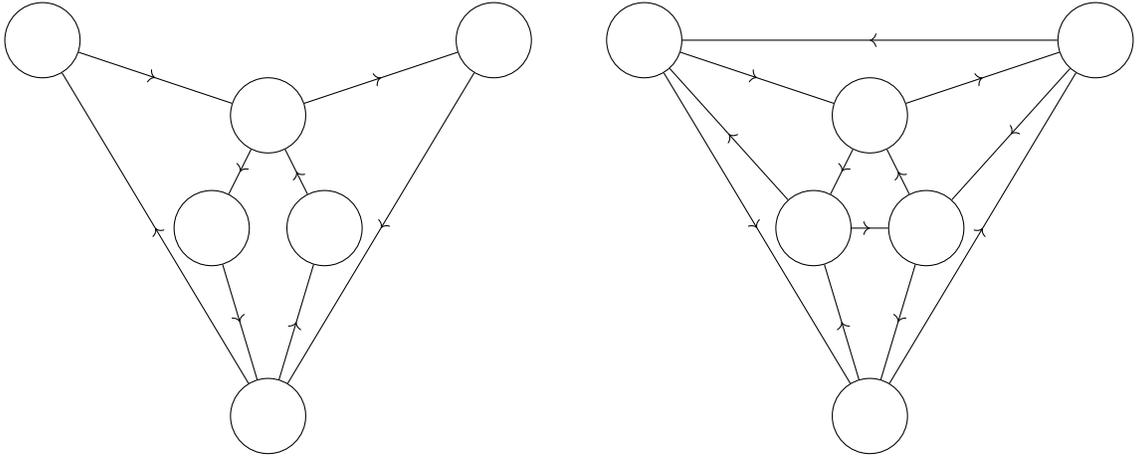

The Wilson loop in this chamber has the same general form as for $N_f=2$ 
\begin{align}
    F'_{D_{0,1}}&= X_{-\frac12 \gamma_1-\frac12 \gamma_2}+ (X_{-\frac12 \gamma_1+\frac12 \gamma_2+ \gamma_3}+ X_{-\frac12 \gamma_1+\frac12 \gamma_2+\gamma_4})+ X_{-\frac12 \gamma_1+\frac32 \gamma_2+ \gamma_3 + \gamma_4}+\cr &+ (X_{-\frac12 \gamma_1+\frac12 \gamma_2}+ X_{\frac12 \gamma_1+\frac32 \gamma_2+ \gamma_3 + \gamma_4})+ X_{\frac12 \gamma_1+\frac12 \gamma_2}
\end{align}
If the expected symmetry holds, we should have two more simple $F$'s:
\begin{align}
    F'_{D_{1,0}}&= X_{- \gamma_2-\frac12 \gamma_3 - \frac12 \gamma_4-\frac12 \gamma_5 - \frac12 \gamma_6}+ (X_{- \gamma_2-\frac12 \gamma_3 - \frac12 \gamma_4+\frac12 \gamma_5 - \frac12 \gamma_6}+ X_{- \gamma_2-\frac12 \gamma_3 - \frac12 \gamma_4-\frac12 \gamma_5 + \frac12 \gamma_6})+\cr &+ X_{- \gamma_2-\frac12 \gamma_3 - \frac12 \gamma_4+\frac12 \gamma_5 + \frac12 \gamma_6}+ (X_{\frac12 \gamma_3 - \frac12 \gamma_4+\frac12 \gamma_5 + \frac12 \gamma_6}+ X_{-\frac12 \gamma_3 + \frac12 \gamma_4+\frac12 \gamma_5 + \frac12 \gamma_6})+ X_{ \gamma_2+\frac12 \gamma_3 + \frac12 \gamma_4+\frac12 \gamma_5 + \frac12 \gamma_6} \cr
    F'_{D_{1,1}}&= X_{-\frac12 \gamma_1+\frac12 \gamma_2+\frac12 \gamma_3+\frac12 \gamma_4-\frac12 \gamma_5 - \frac12 \gamma_6}+ (X_{\frac12 \gamma_1+\frac12 \gamma_2+\frac12 \gamma_3+\frac12 \gamma_4-\frac12 \gamma_5 - \frac12 \gamma_6}+ X_{-\frac12 \gamma_1-\frac12 \gamma_2-\frac12 \gamma_3-\frac12 \gamma_4-\frac12 \gamma_5 - \frac12 \gamma_6})+\cr &+ X_{\frac12 \gamma_1-\frac12 \gamma_2-\frac12 \gamma_3-\frac12 \gamma_4-\frac12 \gamma_5 - \frac12 \gamma_6}+ (X_{\frac12 \gamma_1-\frac12 \gamma_2-\frac12 \gamma_3-\frac12 \gamma_4+\frac12 \gamma_5 - \frac12 \gamma_6}+ X_{\frac12 \gamma_1-\frac12 \gamma_2-\frac12 \gamma_3-\frac12 \gamma_4-\frac12 \gamma_5 + \frac12 \gamma_6})+\cr &+ X_{\frac12 \gamma_1-\frac12 \gamma_2-\frac12 \gamma_3-\frac12 \gamma_4+\frac12 \gamma_5 + \frac12 \gamma_6}
\end{align}
As a test, we can mutate back,
\begin{align}
    F^\UV_{D_{1,1}}&= (\mu_1 \mu_2 \mu_3 \mu_4)^{-\frac12}X_{\frac12 \gamma_1+\frac12 \gamma_2}+  (\mu_1 \mu_2 \mu_3 \mu_4)^{-\frac12} X_{\frac32 \gamma_1+\frac12 \gamma_2}+\cr &+ (\mu_1 \mu_2 \mu_3 \mu_4)^{-\frac12}X_{\gamma_1}(\mu_1 + \mu_2 + \mu_3 + \mu_4) + \cr &+(\mu_1 \mu_2 \mu_3 \mu_4)^{-\frac12}X_{\frac12 \gamma_1-\frac12 \gamma_2} \left(\sum_{i<j} \mu_i \mu_j\right)+ \cr &+(\mu_1 \mu_2 \mu_3 \mu_4)^{\frac12} X_{- \gamma_2}\left(\mu_1^{-1}+\mu_2^{-1}+\mu_3^{-1}+\mu_4^{-1}\right) +\cr &+ (\mu_1 \mu_2 \mu_3 \mu_4)^{\frac12} X_{-\frac12 \gamma_1-\frac32 \gamma_2}+ (\mu_1 \mu_2 \mu_3 \mu_4)^{\frac12} X_{-\frac12 \gamma_1-\frac12 \gamma_2}
\end{align}
i.e. 
\begin{align}
    \RG(D_{1,1})&= (\mu_1 \mu_2 \mu_3 \mu_4)^{-\frac12}\fq^{-\frac12} H_0 H_1+\cr &+ (\mu_1 \mu_2 \mu_3 \mu_4)^{-\frac12}(\mu_1 + \mu_2 + \mu_3 + \mu_4)H_0^2 + \cr &+(\mu_1 \mu_2 \mu_3 \mu_4)^{-\frac12} \left(\sum_{i<j} \mu_i \mu_j\right)\fq^{-\frac12} H_{-1} H_0+ \cr &+(\mu_1 \mu_2 \mu_3 \mu_4)^{\frac12} \left(\mu_1^{-1}+\mu_2^{-1}+\mu_3^{-1}+\mu_4^{-1}\right)H_{-1}^2 +\cr &+ (\mu_1 \mu_2 \mu_3 \mu_4)^{\frac12} \fq^{-\frac12} H_{-2} H_{-1}
\end{align}
as expected \cite{Gaiotto:2024osr}. 
  
\subsection[$SU(2) \times SU(2)$ quiver gauge theory]{\boldmath $SU(2) \times SU(2)$ quiver gauge theory}
\begin{figure}[h]
    \centering
    \begin{tikzpicture}
        % Define Nodes
        \node (A) at (0,1.5) [circle, draw, minimum size=1cm] {}; % Top-left
        \node (B) at (3,1.5) [circle, draw, minimum size=1cm] {}; % Top-right
        \node (C) at (0,-1.5) [circle, draw, minimum size=1cm] {}; % Bottom-left
        \node (D) at (3,-1.5) [circle, draw, minimum size=1cm] {}; % Bottom-right
        \node (M) at (1.5,0) [circle, draw, minimum size=1cm] {}; % Middle
        
        % Double vertical arrows (left)
        \draw[postaction={decorate},
              decoration={markings, 
              mark=at position 0.45 with {\arrow{>}},
              mark=at position 0.55 with {\arrow{>}}}] 
              (A) -- (C);
              
        % Double vertical arrows (right)
        \draw[postaction={decorate},
              decoration={markings, 
              mark=at position 0.45 with {\arrow{>}},
              mark=at position 0.55 with {\arrow{>}}}] 
              (B) -- (D);

        % Diagonal arrows to/from middle
        \draw[postaction={decorate},
              decoration={markings, mark=at position 0.5 with {\arrow{>}}}] 
              (M) -- (A);
              
        \draw[postaction={decorate},
              decoration={markings, mark=at position 0.5 with {\arrow{>}}}] 
              (M) -- (B);

        \draw[postaction={decorate},
              decoration={markings, mark=at position 0.5 with {\arrow{>}}}] 
              (C) -- (M);

        \draw[postaction={decorate},
              decoration={markings, mark=at position 0.5 with {\arrow{>}}}] 
              (D) -- (M);
    \end{tikzpicture}
    \caption{BPS quiver for $SU(2) \times SU(2)$ with bifundamental matter.}
    \label{fig:BPS_quiver_SU2xSU2}
\end{figure}
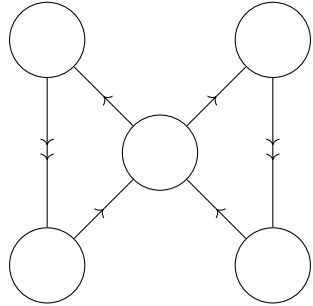

As a further example, we can consider an $SU(2)\times SU(2)$ gauge theory coupled to bi-fundamental matter. To do so, we need to expand 
\begin{equation}
    E_\fq(\mu \zeta_1 \zeta_2)E_\fq(\mu \zeta_1/\zeta_2)E_\fq(\mu/\zeta_1 \zeta_2)E_\fq(\mu/\zeta_1/\zeta_2)
\end{equation}
into characters of the two gauge groups and replace them with two sets of Wilson loops. We can do it in two steps. First as in $SU(2)$ $N_f=2$:
\begin{align} 
&E_{\fq}(\mu \zeta_2 X_{-\frac12 \gamma_1-\frac12 \gamma_2}) E_{\fq}(\mu \zeta_2 X_{-\frac12 \gamma_1+\frac12 \gamma_2})E_{\fq}(\mu \zeta_2 X_{\frac12 \gamma_1+\frac12 \gamma_2}) \cdot \cr &\cdot E_{\fq}(\mu/\zeta_2 X_{-\frac12 \gamma_1-\frac12 \gamma_2}) E_{\fq}(\mu/\zeta_2 X_{-\frac12 \gamma_1+\frac12 \gamma_2})E_{\fq}(\mu/\zeta_2 X_{\frac12 \gamma_1+\frac12 \gamma_2})
\end{align}
We reorganize it to 
\begin{align} 
&E_{\fq}(\mu \zeta_2 X_{-\frac12 \gamma_1-\frac12 \gamma_2}) E_{\fq}(\mu/\zeta_2 X_{-\frac12 \gamma_1-\frac12 \gamma_2})E_{\fq}(\mu^2 X_{- \gamma_1})E_{\fq}(\mu \zeta_2 X_{-\frac12 \gamma_1+\frac12 \gamma_2}) \cdot \cr &\cdot  E_{\fq}(\mu/\zeta_2 X_{-\frac12 \gamma_1+\frac12 \gamma_2})E_{\fq}(\mu^2 X_{ \gamma_2})E_{\fq}(\mu \zeta_2 X_{\frac12 \gamma_1+\frac12 \gamma_2})E_{\fq}(\mu/\zeta_2 X_{\frac12 \gamma_1+\frac12 \gamma_2})
\end{align}
and then we can readily do the replacement with a second set of Wilson lines:
\begin{align} 
&E_{\fq}(\mu X_{-\frac12 \gamma_1-\frac12 \gamma_2-\frac12 \gamma_3-\frac12 \gamma_4})E_{\fq}(\mu X_{-\frac12 \gamma_1-\frac12 \gamma_2-\frac12 \gamma_3+\frac12 \gamma_4})E_{\fq}(\mu X_{-\frac12 \gamma_1-\frac12 \gamma_2+\frac12 \gamma_3+\frac12 \gamma_4}) E_{\fq}(\mu^2 X_{- \gamma_1})\cdot \cr &\cdot
E_{\fq}(\mu X_{-\frac12 \gamma_1+\frac12 \gamma_2-\frac12 \gamma_3-\frac12 \gamma_4}) E_{\fq}(\mu  X_{-\frac12 \gamma_1+\frac12 \gamma_2-\frac12 \gamma_3+\frac12 \gamma_4}) E_{\fq}(\mu X_{-\frac12 \gamma_1+\frac12 \gamma_2+\frac12 \gamma_3+\frac12 \gamma_4})  E_{\fq}(\mu^2 X_{ \gamma_2})\cdot \cr &\cdot E_{\fq}(\mu X_{\frac12 \gamma_1+\frac12 \gamma_2-\frac12 \gamma_3-\frac12 \gamma_4})E_{\fq}(\mu  X_{\frac12 \gamma_1+\frac12 \gamma_2-\frac12 \gamma_3+\frac12 \gamma_4})E_{\fq}(\mu  X_{\frac12 \gamma_1+\frac12 \gamma_2+\frac12 \gamma_3+\frac12 \gamma_4})
\end{align}
and possibly simplify a bit:
\begin{align} 
&E_{\fq}(\mu X_{-\frac12 \gamma_1-\frac12 \gamma_2-\frac12 \gamma_3-\frac12 \gamma_4})E_{\fq}(\mu X_{-\frac12 \gamma_1-\frac12 \gamma_2-\frac12 \gamma_3+\frac12 \gamma_4})E_{\fq}(\mu X_{-\frac12 \gamma_1+\frac12 \gamma_2-\frac12 \gamma_3-\frac12 \gamma_4})\cdot \cr &\cdot E_{\fq}(\mu X_{-\frac12 \gamma_1-\frac12 \gamma_2+\frac12 \gamma_3+\frac12 \gamma_4}) 
 E_{\fq}(\mu X_{-\frac12 \gamma_1+\frac12 \gamma_2-\frac12 \gamma_3+\frac12 \gamma_4}) E_{\fq}(\mu X_{\frac12 \gamma_1+\frac12 \gamma_2-\frac12 \gamma_3-\frac12 \gamma_4}) \cdot \cr &\cdot  E_{\fq}(\mu X_{-\frac12 \gamma_1+\frac12 \gamma_2+\frac12 \gamma_3+\frac12 \gamma_4}) E_{\fq}(\mu  X_{\frac12 \gamma_1+\frac12 \gamma_2-\frac12 \gamma_3+\frac12 \gamma_4})E_{\fq}(\mu  X_{\frac12 \gamma_1+\frac12 \gamma_2+\frac12 \gamma_3+\frac12 \gamma_4})
\end{align}
and combine into the final spectrum generator:
\begin{align} 
S&=E_{\fq}(\mu X_{-\frac12 \gamma_1-\frac12 \gamma_2-\frac12 \gamma_3-\frac12 \gamma_4})E_{\fq}(\mu X_{-\frac12 \gamma_1-\frac12 \gamma_2-\frac12 \gamma_3+\frac12 \gamma_4})E_{\fq}(\mu X_{-\frac12 \gamma_1+\frac12 \gamma_2-\frac12 \gamma_3-\frac12 \gamma_4})\cdot \cr &\cdot E_{\fq}(\mu X_{-\frac12 \gamma_1-\frac12 \gamma_2+\frac12 \gamma_3+\frac12 \gamma_4}) 
 E_{\fq}(\mu X_{-\frac12 \gamma_1+\frac12 \gamma_2-\frac12 \gamma_3+\frac12 \gamma_4}) E_{\fq}(\mu X_{\frac12 \gamma_1+\frac12 \gamma_2-\frac12 \gamma_3-\frac12 \gamma_4}) \cdot \cr &\cdot  E_{\fq}(\mu X_{-\frac12 \gamma_1+\frac12 \gamma_2+\frac12 \gamma_3+\frac12 \gamma_4}) E_{\fq}(\mu  X_{\frac12 \gamma_1+\frac12 \gamma_2-\frac12 \gamma_3+\frac12 \gamma_4})E_{\fq}(\mu  X_{\frac12 \gamma_1+\frac12 \gamma_2+\frac12 \gamma_3+\frac12 \gamma_4})\cdot \cr &\cdot E_\fq(X_{\gamma_1})E_\fq(X_{\gamma_2})E_\fq(X_{\gamma_3})E_\fq(X_{\gamma_4})
\end{align}
with $\langle \gamma_1,\gamma_2\rangle = \langle \gamma_3, \gamma_4\rangle =2$, which can be simplified further to 
\begin{align} 
S&=E_{\fq}(\mu X_{-\frac12 \gamma_1-\frac12 \gamma_2-\frac12 \gamma_3-\frac12 \gamma_4})E_{\fq}(\mu X_{-\frac12 \gamma_1-\frac12 \gamma_2-\frac12 \gamma_3+\frac12 \gamma_4})E_{\fq}(\mu X_{-\frac12 \gamma_1+\frac12 \gamma_2-\frac12 \gamma_3-\frac12 \gamma_4})\cdot \cr &\cdot E_{\fq}(\mu X_{-\frac12 \gamma_1-\frac12 \gamma_2+\frac12 \gamma_3+\frac12 \gamma_4}) 
  E_{\fq}(\mu X_{\frac12 \gamma_1+\frac12 \gamma_2-\frac12 \gamma_3-\frac12 \gamma_4}) E_\fq(X_{\gamma_1})E_\fq(X_{\gamma_3}) E_{\fq}(\mu X_{-\frac12 \gamma_1+\frac12 \gamma_2-\frac12 \gamma_3+\frac12 \gamma_4})  \cdot \cr &\cdot E_\fq(X_{\gamma_2})E_\fq(X_{\gamma_4})
\end{align}
The ``simple'' charges in the support of this spectrum generator appear to be $\gamma_i$ and the charge $\gamma_5$ of $\mu X_{-\frac12 \gamma_1-\frac12 \gamma_2-\frac12 \gamma_3-\frac12 \gamma_4}$, which satisfies 
\begin{equation}
    \langle \gamma_5, \gamma_1\rangle = \langle \gamma_2, \gamma_5\rangle=\langle \gamma_5, \gamma_3\rangle = \langle \gamma_4, \gamma_5\rangle=1
\end{equation}
which is indeed the known BPS quiver for this theory. See Figure \ref{fig:BPS_quiver_SU2xSU2}. 

For later use, we can also build an alternative conjectural spectrum generator from the $\mathrm{Spin}(4)$-covariant spectrum generator of $SU(2)$ with $N_f=2$, rewritten as 
\begin{equation}
      E_{\fq}(\mu^{-1} X_{\gamma_1})E_{\fq}(\mu X_{\gamma_1})E_{\fq}(\zeta X_{\gamma_2}) E_{\fq}(\zeta^{-1} X_{\gamma_2})
\end{equation}
with $\langle \gamma_1, \gamma_2 \rangle =1$.

\begin{figure}[h]
    \centering
    \begin{tikzpicture}
        % Define Nodes
        \node (A) at (0,1.5) [circle, draw, minimum size=1cm] {}; % Top-left
        \node (B) at (3,1.5) [circle, draw, minimum size=1cm] {}; % Top-right
        \node (C) at (0,-1.5) [circle, draw, minimum size=1cm] {}; % Bottom-left
        \node (D) at (3,-1.5) [circle, draw, minimum size=1cm] {}; % Bottom-right
        \node (M) at (1.5,0) [circle, draw, minimum size=1cm] {}; % Middle
        
        % Double vertical arrows (left)
        \draw[postaction={decorate},
              decoration={markings, 
              mark=at position 0.45 with {\arrow{>}},
              mark=at position 0.55 with {\arrow{>}}}] 
              (A) -- (C);

        % Diagonal arrows to/from middle
        \draw[postaction={decorate},
              decoration={markings, mark=at position 0.5 with {\arrow{>}}}] 
              (M) -- (A);
              
        \draw[postaction={decorate},
              decoration={markings, mark=at position 0.5 with {\arrow{>}}}] 
              (M) -- (B);

        \draw[postaction={decorate},
              decoration={markings, mark=at position 0.5 with {\arrow{>}}}] 
              (C) -- (M);

        \draw[postaction={decorate},
              decoration={markings, mark=at position 0.5 with {\arrow{>}}}] 
              (M) -- (D);
    \end{tikzpicture}
    \caption{BPS quiver for $SU(2) \times SU(2)$ with bifundamental matter.}
    \label{fig:su2su2two}
\end{figure}
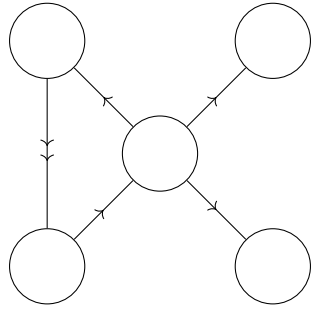

Promoting as usual the $\zeta$ dependence to a Wilson line expression for the new $SU(2)$ gauge theory: 
\begin{equation}
      S' = E_{\fq}(\mu^{-1} X_{\gamma_1})E_{\fq}(\mu X_{\gamma_1})E_{\fq}(X_{\gamma_2-\frac12 \gamma_3 - \frac12 \gamma_4}) E_{\fq}(X_{\gamma_2-\frac12 \gamma_3 + \frac12 \gamma_4}) E_{\fq}(X_{\gamma_2+\frac12 \gamma_3 + \frac12 \gamma_4}) E_\fq(X_{\gamma_3})E_\fq(X_{\gamma_4})
\end{equation}
with $\langle \gamma_3, \gamma_4 \rangle =2$, which makes manifest the additional $SU(2)$ flavour symmetry of the $SU(2) \times SU(2)$ gauge theory. The elementary charges are now $\mu^{\pm 1} \gamma_1$, $\gamma_2-\frac12 \gamma_3 - \frac12 \gamma_4$, $\gamma_3$ and $\gamma_4$.
This new BPS quiver is related to the original one by a lengthy sequence of mutations, which we will not attempt to record in detail. See Figure \ref{fig:su2su2two}.

\subsection[The trinion $SU(2)^3$ theory]{\boldmath The trinion $SU(2)^3$ theory}
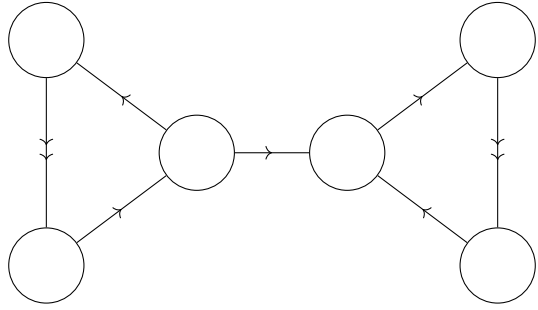
\begin{figure}[h]
    \centering
    \begin{tikzpicture}
        % Define Nodes
        \node (A) at (0,1.5) [circle, draw, minimum size=1cm] {}; % Top-left
        \node (B) at (6,1.5) [circle, draw, minimum size=1cm] {}; % Top-right
        \node (C) at (0,-1.5) [circle, draw, minimum size=1cm] {}; % Bottom-left
        \node (D) at (6,-1.5) [circle, draw, minimum size=1cm] {}; % Bottom-right
        \node (M1) at (2,0) [circle, draw, minimum size=1cm] {}; % Upper middle
        \node (M2) at (4,0) [circle, draw, minimum size=1cm] {}; % Lower middle

        % Double vertical arrows (left)
        \draw[postaction={decorate},
              decoration={markings, 
              mark=at position 0.45 with {\arrow{>}},
              mark=at position 0.55 with {\arrow{>}}}] 
              (A) -- (C);
              
        % Double vertical arrows (right)
        \draw[postaction={decorate},
              decoration={markings, 
              mark=at position 0.45 with {\arrow{>}},
              mark=at position 0.55 with {\arrow{>}}}] 
              (B) -- (D);

        % Middle nodes connected to each other
        \draw[postaction={decorate},
              decoration={markings, mark=at position 0.5 with {\arrow{>}}}] 
              (M1) -- (M2);

        % Diagonal arrows from sides to middle nodes
        \draw[postaction={decorate},
              decoration={markings, mark=at position 0.5 with {\arrow{>}}}] 
              (M1) -- (A);

        \draw[postaction={decorate},
              decoration={markings, mark=at position 0.5 with {\arrow{>}}}] 
              (C) -- (M1);

        \draw[postaction={decorate},
              decoration={markings, mark=at position 0.5 with {\arrow{>}}}] 
              (M2) -- (B);

        \draw[postaction={decorate},
              decoration={markings, mark=at position 0.5 with {\arrow{>}}}] 
              (D) -- (M2);

    \end{tikzpicture}
    \caption{The BPS quiver for the $SU(2)^3$ theory obtained from our general strategy.}
    \label{fig:su23}
\end{figure}

\begin{figure}[h]
    \centering
    \begin{tikzpicture}
        % Define Nodes
        \node (A) at (0,1.5) [circle, draw, minimum size=1cm] {}; % Top-left
        \node (B) at (6,1.5) [circle, draw, minimum size=1cm] {}; % Top-right
        \node (C) at (0,-1.5) [circle, draw, minimum size=1cm] {}; % Bottom-left
        \node (D) at (6,-1.5) [circle, draw, minimum size=1cm] {}; % Bottom-right
        \node (M1) at (2,0) [circle, draw, minimum size=1cm] {}; % Upper middle
        \node (M2) at (4,0) [circle, draw, minimum size=1cm] {}; % Lower middle

        % Double vertical arrows (left)
        \draw[postaction={decorate},
              decoration={markings, 
              mark=at position 0.5 with {\arrow{>}}}] 
              (A) -- (C);
              
        % Double vertical arrows (right)
        \draw[postaction={decorate},
              decoration={markings, 
              mark=at position 0.5 with {\arrow{>}}}] 
              (D) -- (B);

        % Middle nodes connected to each other
        \draw[postaction={decorate},
              decoration={markings, mark=at position 0.5 with {\arrow{>}}}] 
              (M1) -- (M2);

        % Diagonal arrows from sides to middle nodes
        \draw[postaction={decorate},
              decoration={markings, mark=at position 0.5 with {\arrow{>}}}] 
              (M1) -- (A);

        \draw[postaction={decorate},
              decoration={markings, mark=at position 0.5 with {\arrow{>}}}] 
              (C) -- (M1);

        \draw[postaction={decorate},
              decoration={markings, mark=at position 0.5 with {\arrow{>}}}] 
              (B) -- (M2);

        \draw[postaction={decorate},
              decoration={markings, mark=at position 0.5 with {\arrow{>}}}] 
              (M2) -- (D);
        \draw[postaction={decorate},
              decoration={markings, mark=at position 0.5 with {\arrow{>}}}] 
              (A) -- (B);
        \draw[postaction={decorate},
              decoration={markings, mark=at position 0.5 with {\arrow{>}}}] 
              (C) -- (D);

    \end{tikzpicture}
    \caption{A more symmetric quiver for the $SU(2)^3$ theory.}
    \label{fig:su23b}
\end{figure}
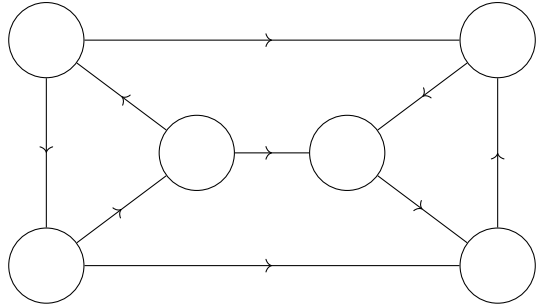

We can now complete our sequence of examples by gauging the final $SU(2)$ flavour symmetry in the previous example:
\begin{align}
      S &= E_{\fq}(X_{\gamma_1-\frac12 \gamma_5 - \frac12 \gamma_6})E_{\fq}(X_{\gamma_1-\frac12 \gamma_5 + \frac12 \gamma_6})E_{\fq}(X_{\gamma_1+\frac12 \gamma_5 + \frac12 \gamma_6}) E_{\fq}(X_{\gamma_2-\frac12 \gamma_3 - \frac12 \gamma_4})\cdot\cr &\cdot E_{\fq}(X_{\gamma_2-\frac12 \gamma_3 + \frac12 \gamma_4}) E_{\fq}(X_{\gamma_2+\frac12 \gamma_3 + \frac12 \gamma_4}) E_\fq(X_{\gamma_3})E_\fq(X_{\gamma_4})E_\fq(X_{\gamma_5})E_\fq(X_{\gamma_6})
\end{align}
This theory has three $SU(2)$ gauge groups acting symmetrically on eight half-hypermultiplets. The $S_3$ symmetry is not manifest 
in this spectrum generator, see Figure \ref{fig:su23}, but a judicious sequence of mutations can bring the quiver more symmetric forms, such as a triangular prism with all side edges parallel and face edges ordered in opposite cyclic order. See Figure \ref{fig:su23b}.

\section[Towards higher ranks: examples with $SU(3)$ gauge group]{\boldmath Towards higher ranks: examples with $SU(3)$ gauge group}
\label{sec:su3}
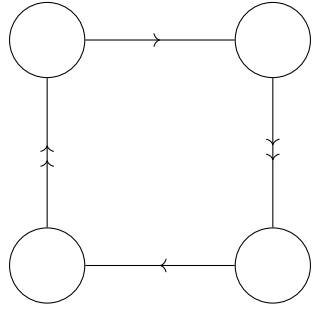
\begin{figure}[h]
    \centering
    \begin{tikzpicture}
        % Define Nodes
        \node (A) at (0,1.5) [circle, draw, minimum size=1cm] {}; % Top-left
        \node (B) at (3,1.5) [circle, draw, minimum size=1cm] {}; % Top-right
        \node (C) at (0,-1.5) [circle, draw, minimum size=1cm] {}; % Bottom-left
        \node (D) at (3,-1.5) [circle, draw, minimum size=1cm] {}; % Bottom-right

        % Double vertical arrows (left side)
        \draw[postaction={decorate},
              decoration={markings, 
              mark=at position 0.45 with {\arrow{>}},
              mark=at position 0.55 with {\arrow{>}}}] 
              (C) -- (A);
              
        % Double vertical arrows (right side)
        \draw[postaction={decorate},
              decoration={markings, 
              mark=at position 0.45 with {\arrow{>}},
              mark=at position 0.55 with {\arrow{>}}}] 
              (B) -- (D);

        % Single arrows forming a cyclic loop
        \draw[postaction={decorate},
              decoration={markings, mark=at position 0.5 with {\arrow{>}}}] 
              (A) -- (B);

        \draw[postaction={decorate},
              decoration={markings, mark=at position 0.5 with {\arrow{>}}}] 
              (D) -- (C);

    \end{tikzpicture}
    \caption{BPS quiver for pure $SU(3)$. }
    \label{fig:BPS_quiver_SU3}
\end{figure}

In the literature, the BPS spectrum of pure $SU(3)$ gauge theory is encoded into a BPS quiver with four nodes, with charges $\gamma_1, \cdots, \gamma_4$ such that
\begin{align}
    &\langle \gamma_1, \gamma_2 \rangle = 2 \cr
    &\langle \gamma_2, \gamma_3 \rangle = 1 \cr
    &\langle \gamma_3, \gamma_4 \rangle = 2 \cr
    &\langle \gamma_4, \gamma_1 \rangle = 1 
\end{align}
and quantum spectrum generator
\begin{equation}
    S= E_\fq(X_{\gamma_1}) E_\fq(X_{\gamma_3}) E_\fq(X_{\gamma_1+\gamma_4}) E_\fq(X_{\gamma_2 + \gamma_3}) E_\fq(X_{\gamma_2}) E_\fq(X_{\gamma_4}) 
\end{equation}
See Figure \ref{fig:BPS_quiver_SU3}.

Observe that a double mutation maps this to 
\begin{equation}
     E_\fq(X_{\gamma_1+\gamma_4}) E_\fq(X_{\gamma_2 + \gamma_3}) E_\fq(X_{\gamma_2}) E_\fq(X_{\gamma_4}) E_\fq(X_{-\gamma_1}) E_\fq(X_{-\gamma_3})
\end{equation}
which is actually the same expression up to a charge re-definition. This is analogous to the $\gamma_1$ mutation in pure $SU(2)$ gauge theory. 

Consider now a charge which has non-zero pairing only with $\gamma_1$ and $\gamma_2$ of the same form as the $SU(2)$ Wilson loop, such as $-\frac{2}{3}\gamma_1 - \frac{2}{3}\gamma_2- \frac{1}{3}\gamma_4- \frac{1}{3}\gamma_3$. We will look for a candidate $F$ of the form 
\begin{equation}
    X_{-\frac{2}{3}\gamma_1 - \frac{2}{3}\gamma_2- \frac{1}{3}\gamma_4- \frac{1}{3}\gamma_3} + \cdots 
\end{equation}
This behaves well under $\gamma_1$ and $\gamma_3$ mutations. The inverse $\gamma_2$ mutation requires an extra term  
\begin{equation}
    X_{-\frac{2}{3}\gamma_1 - \frac{2}{3}\gamma_2- \frac{1}{3}\gamma_4- \frac{1}{3}\gamma_3} + X_{-\frac{2}{3}\gamma_1 + \frac{1}{3}\gamma_2- \frac{1}{3}\gamma_4- \frac{1}{3}\gamma_3} +\cdots 
\end{equation}
and then the $\gamma_1$ mutation requires 
\begin{equation}
    X_{-\frac{2}{3}\gamma_1 - \frac{2}{3}\gamma_2- \frac{1}{3}\gamma_4- \frac{1}{3}\gamma_3} + X_{-\frac{2}{3}\gamma_1 + \frac{1}{3}\gamma_2- \frac{1}{3}\gamma_4- \frac{1}{3}\gamma_3} +X_{\frac{1}{3}\gamma_1 + \frac{1}{3}\gamma_2- \frac{1}{3}\gamma_4- \frac{1}{3}\gamma_3} +\cdots 
\end{equation}
then inverse $\gamma_4$ requires 
\begin{equation}
    X_{-\frac{2}{3}\gamma_1 - \frac{2}{3}\gamma_2- \frac{1}{3}\gamma_3- \frac{1}{3}\gamma_4} + X_{-\frac{2}{3}\gamma_1 + \frac{1}{3}\gamma_2- \frac{1}{3}\gamma_3- \frac{1}{3}\gamma_4} +X_{\frac{1}{3}\gamma_1 + \frac{1}{3}\gamma_2- \frac{1}{3}\gamma_3- \frac{1}{3}\gamma_4} +X_{\frac{1}{3}\gamma_1 + \frac{1}{3}\gamma_2- \frac{1}{3}\gamma_3+\frac{2}{3}\gamma_4} +\cdots 
\end{equation}
and $\gamma_3$ requires the final correction:
\begin{align}
    F_{w_3} &= X_{-\frac{2}{3}\gamma_1 - \frac{2}{3}\gamma_2- \frac{1}{3}\gamma_3- \frac{1}{3}\gamma_4} + X_{-\frac{2}{3}\gamma_1 + \frac{1}{3}\gamma_2- \frac{1}{3}\gamma_3- \frac{1}{3}\gamma_4} +X_{\frac{1}{3}\gamma_1 + \frac{1}{3}\gamma_2- \frac{1}{3}\gamma_3- \frac{1}{3}\gamma_4} + \cr &+ X_{\frac{1}{3}\gamma_1 + \frac{1}{3}\gamma_2- \frac{1}{3}\gamma_3+\frac{2}{3}\gamma_4}+X_{\frac{1}{3}\gamma_1 + \frac{1}{3}\gamma_2+ \frac{2}{3}\gamma_3+\frac{2}{3}\gamma_4} 
\end{align}
which we identify as a fundamental Wilson line. We also have an anti-fundamental candidate:
\begin{align}
    F_{w_{\bar 3}} &= X_{-\frac{1}{3}\gamma_1 - \frac{1}{3}\gamma_2- \frac{2}{3}\gamma_3- \frac{2}{3}\gamma_4} + X_{-\frac{1}{3}\gamma_1 - \frac{1}{3}\gamma_2- \frac{2}{3}\gamma_3+ \frac{1}{3}\gamma_4} +X_{-\frac{1}{3}\gamma_1 - \frac{1}{3}\gamma_2+ \frac{1}{3}\gamma_3+ \frac{1}{3}\gamma_4} + \cr &+X_{-\frac{1}{3}\gamma_1 + \frac{2}{3}\gamma_2+ \frac{1}{3}\gamma_3+ \frac{1}{3}\gamma_4}+X_{\frac{2}{3}\gamma_1 + \frac{2}{3}\gamma_2+ \frac{1}{3}\gamma_3+ \frac{1}{3}\gamma_4} 
\end{align}
Another natural candidate for an $F$ is the monomial \begin{equation}
    F_{H_{0}} = X_{\frac23\gamma_1+ \frac13 \gamma_3}
\end{equation}
We can combine it with Wilson loops to get other candidates:
\begin{align}
    F_{H_{0}}\cdot F_{w_3} &= \fq^{-\frac23} X_{- \frac{2}{3}\gamma_2- \frac{1}{3}\gamma_4} + \fq^{\frac13}(X_{\frac{1}{3}\gamma_2- \frac{1}{3}\gamma_4} +X_{\gamma_1 + \frac{1}{3}\gamma_2- \frac{1}{3}\gamma_4} +X_{\gamma_1 + \frac{1}{3}\gamma_2+\frac{2}{3}\gamma_4}+X_{\gamma_1 + \frac{1}{3}\gamma_2+ \gamma_3+\frac{2}{3}\gamma_4}) \cr
    F_{H_{0}}\cdot F_{w_{\bar 3}} &= \fq^{-\frac13}(X_{\frac{1}{3}\gamma_1 - \frac{1}{3}\gamma_2- \frac{1}{3}\gamma_3- \frac{2}{3}\gamma_4} + X_{\frac{1}{3}\gamma_1 - \frac{1}{3}\gamma_2- \frac{1}{3}\gamma_3+ \frac{1}{3}\gamma_4} +X_{\frac{1}{3}\gamma_1 - \frac{1}{3}\gamma_2+ \frac{2}{3}\gamma_3+ \frac{1}{3}\gamma_4} )+ \cr &+ \fq^{\frac23}( X_{\frac{1}{3}\gamma_1 + \frac{2}{3}\gamma_2+ \frac{2}{3}\gamma_3+ \frac{1}{3}\gamma_4}+X_{\frac{4}{3}\gamma_1 + \frac{2}{3}\gamma_2+ \frac{2}{3}\gamma_3+ \frac{1}{3}\gamma_4}) 
\end{align}
etcetera. We can define a sequence of dyonic operators as 
\begin{equation}
    (\fq-\fq^{-1}) H_{a+1} = \fq^{\frac13} w_3 H_a - \fq^{-\frac13} H_a w_3 \qquad \qquad (\fq-\fq^{-1}) H_{a-1} = \fq^{\frac13} H_a w_{\bar 3} - \fq^{-\frac13} w_{\bar 3} H_a 
\end{equation}
These satisfy $H_{a+1}H_a = \fq^{\frac43} H_a H_{a+1}$. We can also define 
\begin{equation}
    \bar H_a = \fq^{\frac13} H_{a-1} H_{a+1}- \fq^{-1} H_a^2
\end{equation}
and more. It would be nice to describe the full set of tropical generators. 

We can also consider the RG flow from $SU(3)$ to $SU(2)$ gauge theory, mutating at $\gamma_1$ to 
\begin{equation}
    S'_{SU(3)} = E_\fq(X_{\gamma_3}) E_\fq(X_{\gamma_1+\gamma_4}) E_\fq(X_{\gamma_2 + \gamma_3}) E_\fq(X_{\gamma_4}) E_\fq(X_{\gamma_2}) E_\fq(X_{-\gamma_1}) 
\end{equation}
and setting $S_{SU(2)} = E_\fq(X_{\gamma_2}) E_\fq(X_{-\gamma_1})$, 
so that the RG flow is associated to the factors 
\begin{equation}
    E_\fq(X_{\gamma_3}) E_\fq(X_{\gamma_2 + \gamma_3}) E_\fq(X_{\gamma_1+\gamma_4})  E_\fq(X_{\gamma_4}) = E_\fq(X_{\gamma_3}+ X_{\gamma_2 + \gamma_3}) E_\fq(X_{\gamma_1+\gamma_4}+X_{\gamma_4})
\end{equation}
which is the IR image of 
\begin{equation}
    S^{SU(2)}_{SU(3)} = E_\fq(X_{\gamma_3+\frac12 \gamma_1} H^{SU(2)}_1) E_\fq(X_{\gamma_4+\frac12 \gamma_2} H^{SU(2)}_{-2}) \, .
\end{equation}
The Wilson lines mutate to 
\begin{align}
    F_{w_3} &= X_{-\frac{1}{6}\gamma_1 - \frac{1}{6}\gamma_2- \frac{1}{3}\gamma_3- \frac{1}{3}\gamma_4}F^\IR_{w_1^{SU(2)}}  +X_{-\frac{1}{6}\gamma_1 + \frac{1}{3}\gamma_2- \frac{1}{3}\gamma_3+\frac{2}{3}\gamma_4}F^\IR_{H_{-1}^{SU(2)}} +X_{\frac{1}{3}\gamma_1 + \frac{1}{3}\gamma_2+ \frac{2}{3}\gamma_3+\frac{2}{3}\gamma_4} \cr
    F_{w_{\bar 3}} &= X_{-\frac{1}{3}\gamma_1 - \frac{1}{3}\gamma_2- \frac{2}{3}\gamma_3- \frac{2}{3}\gamma_4} +  X_{-\frac{1}{3}\gamma_1 + \frac{1}{6}\gamma_2- \frac{2}{3}\gamma_3+ \frac{1}{3}\gamma_4}F^\IR_{H_{-2}^{SU(2)}} +X_{\frac{1}{6}\gamma_1 + \frac{1}{6}\gamma_2+ \frac{1}{3}\gamma_3+ \frac{1}{3}\gamma_4}F^\IR_{w_1^{SU(2)}}  
\end{align}
It would be interesting to discuss a partial diagonalization of these operators in the $SU(2)$ theory. It should be implemented by the superconformal index for $3 \times 2$ 3d chiral multiplets, a wavefunction written as the ratio of six $\fq^2$-factorials at numerator and six at denominator. 

\subsection{Adding fundamental flavour}
We verified that the contribution of a fundamental representation, 
\begin{equation}
    E_\fq(\mu \zeta_1^{-\frac23}\zeta_2^{-\frac13})E_\fq(\mu \zeta_1^{\frac13}\zeta_2^{-\frac13})E_\fq(\mu \zeta_1^{\frac13}\zeta_2^{\frac23}) = g_3(\mu, \zeta_1^{-\frac23}\zeta_2^{-\frac13}+\zeta_1^{\frac13}\zeta_2^{-\frac13}+\zeta_1^{\frac13}\zeta_2^{\frac23},\zeta_1^{-\frac13}\zeta_2^{-\frac23}+\zeta_1^{-\frac13}\zeta_2^{\frac13}+\zeta_1^{\frac23}\zeta_2^{\frac13})
\end{equation}
factors nicely:
\begin{align}
     g_3(\mu,w_3,w_{\bar 3}) &= E_\fq(X_{-\frac{2}{3}\gamma_1 - \frac{2}{3}\gamma_2- \frac{1}{3}\gamma_3- \frac{1}{3}\gamma_4})E_\fq(X_{-\frac{2}{3}\gamma_1 + \frac{1}{3}\gamma_2- \frac{1}{3}\gamma_3- \frac{1}{3}\gamma_4}) \cdot\cr &\cdot E_\fq(X_{\frac{1}{3}\gamma_1 + \frac{1}{3}\gamma_2- \frac{1}{3}\gamma_3- \frac{1}{3}\gamma_4})E_\fq(X_{\frac{1}{3}\gamma_1 + \frac{1}{3}\gamma_2- \frac{1}{3}\gamma_3+ \frac{2}{3}\gamma_4})E_\fq(X_{\frac{1}{3}\gamma_1 + \frac{1}{3}\gamma_2+ \frac{2}{3}\gamma_3+ \frac{2}{3}\gamma_4}) \,.
\end{align}
This leads to a prediction for the spectrum generator of $SU(3)$ with $N_f=1$:
\begin{align}
    S&= E_\fq(X_{-\frac{2}{3}\gamma_1 - \frac{2}{3}\gamma_2- \frac{1}{3}\gamma_3- \frac{1}{3}\gamma_4})E_\fq(X_{-\frac{2}{3}\gamma_1 + \frac{1}{3}\gamma_2- \frac{1}{3}\gamma_3- \frac{1}{3}\gamma_4})  E_\fq(X_{\frac{1}{3}\gamma_1 + \frac{1}{3}\gamma_2- \frac{1}{3}\gamma_3- \frac{1}{3}\gamma_4}) \cdot\cr &\cdot E_\fq(X_{\frac{1}{3}\gamma_1 + \frac{1}{3}\gamma_2- \frac{1}{3}\gamma_3+ \frac{2}{3}\gamma_4})E_\fq(X_{\frac{1}{3}\gamma_1 + \frac{1}{3}\gamma_2+ \frac{2}{3}\gamma_3+ \frac{2}{3}\gamma_4}) \cdot\cr &\cdot E_\fq(X_{\gamma_1}) E_\fq(X_{\gamma_3}) E_\fq(X_{\gamma_1+\gamma_4}) E_\fq(X_{\gamma_2 + \gamma_3}) E_\fq(X_{\gamma_2}) E_\fq(X_{\gamma_4}) \,.
\end{align}

\subsection{Wild chambers}
We now go back to pure $SU(3)$ and explore the potential mismatch between the mutation orbit of the original BPS quiver and the collection of physically realizable RG flows. The key observation is that 
if we mutate at $\gamma_1$ and then $\gamma_2$, the new BPS quiver has three edges between the first and last nodes. If an RG flow existed which allows decoupling of the remaining nodes, we will land on a BPS quiver which does not seem to correspond to a physical theory. 

The node $\gamma_3$ does not really play a role in the sequence of mutations, so the problem would already manifest in a simpler model where we decoupled $\gamma_3$ from the very beginning. This leads to the question: is there a partial RG flow which eliminates a single node of the $SU(3)$ quiver? 

The geometry of the space of vacua of pure $SU(3)$ gauge theory is intricate. The space of vacua is parameterized by two complex variables $u$ and $v$. When $u$ and $v$ are both very large, instanton effects are suppressed and the space of vacua has an intuitive characterization: there is a weak-coupling discriminant $v^2- u^3$ and near the vanishing locus of the discriminant the $SU(3)$ theory flows to a pure $SU(2) \times U(1)$ gauge theory first, and then to the IR description of the latter. This the RG flow we already discussed. The proposal of the partial spectrum generator capturing two BPS particles which behave as $H_1^{SU(2)}$ and $H_{-2}^{SU(2)}$ dressed by Abelian charges appears to be novel. 

In order to find other interesting RG flows, we need to make one of these BPS particles light, possibly by exploring smaller values of $u$ and $v$. The instanton-corrected discriminant, in units of the instanton action, is a more complicated polynomial in $u^3$ and $v^2$ which vanishes at loci where a BPS particle of the Seiberg-Witten description has vanishing mass. The loci are simply described as $v(u) = \pm u^{\frac32} \pm 1$.

Another well-known RG flow goes to the $[A_1,A_2]$ Argyres-Douglas theory, say by dropping $\gamma_2$ and $\gamma_3$. This happens when, say, $1-u^{\frac32} = -1+u^{\frac32}$ and two branches of the vanishing locus meet. A partial RG flow dropping $\gamma_3$ would be a partial step in both known RG flows. It would essentially require one to keep three of the $v(u)$ values closer to each other than the fourth. This is manifestly impossible. 

We conclude that an RG flow such as described by 
\begin{equation}
    S'_{SU(3)} = E_\fq(X_{\gamma_3}) E_\fq(X_{\gamma_2 + \gamma_3}) S'_\IR
\end{equation}
with 
\begin{equation}
    S'_\IR = E_\fq(X_{\gamma_1+\gamma_4}) E_\fq(X_{\gamma_4}) E_\fq(X_{\gamma_2}) E_\fq(X_{-\gamma_1}) 
\end{equation}
should be unphysical. The candidate IR theory can be more concisely described in terms of the spectrum generator 
\begin{equation}
    S_\IR = E_\fq(X_{\gamma_4}) E_\fq(X_{\gamma_1})E_\fq(X_{\gamma_2})  
\end{equation}
This should be another example of BPS quiver which fails to lead to a consistent Schur quantization. As for the simpler example of the quiver with three arrows, we expect this to happen because some $\langle a|b\rangle$ inner products diverge. 

\section{Applications to the theory of quantum groups}\label{sec:Uq}
The (central quotient of) the quantum groups $U_\fq(\fg)$, presented as an algebra over $\bZ[\fq, \fq^{-1}]$ \footnote{in a manner which does not seem to coincide with the Lusztig form}, is a special case of an algebra $\CA$ which is expected to be associated to a 4d ${\cal N}=2$ gauge theory and thus amenable of our analysis. We will denote the theory as $T^{4d}[\fg]$, as it is a close relative of a 3d theory which plays an important role in S-duality. 

In particular, there is a conjectural definition of $T^{4d}[\sl_N]$ as a balanced quiver gauge theory with $N-1$ unitary nodes with rank increasing linearly from $1$ to $N-1$. The identification with the central quotient of $U_\fq(\sl_N)$ is well understood. It also has a class $\CS$ description 
which makes contact with a known relation between $U_\fq(\sl_N)$ and quantized character varieties. 

Our results thus lead to a simple conjecture: the quantum spectrum generators $S$ associated to such cluster parameterizations of $U_\fq(\fg)$
must be normalizable and there is a well-defined complex cluster quantization of $U_\fq(\fg)$ leading to spherical unitary representations
with unitary central character $\mu$. Gauging the Abelian flavour symmetry associated to $\mu$ should lead to representations of a larger algebra which decomposes into sectors with arbitrary $\mu$ with $|\mu| = \fq^k$
with integral $k$. 

These theories are expected to be part of a more general family of theories $T_\rho^{4d}[\fg]$
for which we have a map $U_\fq(\fg) \to \CA$ with special values of the Casimir generators of the quantum group. In particular, there is a conjectural definition of $T_\rho^{4d}[\sl_N]$ as a balanced quiver gauge theories with $N-1$ unitary nodes but more general choices of rank. 

These theories do have a class $\CS$ description, but the corresponding character varieties do not have a known systematic cluster definition. The gauge theory description is thus a valuable tool to explore quantum group representations.

In this Section we would like to spell out some concrete examples of these statements. A full discussion would deserve a separate publication. 

\subsection[The central quotient of $U_\fq(\sl_2)$ and SQED$_2$.]{\boldmath The central quotient of $U_\fq(\sl_2)$ and SQED$_2$.}
This case was discussed at some length in previous work. As the $U(1)$ gauge group is Abelian, 
the RG flow to a pure $U(1)$ gauge theory is very simple and the spectrum generator is written as
\begin{equation}
    S = E_\fq(X_{0,1,1})E_\fq(X_{0,1,-1}) 
\end{equation}
with $\mu = X_{0,0,1}$ being a flavour parameter which will control the Casimir of $U_\fq(\sl_2)$ and $\langle (1,0,0),(0,1,0)\rangle = 1$.

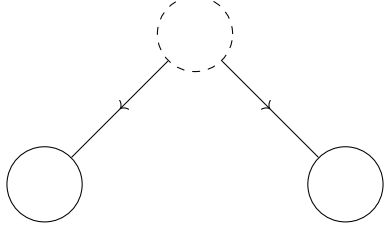
\begin{figure}[h]
    \centering
    \begin{tikzpicture}
        % Normal nodes
        \node (A) at (-2,0) [circle, draw, minimum size=1cm] {};
        \node (B) at (2,0) [circle, draw, minimum size=1cm] {};

        % Dashed node
        \node (C) at (0,2) [circle, draw, dashed, minimum size=1cm] {};

        % Arrows from dashed node to normal nodes
        \draw[postaction={decorate},
              decoration={markings, mark=at position 0.5 with {\arrow{>}}}] 
              (C) -- (A);
        \draw[postaction={decorate},
              decoration={markings, mark=at position 0.5 with {\arrow{>}}}] 
              (C) -- (B);

    \end{tikzpicture}
    \caption{The BPS quiver for SQED${}_2$.}
    \label{fig:simple_quiver}
\end{figure}

The basic loop operators $D_{a,b}$ give the standard (divided powers of) quantum group generators up to a rescaling which makes the algebra over $\bZ[\fq,\fq^{-1}]$. In particular, we have multiplicative generators $u_+ = D_{1,0}$, $u_-=D_{-1,-1}$, $v^\pm = D_{0,\pm 1}$ and relations 
\begin{align}
    u_+ u_- &= \fq^{-1} v^{-1} + \mu + \mu^{-1} + \fq v \cr
    u_- u_+ &= \fq v^{-1} + \mu + \mu^{-1} + \fq^{-1} v
\end{align}
with quadratic Casimir proportional to $\mu + \mu^{-1}$ and 
\begin{align}
    F_v &= X_{0,1,0} \cr
    F_{u_+} &= X_{1,0,0} \cr
    F_{u_-} &= X_{-1,-1,0} + X_{-1,0,1} + X_{-1,0,-1} +  X_{-1,1,0}
\end{align}
In particular, $u_- = \ell_{-1,-1,0} = \rho(\fq u_+ v^{-1})$ and $u_+ = \rho(\fq u_- v)$, $\rho(v) = v^{-1}$. So $\rho(u_+) = \fq v^{-1} u_-$ and $\rho(u_-) = \fq v u_+$.

We denote the associated representation on $\ell^2(\bZ^2)$ as a spherical principal series representation.

\subsection[The $\fq$ analogue of the $L^2(\mathbb{C}^2)$ spectral decomposition]{\boldmath The $\fq$ analogue of the $L^2(\mathbb{C}^2)$ spectral decomposition}
If we gauge the $U(1)$ symmetry associated to $\mu$, we embed $A$ in a larger algebra $A'$ where $\mu$ is now a non-trivial generator, a close relative of $\CW^2$. We can roughly think about this as a $\fq$-oscillator representation of the quantum group. 

In order to simplify the presentation, we will extend $\CW$ to an algebra $\CW_2$ over $\bZ[\fq^{\frac12},\fq^{-\frac12}]$. We will denote the generators of $\CW_2$ as $p_\pm$ and $x$, with 
\begin{align}
    p_\pm v &= \fq^{\pm} x p_\pm \cr
    p_+ p_- &= \fq^{\frac12} x + \fq^{-\frac12} x^{-1} \cr
    p_- p_+ &= \fq^{-\frac12} x + \fq^{\frac12} x^{-1} \cr
\end{align}
Then given two copies of the algebra we write
\begin{align}
    v &= x_1 x_2 \cr
    \mu &= x_1 x_2^{-1} \cr
    u_+ &= p_{1,+} p_{2,+} \cr
    u_- &= p_{1,-} p_{2,-}
\end{align}

Then $\CW_2$ can be represented on $\ell^2(\bZ \oplus \frac12 \bZ)$,
say by 
\begin{align}
    F_x &= X_{0,\frac12} \cr
    F_{p_+} &= X_{1,0} \cr
    F_{p_-} &= X_{-1,\frac12}+X_{-1,-\frac12}
\end{align}
Accordingly, we represent $\CA$ on $\ell^2(\bZ \oplus \frac12 \bZ)\otimes \ell^2(\bZ \oplus \frac12 \bZ)$.

The image of the spherical vector under $A$ is obviously supported on $(m,a;m,b)$. Summands of the Hilbert space supported on $(m,a;m+k,b)$
are eigenspaces of $|\mu|^2 = \fq^{k}$. Each can be further decomposed into distributional eigenspaces for $\mu$ which can be identified as non-spherical principal series representations of $\CA$. These are literally $\fq$-deformed versions of familiar representations of $SL(2,\bC)$ 
appearing in a spectral decomposition of $L^2(\bC^2)$, wavefunctions on $\bC^2$ with $GL(2)$ acting as holomorphic rotations. 

\subsection[$U(2)$ with $N_f=1$]{\boldmath $U(2)$ with $N_f=1$}
The $SU(2)$ $N_f=1$ theory has a $U(1)$ flavour symmetry acting on 
two hypermultiplets. Gauging the $U(1)$ will thus enlarge the operator algebra to include a copy of $U_\fq(\sl_2)$, with Casimir equal to the Wilson loop $D_{0,1}$. 

A possible application of this structure is to give a first example of 
the relation between quantum group representation theory and the  diagonalizing $D_{0,1}$. 

Add an Abelian magnetic charge $\gamma_m$ with $\langle \gamma_m,\gamma_3\rangle =1$ and other vanishing. In the $\RG$ chamber, we can set $F_v = \mu = X_{\frac12 \gamma_1 + \frac12 \gamma_2 +\gamma_3}$, 
\begin{equation}
    F_{u_+} = X_{\gamma_m} \, .
\end{equation}
Then $u_+ u_- = \fq v+D_{0,1}+\fq^{-1}v^{-1}$ gives 
\begin{equation}
    F_{u_-} = X_{-\gamma_m -\frac12 \gamma_1 - \frac12 \gamma_2 -\gamma_3}+X_{-\gamma_m -\frac12 \gamma_1 - \frac12 \gamma_2}+ X_{-\gamma_m-\frac12 \gamma_1 + \frac12 \gamma_2 }+X_{-\gamma_m+\frac12 \gamma_1 +\frac12 \gamma_2}+X_{-\gamma_m+\frac12 \gamma_1 + \frac12 \gamma_2 +\gamma_3}
\end{equation}
We thus get a spherical representation of (the $*$-double of) $U_\fq(\sl_2)$ on $\ell^2(\bZ \gamma_m \oplus \Gamma)$ which can be decomposed into principal series representations by diagonalizing $F_{D_{1,0}}$. The diagonalization gives the UV presentation of Schur quantization, so each representation should occur once (in the distributional sense) and the UV presentation makes the decomposition of individual $|a\rangle$ vectors very explicit. Conversely, diagonalizing $v$ gives representations of the $N_f=1$ algebra. 

In the primed chamber, the quantum group generators are simpler:
\begin{align}
    F'_v &= X_{\frac12 \gamma_1 + \frac12 \gamma_2 +\gamma_3} \cr
    F'_{u_+} &= X_{\gamma_m}+X_{\gamma_m+\gamma_3} \cr
    F'_{u_-} &= X_{-\gamma_m -\frac12 \gamma_1 - \frac12 \gamma_2 -\gamma_3}+ X_{-\gamma_m-\frac12 \gamma_1 + \frac12 \gamma_2 }+X_{-\gamma_m+\frac12 \gamma_1 +\frac12 \gamma_2}
\end{align}
but the Casimir generator is a bit more complicated. 

Again, formulae will be a bit simpler if we enlarge the relevant algebras slightly at the price of allowing powers of $\fq^{\frac12}$. We can include operators which have half-integral magnetic charges for both gauge groups:
\begin{align}
    F'_{D_{\frac12,\frac12;0,0}}&= X_{\frac12 \gamma_m + \frac12 \gamma_1}\cr
    F'_{D_{\frac12,\frac12;-1,0}}&= X_{\frac12 \gamma_m - \frac12 \gamma_2}\cr
     F'_{D_{\frac12,-\frac12;0,0}}&= X_{-\frac12 \gamma_m + \frac12 \gamma_1}\cr
    F'_{D_{\frac12,-\frac12;-1,0}}&= X_{-\frac12 \gamma_m - \frac12 \gamma_2}+X_{-\frac12 \gamma_m + \frac12 \gamma_2+\gamma_3}
\end{align}
which have a nice interplay with the quantum group generators $u_\pm$. 

\subsection{Glued quantum groups}
We will only look briefly at the next example, mostly in light of a possible application to real versions of Schur quantization. The idea is to start from $SU(2)$ with $N_f=2$ and gauge both $U(1)$ flavour symmetries, to get an extended algebra with two $U_\fq(\sl_2)$ sub-algebras. The Casimir of both sub-algebras coincide with the Wilson loop $D_{0,1}$. 

If we add charges $\gamma_m$ and $\gamma_n$ dual to $\gamma_3$ and $\gamma_4$ respectively, the $F$'s for the two sets of quantum group generators are identical to these in the previous example, one literally and the other under $\gamma_3 \to \gamma_4$, $\gamma_m \to \gamma_n$.

Mutating to the primed chamber, one can do a partial RG flow preserving a doublet of simple charges, mapping the algebra to a combination of two $\CW$'s and a quantum torus algebra. This does not seem particularly illuminating, so we will omit it. 

We expect this algebra to be a $\fq$-deformed version of the algebra of holomorphic vectorfields on $SL(2,\bC)$, with $U(\sl_2)$ acting by left- and right- vectorfields. Again, we leave the details to the enthusiastic reader. 

By construction, the diagonalization of the Wilson line $D_{0,1}$ will decompose the Schur quantization representation into a direct integral of products of principal series representations for the two $U_\fq(\sl_2)$'s with the same character, each appearing once. 

\subsection[The $U_\fq(\sl_2)$ coproduct as a partial RG flow]{\boldmath The $U_\fq(\sl_2)$ coproduct as a partial RG flow}
The $SU(2)$ $N_f=3$ theory can be maximally enlarged by gauging three $U(1)$ flavour groups. The resulting algebra has three $U_\fq(\sl_2)$ sub-algebras, with coincident Casimirs given by $D_{0,1}$. 

A crucial feature is the possibility of an RG flow to four copies of $\CW_2$, represented by the four symmetric nodes in the last BPS quiver presentation of the theory given above. This should be an $\fq$-deformation of the embedding of three $U(\sl_2)$ in a Weyl algebra with $8$ generators, which has interesting Langlands applications. 

Furthermore, we will argue that if we focus on a single copy of $U_\fq(\sl_2)$ in the extended algebra, the map to $\CW_2^4$ coincides with a well-known realization of the $U_\fq(\sl_2)$ coproduct, composed with the maps to $\CW_2^2$ we discussed above. Accordingly, if we diagonalize $\mu_2$ and $\mu_3$ we will find the coproduct of principal series representations with the corresponding Casimirs. 

The diagonalization of $D_{1,0}$ then solves the problem of decomposing the coproduct of two principal series representations of $U_\fq(\sl_2)$ into principal series representations and the UV formulae for the Scur index essentially provide the Clebsh-Gordon coefficients. This is an $\fq$-deformation of an analogous statement about $SL(2,\bC)$ actions on $L^2(\bC^4)$.

In order to builg the appropriate lattice for this example, we add charges $\gamma_m$, $\gamma_n$, $\gamma_o$ dual to $\gamma_{4,5,6}$ respectively, as we did in simpler examples. Hence
\begin{align}
    F_{v_1} &= \mu_1 = X_{\frac12 \gamma_1 + \frac12 \gamma_2 +\gamma_3}\cr 
    F_{u_{1,+}} &= X_{\gamma_m} \cr
    F_{u_{1,-}} &= X_{-\gamma_m -\frac12 \gamma_1 - \frac12 \gamma_2 -\gamma_3}+X_{-\gamma_m -\frac12 \gamma_1 - \frac12 \gamma_2}+ X_{-\gamma_m-\frac12 \gamma_1 + \frac12 \gamma_2 }+X_{-\gamma_m+\frac12 \gamma_1 +\frac12 \gamma_2}\nonumber \\
    &\quad +X_{-\gamma_m+\frac12 \gamma_1 + \frac12 \gamma_2 +\gamma_3}
\end{align}
etcetera. We can now perform a series of mutations:
\begin{align}
    F'_{v_1} &= \mu_1 = X_{\frac12 \gamma_1 + \frac12 \gamma_2 +\gamma_3}\cr 
    F'_{u_{1,+}} &= X_{\gamma_m}+X_{\gamma_m+\gamma_3} \cr
    F'_{u_{1,-}} &= X_{-\gamma_m -\frac12 \gamma_1 - \frac12 \gamma_2 -\gamma_3}+ X_{-\gamma_m-\frac12 \gamma_1 + \frac12 \gamma_2 }+ X_{-\gamma_m-\frac12 \gamma_1 + \frac12 \gamma_2 +\gamma_4}+ X_{-\gamma_m-\frac12 \gamma_1 + \frac12 \gamma_2 +\gamma_5}  \nonumber \\
    &\quad +X_{-\gamma_m-\frac12 \gamma_1 + \frac12 \gamma_2+\gamma_4 + \gamma_5 }+X_{-\gamma_m+\frac12 \gamma_1 +\frac12 \gamma_2}
\end{align}
and
\begin{align}
    F''_{v_1} &= \mu_1 = X_{\frac12 \gamma_1 + \frac12 \gamma_2 +\gamma_3}\cr 
    F''_{u_{1,+}} &= X_{\gamma_m}+X_{\gamma_m+\gamma_2 + \gamma_3 + \gamma_4+\gamma_5}+X_{\gamma_m+\gamma_3} \cr
    F''_{u_{1,-}} &= X_{-\gamma_m -\frac12 \gamma_1 - \frac12 \gamma_2 -\gamma_3}+ X_{-\gamma_m-\frac12 \gamma_1 + \frac12 \gamma_2 }+ X_{-\gamma_m-\frac12 \gamma_1 + \frac32 \gamma_2+\gamma_3 + \gamma_4 + \gamma_5 }+ X_{-\gamma_m-\frac12 \gamma_1 + \frac12 \gamma_2 +\gamma_4}\nonumber\\ 
    &\quad + X_{-\gamma_m-\frac12 \gamma_1 + \frac12 \gamma_2 +\gamma_5}+X_{-\gamma_m+\frac12 \gamma_1 +\frac12 \gamma_2}
\end{align}
to finally obtain: 
\begin{align}
    F'''_{v_1} &= \mu_1 = X_{\frac12 \gamma_1 + \frac12 \gamma_2 +\gamma_3}\cr 
    F'''_{u_{1,+}} &= X_{\gamma_m}+X_{\gamma_m+\gamma_2 + \gamma_3 + \gamma_4+\gamma_5}+X_{\gamma_m+\gamma_1 + \gamma_2 + \gamma_3 + \gamma_4+\gamma_5}+X_{\gamma_m+\gamma_3}+X_{\gamma_m+\gamma_1 +\gamma_3} \cr
    F'''_{u_{1,-}} &= X_{-\gamma_m -\frac12 \gamma_1 - \frac12 \gamma_2 -\gamma_3}+ X_{-\gamma_m-\frac12 \gamma_1 + \frac12 \gamma_2 }+ X_{-\gamma_m-\frac12 \gamma_1 + \frac32 \gamma_2+\gamma_3 + \gamma_4 + \gamma_5 }+ \\
  &\quad+  X_ {-\gamma_m-\frac12 \gamma_1 + \frac12 \gamma_2 +\gamma_4}+ X_{-\gamma_m-\frac12 \gamma_1 + \frac12 \gamma_2 +\gamma_5}\period 
\end{align}
In this chamber, we would like to have an RG flow from
\begin{align}
    S'''&= E_{\fq}(X_{-\gamma_2-\gamma_3-\gamma_4-\gamma_5})E_{\fq}(X_{\gamma_2+\gamma_4+\gamma_5}) E_{\fq}(X_{\gamma_2+\gamma_3+\gamma_5}) E_{\fq}(X_{\gamma_2+\gamma_3+\gamma_4}) E_{\fq}(X_{-\gamma_1})
\end{align}
to 
\begin{align}
    S_\IR&=E_{\fq}(X_{\gamma_2+\gamma_4+\gamma_5}) E_{\fq}(X_{\gamma_2+\gamma_3+\gamma_5}) E_{\fq}(X_{\gamma_2+\gamma_3+\gamma_4}) E_{\fq}(X_{-\gamma_1})
\end{align}
The arguments will be the $x_i^2$ for the four copies of $\CW_2$. We can identify dual variables:
\begin{align}
    F^\IR_{p_{1,+}} &= X_{-\frac12 \gamma_m +\frac12 \gamma_n +\frac12 \gamma_o} \cr
    F^\IR_{p_{1,-}} &= X_{\frac12 \gamma_m -\frac12 \gamma_n -\frac12 \gamma_o-\frac12 \gamma_2-\frac12 \gamma_4-\frac12 \gamma_5}+X_{\frac12 \gamma_m -\frac12 \gamma_n -\frac12 \gamma_o+\frac12 \gamma_2+\frac12 \gamma_4+\frac12 \gamma_5}\cr
    F^\IR_{p_{2,+}} &= X_{\frac12 \gamma_m -\frac12 \gamma_n +\frac12 \gamma_o} \cr
    F^\IR_{p_{2,-}} &= X_{-\frac12 \gamma_m +\frac12 \gamma_n -\frac12 \gamma_o-\frac12 \gamma_2-\frac12 \gamma_3-\frac12 \gamma_5}+X_{-\frac12 \gamma_m +\frac12 \gamma_n -\frac12 \gamma_o+\frac12 \gamma_2+\frac12 \gamma_3+\frac12 \gamma_5}\cr
    F^\IR_{p_{3,+}} &= X_{\frac12 \gamma_m +\frac12 \gamma_n -\frac12 \gamma_o} \cr
    F^\IR_{p_{3,-}} &= X_{-\frac12 \gamma_m -\frac12 \gamma_n +\frac12 \gamma_o-\frac12 \gamma_2-\frac12 \gamma_3-\frac12 \gamma_4}+X_{-\frac12 \gamma_m -\frac12 \gamma_n +\frac12 \gamma_o+\frac12 \gamma_2+\frac12 \gamma_3+\frac12 \gamma_4}\cr
    F^\IR_{p_{4,+}} &= X_{-\frac12\gamma_m-\frac12\gamma_n-\frac12\gamma_o-\frac12 \gamma_1 +\frac12 \gamma_2} \cr
    F^\IR_{p_{4,-}} &= X_{\frac12\gamma_m+\frac12\gamma_n+\frac12\gamma_o + \gamma_1-\frac12 \gamma_2}+X_{\frac12\gamma_m+\frac12\gamma_n+\frac12\gamma_o -\frac12 \gamma_2}\cr
\end{align}
so that e.g.
\begin{align}
    \RG(v_1) &= x_1^{-1} x_2 x_3 x_4^{-1} \cr 
    \RG(u_{1,+}) &= p_{2,+} p_{3,+}+x_2 x_3 p_{4,-} p_{1,-} \cr
    \RG(u_{1,-}) &= p_{1,+}p_{4,+} + x_1 x_4 p_{2,-} p_{3,-} 
\end{align}
which is a nice coproduct of two sets of bilinear generators embedded in two copies of $\CW_2^2$. The relative spectrum generator is $E_\fq(\fq^2\prod_i x_i^{-1} p_{i,+})$.

Notice that the Casimirs of the two sets of bilinear quantum group generators are controlled by $\mu_2 \mu_3$ and $\mu_2/\mu_3$, as expected from the character variety interpretation of the coproduct.

\section{Partial RG flows and $SL(2)$ character varieties}\label{sec:cut}
In this Section we will very briefly discuss a geometric perspective on partial RG flows in for theories of class $\CS$ associated to $\sl_2$.  Recall that the theories are labeled by a compact surface $C$, possibly decorated by punctures which can be regular or irregular. 

Regular punctures are associated to an $SU(2)$ flavour symmetry factor. 
Irregular punctures are labeled by an half-integral rank and for integral rank they are associated to a $U(1)$ factor of the flavour group, which we will often weakly gauge. 

As long as ranks are bounded by $1$, the class $\CS$ theories will have multiple equivalent Lagrangian descriptions. Higher rank punctures can be defined by partial RG flow from multiple punctures of lower rank.

When discussing the associated algebra $\CA$, it is useful to formally replace irregular singularities of rank $k/2$ with small holes with $k$ marked points. The algebra is a variant of the familiar Skein algebra for $\sl_2$ Chern-Simons theory, where lines are allowed to end at marked points. The roles governing these marked points are a bit complicated and we will not attempt to review them here. 

As long as $C$ has at least one puncture, typical Coulomb RG flows are labelled by triangulations of $C$, with vertices at punctures or marked points \cite{Gaiotto:2009hg}.

In that context, the resulting BPS quiver has nodes associated to internal edges and arrows associated to vertices of individual triangles, joining the corresponding pairs of edges counterclockwise. Each of the $SU(2)$ examples discussed in the previous Sections can be recovered in this manner. 

It was observed (see e.g. \cite{Gabella:2017hpz}) that each edge of the triangulation can be thought of as capturing some part of the RG flow. Indeed, from the perspective of this paper we would consider partial triangulations of $C$ as representing partial RG flows to an IR theory which includes both $U(1)$ gauge fields and matter theories associated to the surface $C_\IR$ obtained from $C$ by cutting along edges. We also expect the RG flow map on the Skein algebra to be ``local'', with precise rules on how to break lines which cross the edges. 

If this perspective is correct, the minimal RG flow step consists of adding an edge and cutting along it, and more general RG flows can be expressed in multiple ways as sequences of minimal flows.

We can give some concrete justifications for this proposal. For example, consider a surface with two regular punctures and an edge between them. 
The two theories have very similar Lagrangian descriptions, when available: the UV theory includes an $SU(2)$ gauge group acting on $T^*\bC^4$, the IR
theory is obtained by replacing $T^*\bC^4$ with $T^*\bC^2$. The corresponding RG flow is naturally triggered by giving a large mass to the 
$T^*\bC^2$ matter fields we want to remove. 

Skeins with lines which pass between the two punctures map to line defects with magnetic charge under $SU(2)$ and the ``cutting'' rules are associated to the behavior of these 't Hooft operators when some matter is removed. If we use our ``gauging'' technology to decompose $S_\UV$ and $S_\IR$ in terms of the spectrum generator of pure $SU(2)$ gauge theory, we see that they will differ by a simple factor of $f_2(\mu,w_1)$ as in our examples. Here $w_1$ is the algebra element associated to a loop around the two punctures. 

Similarly, we can consider an edge which joins a puncture to itself, going around a non-trivial loop on the surface. Then cutting along the edge leaves behind two irregular punctures of rank $\frac12$ on $C_\IR$. In a gauge theory description, we have two $SU(2)$ gauge fields acting on bi-fundamental hypermultiplets. Giving a mass to these and removing them leaves the expected gauge theory description associated to $C_\IR$. 
Conversely, we would build $S_\UV$ from $S_\IR$ by adding $f_{2 \times 2}(\mu, w_1,w_1')$. Here $w_1$ and $w_1'$ are the loops just to the left and just to the right of the cut. 

A progressive decomposition of $C$ will at some point require cutting along edges with at least one end on a marked point. In turn, this will require increasingly complex RG flows which we have not explored at this point. 

An interesting exception is an edge which joins the two marked points of a rank 1 singularity along some non-trivial path. We can pick a duality frame such that the whole gadget is captured by an $SU(2)$ $N_f=3$ theory coupled to the rest by gauging an $SO(4)$ subgroup of the flavour symmetry. The RG flow we studied in the context of the quantum group coproduct then breaks the curve into two IR curves both with a regular puncture, coupled by a diagonal $U(1)$ gauge field. The spectrum generator for this flow was described above. 

Accordingly, as soon as a singe rank 1 irregular singularity appears, the curve can be rapidly decomposed until no non-trivial loops connecting the two marked points of a rank 1 singularity exist. This leaves building blocks which we have already studied as $SU(2)$ gauge theory examples. 
In all these situations, we can thus verify the expected IR BPS quiver and even build the spectrum generator systematically. It would be interesting to compare the result with other approaches, such as the construction of Strebel spectral networks \cite{Longhi:2016wtv}. 

\section{Quantization of the Teichm\"uller space }
\label{sec:teichquant}
In this section we address the quantization of the Teichm\"uller space associated to punctured Riemann 
surfaces.  In the regime $\abs{\fq}^2 = 1$ this subject has a long history \cite{Verlinde:1989ua, kashaev1997,Kashaev:1998xp,Kashaev:2000ku,Teschner:2003em,Teschner:2005bz,Chekhov:1999tn,Aghaei:2015bqi,Teschner:2003em}, both in the infinite dimensional representations for generic $\fq$, and more recently in the finite dimensional case where $\fq$ is a root of unity \cite{ishibashi2025}. Our goal here is extend this construction to the Schur quantization regime, where the moduli space of flat connection is quantized for $0<\abs{\fq}^2 < 1$. This question naturally arises in our work as the Teichm\"uller space corresponds to the phase space of $\mathfrak{sl}_2(\bC)$ Chern-Simons theory in its Schur quantization\cite{Gaiotto:2024osr}, and therefore to the phase space \cite{Witten:1989ip} of Lorentzian de Sitter 3d gravity\footnote{In this regards, it would be very interesting to compare our approach to the one advocated in \cite{Collier:2025pbm,Collier:2024kmo}}. As we see shortly, the construction of the Teichm\"uller space follows from the IR Schur quantization applied to theories of class $\cS$, and it is deeply connected to the theory of quantum groups and cluster algebras \cite{fock2006modulispaceslocalsystems,Fock_2008}. 

Let us briefly summarize the construction before delving into the details.
Providing a quantization of the Teichm\"uller space accounts to constructive a (projective) representation of the (dotted) Ptolemy groupoid, whose objects consist of ideal (decorated) triangulations and its morphisms are \textit{flips} among them corresponding to the Mapping Class Group action. 
Concretely in Kashaev's construction, one defines coordinates on the Teichmuller space associating to each (marked) triangulations a vector space, spanned by a quantization of Kashaev's coordinates, onto which the flip morphisms act as linear operators generating quantum coordinate change. For $\abs{\fq} = 1$, these linear operators are constructed based on the Faddeev non-compact quantum dilogarithm \cite{Faddeev:1993rs}.  In our $-1<\fq<1$ story, we find that this crucial ingredients is replaced by (a version of) the corresponding quantum exponential function\eqref{quantumdil}\cite{Woronowicz1992}:
\begin{equation} 
\label{qexp}
\Phi_\fq(z) = \frac{\left(\fq^2 z; \fq^2 \right)_\infty}{\left( \overline{z}; \fq^2 \right)_\infty}\comma\qquad \abs{z} \in \fq^\bZ
\end{equation}at values $\abs{\fq} <1$, that, as expected, produces the intertwiner for the diagonalization of the Wilson Line on the annulus\footnote{That is the diagonalization of the Dehn Twist for this surface.}. We find that this special function is the one giving the multiplicative unitary operator associated with the $U_\fq(\sl_2)$ quantum group in the Schur regime. This elucidates the relation to the Teichm\"uller quantization\cite{Teschner:2005bz} based on the quantization of cluster varieties proposed by Fock and Goncharov \cite{fock2006modulispaceslocalsystems,Fock_2008,Chekhov:1999tn,Teschner:2003em}
 for the case $\abs{\fq} =1$, that we extended here for the Schur regime.  
 
\medskip
The quantization of Teichm\"uller space provides a realization of the Schur quantization developed in Section \ref{Sec:alg} for the case of Class $\cS$ theories.  The Hilbert space is realized explicitly by $L^2(\Gamma/\Gamma_f)$ where the generators are naturally associated with the cluster basis given by Fock-Goncharov coordinates associated to the surface $\cS$ as discussed below in \ref{sec:teichcluster}. Indeed, the Fock-Goncharov quantization produces precisely the cluster structure presented in \ref{introcluster}. The algebra $\cA$ is the Skein algebra $\rm Sk(\mathfrak{g},\cS)$ and is realized canonically within Teichm\"uller theory via the ``traffic rules''  construction\cite{Gabella:2017hpz, Gaiotto:2012rg}.
 
\subsection{Representation of Ptolemy groupoid}
\begin{figure}[ht]
    \centering
\begin{tikzpicture}
\draw(-30:1) -- (90:1) -- (210:1) --cycle;
\node at (0,0) {$v$};
\dast{(90:0.8)};
%\node at (0,-1) {$\dot\Delta_1$};
\draw[->] (1,0.3) --node[midway,above]{$A_v$} (2,0.3);
\draw[shift={(3,0)}] (-30:1) -- (90:1) -- (210:1) --cycle;
\node at (3,0) {$v$};
%\node at (3,-1) {$\dot\Delta_2$};
\dast{(3,0)++(210:0.8)};

\begin{scope}[xshift=6cm,yshift=-0.5cm]
\draw (0,0) -- (1.5,0) -- (1.5,1.5) -- (0,1.5) --cycle; 
\draw (0,1.5) -- (1.5,0);
\node at (0.5,0.5) {$v$};
\node at (1,1) {$w$};
\dast{(0.15,0.15)};
\dast{(1.4,0.25)};
%\node at (0.75,-0.5) {$\dot\Delta_1$};
\draw[->] (2,0.75) --node[midway,above]{$T_{vw}$} (3,0.75);
\end{scope}
\begin{scope}[xshift=9.5cm,yshift=-0.5cm]
\draw (0,0) -- (1.5,0) -- (1.5,1.5) -- (0,1.5) --cycle; 
\draw (1.5,1.5) -- (0,0);
\node at (0.5,1) {$v$};
\node at (1,0.5) {$w$};
\dast{(0.1,0.25)};
\dast{(1.35,0.15)};
%\node at (0.75,-0.5) {$\dot\Delta_2$};
\end{scope}
\end{tikzpicture}
    \caption{Elementary moves.}
    \label{fig:elem_morph}
\end{figure}
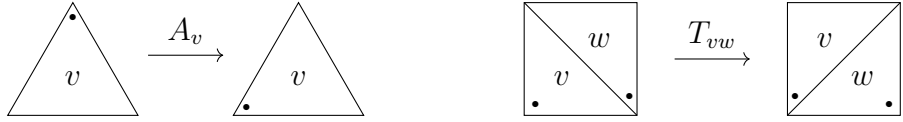

Let us consider an ideal triangulation $\Delta$ of a  compact Riemann surface $\Sigma$ equipped with a set of marked points. A dotted triangulation $\dot\Delta$ is an ideal triangulation equipped with the a map choosing a particular vertex of each triangle $\Delta_v \in \Delta$. 
The dotted Ptolemy groupoid of $\Sigma$ is the full groupoid  over the set of dotted ideal triangulations of $\Sigma$. This decoration of the Ptolemy groupoid accounts to specifying a particular polarization in assigning Weyl commuting variables to each triangle in the triangulation  Its morphisms are generated by the two elementary moves\cite{Teschner:2005bz,kashaev1997,Kashaev:2000ku,Teschner:2003em}: the dot rotation $A_v$  and the flip T$_{vw}$ reported in Figure \ref{fig:elem_morph}. $A_v$ acts on a triangulation by rotating the marked vertex of the $v$-th triangle, and corresponds to the elementary move generating the change of polarization. The $T_{vw}$ morphism acts by \textit{flipping} the diagonal of two triangles in as in Figure \ref{fig:elem_morph}.

\begin{figure}
    \centering
\includegraphics[width=0.5\linewidth]{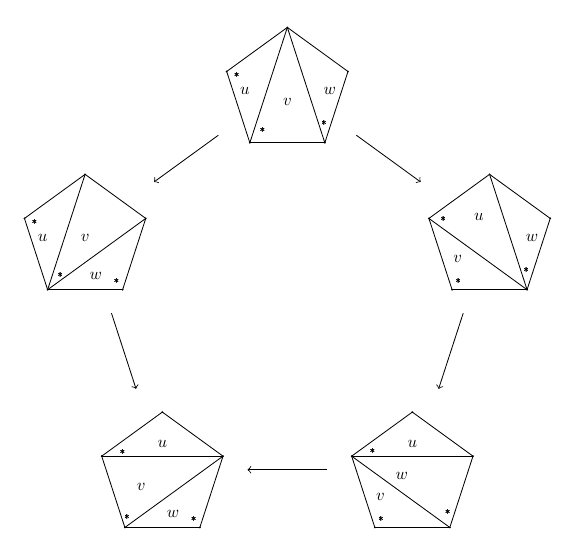}
    \caption{Pentagon equation}
    \label{fig:pentagon}
\end{figure}

The elementary moves are subjected to the following set of relations\cite{Kashaev:2000ku}:
\begin{align}
\label{ptolemy1}
    A_v^3 &= \id \\
    \label{ptolemy2}
 T_{uv} \, T_{uw}\, T_{vw} &= T_{vw}\, T_{uv}\\
   \label{ptoplemy3}
  T_{uv} \, T_{wz}&= T_{wz}\, T_{uv}\\
  \label{ptoplemy4}
    A_v \, T_{vw}\, A_w &= A_w \, T_{wv}\, A_v\\
    \label{ptoplemy5}
    T_{vw} \, A_v\, T_{wv} &= \, A_v\, A_w\, P_{vw}
\end{align}
where $P_{vw}$ denotes the permutation of $v$ and $w$. The second of those identities is the famous pentagon equation in Figure \ref{fig:pentagon}. 
Clearly under the forgetful functor dropping the marked points, $\dot\Delta\to \Delta$, the dot rotation descends to the identity morphism $A_v\to \id$ and the Ptolemy groupoid consist only of flip relation. 

Now, let us fix a polarization by quantizing Kashaev coordinates\footnote{It is standard how to recover Penner/shear/Fock coordinates from the Kashaev ones and we defer the reader to \cite{Aghaei:2015bqi,Teschner:2003em,Teschner:2005bz,Kashaev:2000ku} for the details}  by assigning to each dotted triangle $\dot t_v\in \dot\Delta$ of any given triangulation $\fq$-Weyl commuting coordinates for the $*$-algebra double $\cW_v \times \cW_v^{\rm op}$ satisfying the Schur commutation relations(cfr. with  related discussion in Section 5.1 of \cite{Gaiotto:2024osr})\footnote{In the case of Class $\cS$ the $\rho$ is the trivial automorphism of the doubled algebra and  $\cU_v^\dagger = \ov\cU_v,\cV_v^\dagger = \ov\cV_v $}:
 \begin{equation} \cU_v \,\cV_v = \fq \, \cV_v \, \cU_v \comma \quad \overline{\cU}_v \,\overline{\cV}_v = \fq^{-1}\, \overline{\cV}_v \, \overline{\cU}_v \comma
 \end{equation} 
 and commuting otherwise, 
 in such a way that the variables are associated to the edges as:
\begin{align}
\label{dualgraph}
\begin{tikzpicture}[baseline=(current bounding box.center)]
\draw(-30:1) -- (90:1) -- (210:1) --cycle;
\node at (0,0) {$v$};
\dast{(90:0.8)};  
 \node at ($(90:1)!0.5!(-30:1) + (0.4,0)$) {$\cV^{-1}_v$};
    \node at ($(210:1)!0.5!(90:1) + (-0.5,0.0)$) {$\cU_v$};
    \node at ($(-30:1)!0.5!(210:1) + (0.1,-0.3)$) {$\cV_v\, \cU^{-1}_v $};
\end{tikzpicture}
\end{align}
and analogously for the adjoint generators.
Then, it is clear that over the Schur representation
the operator $A_v$ is simply the unitary operator implementing the change of polarization as\footnote{The  explicit expression for $A_v$ is identical to the one in e.g.\ \cite{Kashaev:2000ku}. }:
\begin{equation}
    A_v \,  \cU_v \, A^{-1}_v = \cV_v \,\cU^{-1}_v\comma \qquad  A_v \,\cV_v \, A^{-1}_v  = \cU_v^{-1} \period
\end{equation} 
and analogously on the $\overline{\cU}_v, \overline{\cV}_v$ pair. 
We found that the operator:
\begin{equation}
\label{Tuv}
    T_{uv} = \Phi_\fq(\cU_u \, \cV_v \, \cU_v^{-1}) \,\chi(\cU_v, \cV_u)
\end{equation}
where we used the standard tensor product notation, and $\chi(\cU_v, \cV_u)$ is the bi-character:
\begin{equation}
\label{bicaracter}
   \chi(\cU_v, \cV_u) := \chi(\id \otimes \cU_v, \cV_u\otimes \id) =  \left(\frac{\cU_v}{\ov\cU_v}\right)^{-N_u}\, \otimes \id ,\qquad \abs{\cV_u} = \abs{\ov\cV_u} = \fq^{N_u}
\end{equation}
gives a representation of the flip operator (see also related discussions in \cite{woronowicz_2025,Woronowicz1992,lirias1743123,Fock_2008}). This can be proven by showing that \eqref{ptolemy2}-\eqref{ptoplemy5} are satisfied. In particular, the pentagon identity can be shown by observing that:
\begin{equation}
    T_{uv}\, T_{uw}\, T_{vw} = \Phq(\cU_u \, \cV_v \, \cU^{-1}_v) \Phq(\cU_u \, \cU^{-1}_v\,\cU^{-1}_w \,\cV_w ) \, \Phq(\cU_v \, \cV_w \, \cU^{-1}_w) \, \chi(\cU_v, \cV_u)\, \chi(\cU_w, \cV_u)\, \chi(\cU_w, \cV_v)
\end{equation}
a direct computations gives  the pentagon identity for the characters \begin{equation}\chi(\cU_v, \cV_u)\, \chi(\cU_w, \cV_u)\, \chi(\cU_w, \cV_v) = \chi(\cU_w, \cV_v) \, \chi(\cU_v, \cV_u)\comma\end{equation} while the non trivial relation for the special function $\Phi_\fq(-)$ over the Schur representation:
\begin{equation}
    \Phq(\cU_u \, \cV_v \, \cU^{-1}_v) \Phq(\cU_u \, \cU^{-1}_v\,\cU^{-1}_w \,\cV_w ) \, \Phq(\cU_v \, \cV_w \, \cU^{-1}_w) = \Phq(\cU_v \, \cV_w \, \cU^{-1}_w)\,  \Phq(\cU_u \, \cV_v \, \cU^{-1}_v)
\end{equation}
follows from Theorem 2.2 and 4.1 in \cite{Woronowicz1992}.  \eqref{ptoplemy3}, \eqref{ptoplemy4} follow directly from the definitions, while the inversion \eqref{ptoplemy5} is a consequence of the inversion relation for the function $F_\fq(z)$ \eqref{inversion}. Indeed, after basic algebra, one finds that the operator:
\begin{equation}
\begin{split}
  \Pi_{vw}  &= A_w^{-1} \, A^{-1}_v\, T_{vw}\, A_v T_{wv} = A_w^{-1} \,\chi(\cU_2\, \cV_1^{-1},\cU_2\, \cV_1^{-1}) \chi(\cV_1 \, \cU_2^{-1},\cV_2) \, \chi(\cU_2,\cV_1)
    \end{split}
\end{equation} 
coincides with the permutation operator $\Pi_{vw} = P_{vw}$. This is different compared to the case $\abs{\fq}^2=1$ where the representation of the Ptolemy groupoid is projective and extended by the central charge of Liouville theory.

\subsection{Relation to cluster algebra quantization}\label{sec:teichcluster}
In the previous section we found that the representation of the Ptolemy groupoid is not Projective over the Schur quantization Hilbert space. This suggest we could have avoided to first fix a particular polarization, and then showing compatibility with the flip operation. 
This is a rather expected result in lieu of the works\cite{fock2006modulispaceslocalsystems, Chekhov:1999tn, Fock_2008} where it was shown that in the construction of quantum double already in the regime $\abs{\fq}^2=1$ there is no dependence on the polarization choice. Here we will we schematically extend to the Schur case their argument, and we refer to their work, and in particular to \cite{Fock_2008} for more details. From this perspective, the Schur Teichm\"uller quantization can be thought as following from the quantization of the underlying quantum doubled  cluster variety. 

We associate to the triangulation $\Delta$ (not it does not need to have any marking) a quantum torus algebra $\cX_\Delta$ by assigning a generator $\left(X_\alpha^\Delta\right)^{\pm 1}$ for each $\alpha$ being an edge in $\Delta$, satisfying the relation:
\begin{equation}
\label{gen}
    X^\Delta_\A \, X^\Delta_\B = \fq^{2\epsilon^\Delta_{\alpha\beta}} X^\Delta_\B \, X^\Delta_\A
\end{equation} 
where $\epsilon^{\Delta}_{\alpha\beta}$ is a skewsymmetric pairing obtained if the following way: for each vertex between $\A$ and $\B$, $\epsilon^\Delta_{\A\B}$ is the number of angular sectors between the edges $\A$ and $\B$, with positive sign if counterclockwise or with negative sign if clockwise around the vertex. Then a basis for the Schur quantum double is given by the generators \eqref{gen} together with their counterpart in the opposite algebra $\cX_\Delta^{\rm op}$: \be 
   \ov X^\Delta_\A \, \ov X^\Delta_\B = \fq^{-2\epsilon^\Delta_{\alpha\beta}}\ov X^\Delta_\B \, \ov X^\Delta_\A
\ee
with $\ov X_\A \, X_\B  = X_\B \, \ov X_\A$. 

Mutations $\mu_\gamma$ of the quantum double along the direction $\gamma$ correspond to flips of the underlying triangulation at the edge $\gamma$ and the corresponding pairings are related by:
\begin{align}
\mu_\gamma &: \cX_\Delta \times \cX^{\rm op}_\Delta \to  \cX_{\Delta'} \times \cX^{\rm op}_{\Delta'}\comma \\
\mu_\gamma &: \epsilon^\Delta_{\A\B}\mapsto \epsilon^{\Delta'}_{\A\B} = \begin{cases} - \epsilon^\Delta_{\A\B} & {\rm if} \quad  \A = \gamma \quad  {\rm or}\quad  \B = \gamma\\
    \epsilon^{\Delta}_{\A\B} & {\rm if} \quad \epsilon^\Delta_{\A\gamma} \,  \epsilon^\Delta_{\gamma\B} \leq 0 \\
     \epsilon^{\Delta}_{\A\B} + \abs{\epsilon^{\Delta}_{\A\gamma}}\, \epsilon^{\Delta}_{\B\gamma} & {\rm if} \quad \epsilon^\Delta_{\A\gamma} \,  \epsilon^\Delta_{\gamma\B} > 0
    \end{cases}
\end{align}
Note that graphically, these exactly correspond to quiver mutations, where $\epsilon_{\A\B}$ is the number of arrows between the node $\A$, $\B$ with corresponding sign. The mutation induce corresponding rational maps between the cluster tori that we indicate with $\mu^*_\gamma$  by:
\begin{align}
\mu^*_\gamma \, X_\A &= \begin{cases} X_\A^{-1} & {\rm if} \quad \gamma = \A\\
X_\A\left(1+X_\gamma^{-{\rm sgn}(\epsilon^{\Delta}_{\A\gamma})}\right)^{-\epsilon^\Delta_{\A\gamma}} & {\rm if} \quad \gamma \neq \A
\end{cases} \\
\mu^*_\gamma \, \ov X_\A &= \begin{cases} \ov X_\A^{-1} & {\rm if} \quad \gamma = \A\\
\ov X_\A\left(1+\ov X_\gamma^{{\rm sgn}(\epsilon^{\Delta}_{\A\gamma})}\right)^{\epsilon^\Delta_{\A\gamma}} & {\rm if} \quad \gamma \neq \A\end{cases}
\end{align}
and analogously on $ \cX^{\rm op}_\Delta$. As advocated in \cite{Fock_2008} the quantum version of the mutation maps may be decomposed in a composition:
\be 
\mu^{\fq}_\gamma = \mu^\#_\gamma \circ \mu'_k
\ee
while the ``monomial'' part is universal and given by:
\begin{align}
    \mu'_\gamma(X_\A) = \begin{cases}   
  X^{-1}_{\A} & {\rm if}\quad \A = \gamma\\
X_{\A} \left( X_\gamma \right)^{[\epsilon^\Delta_{\A \gamma}]_+} & {\rm if}\quad \A \neq\gamma
\end{cases}\comma\quad     \mu'_\gamma(\ov X_\A) = \begin{cases}   
  \ov X^{-1}_{\A} & {\rm if}\quad \A = \gamma\\
\ov X_{\A} \left(\ov X_\gamma \right)^{-[\epsilon^\Delta_{\A \gamma}]_+} & {\rm if}\quad \A \neq \gamma
\end{cases}
\end{align}
where $[a]_+= \max(a,0)$, as in the case of $\abs{q}^2 = 1$. The ``automorphism'' part $\mu^\#_\gamma$ is instead governed by the quantum dilogarithm.  For the quantization on the unit circle the automorphism part is shown be be conjugation by the Fadeevv non-compact quantum dilog, as expected. In our regime\footnote{It was already observed in \cite{Fock_2008} that this ration of quantum dilogarithm provides the correct qunatum mutation action. Yet, in that work they were interested in the quantization at $\abs{\fq}^2 =1$, where the ratio can merely be understood as a formal power series. They indeed replace it by its convergent version being the Fadeevv one.} of interested, is it given by the corresponding adjoint action by the Schur multiplicative unitarity ``quantum dilogorithm'' $F_\fq(X_\gamma)$:
\begin{equation}
    \mu^\#_\gamma(A) = F^{-1}_\fq(X_\gamma) \, A F_\fq(X_\gamma) \comma\qquad F_\fq(X_\gamma) = \frac{\left( X_\gamma; \fq^2 \right)_\infty}{\left( \ov{X}_\gamma; \fq^2 \right)_\infty} 
\end{equation} 
where $A = X_\alpha, \ov X_\A$.

Consider the $\star$-algebra representation of $(\cX_\Delta \times \cX^{\rm op}_\Delta )$ over a Hilbert space of $L^2$ functions on the quantum torus\footnote{ E.g.\ for a single pair of cluster coordinates this is just a representation\cite{Gaiotto:2024osr} where $X_1 \, X_2 = \fq^2 \, X_2 \, X_1$ and the Hilbert space is just $L^2(\bZ \times S^1)$ where $X_1 g_n(\theta) = g_{n+1}(\theta -i\hbar)\comma$ $X_2 g_n(\theta)= \fq^n\, e^{i\theta} g_n(\theta)$ where $\fq^2 = e^{-\hbar}$.  }. Then mutation $\mu^\fq$ descends to an unitary intertwiner $K^k_{\Delta, \Delta'}:= K^{\#} \circ K'$ between the Schwartz spaces $\cS_\Delta$ and $\cS_{\Delta'}$ introduced in section \ref{Schwartzspace}, where $\Delta$ and $\Delta'$ differ by a mutation along $k$. Explicitly, $K^{\#}$ acts as multiplication operator by $\Phq(X_k)$ and $K'$ acts the bi-character \eqref{bicaracter} $\chi(X_k\otimes \id, \id \otimes X_k)$. Then, the intertwiner satisfies the pentagon relation as follows from the discussion in the previous subsection.

\begin{figure}
    \centering
    \includegraphics[width=0.5\linewidth]{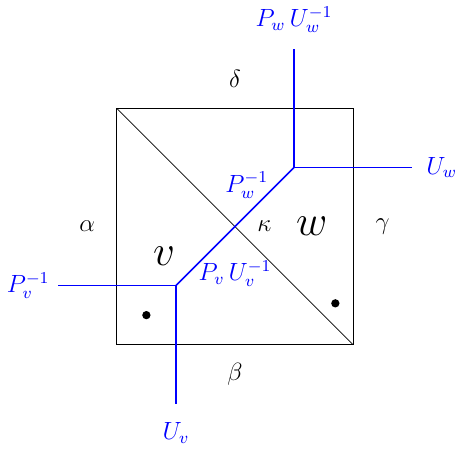}
    \caption{Dual graph associated to a triangulation $\Delta = \Delta_v \oplus \Delta_w$ and coordinates as prescribed in \eqref{dualgraph}. Here $W_{\A,v} = P^{-1}_v$, $W_{\B,v} = U_v$,  $W_{\kappa,v} = P_v\, U^{-1}_v$ etc\dots }
    \label{fig:dualgraph}
\end{figure}

Henceforth, from the quantization of the cluster variety we inherit a quantization of the Teichm\"uller space. To clarify the relation with the discussion in the previous subsection, one can embed $\cX_\Delta$ into the star representation of the dotted Ptolemy groupoid by just associating to each triangle $\Delta_v$ in $\Delta$ coordinates in $\cW_v\times \cW^{\rm op}_v$ denoted by $W_{\alpha, v}$, $\ov W_{\alpha,v}$ for the dual graph \cite{Teschner:2005bz} as in \eqref{dualgraph}  and \ref{fig:dualgraph} , where one of the corners inherits the decoration of $\dot\Delta_v$; explicitly one associates  $W_{1, v} = \cU_v \comma$ $W_{2,v} = \cV_v\, \cU^{-1}_v\comma$  $W_{3,v} = \cV^{-1}_v\comma$ to the triangle in \eqref{dualgraph}, and analogously for $\ov \cW$.  Then, the embedding : 
\be 
X^\Delta_\A \mapsto W_{\A,v}\,  W_{\A,w}\comma \qquad \ov X^\Delta_\A \mapsto \ov W_{\A,v} \,\ov W_{\A,w}
\ee
is the unique\cite{ishibashi2025} algebra embedding of $\cX_\Delta$ in the dotted torus algebra,  
 where $W_{\A,w}$, $W_{\A,w}$ are the coordinates in $\Delta_v$ and $\Delta_w$ associated to the same edge $\alpha$ of $\Delta$. It is then straightforward to verify that the quantum mutation maps constructed above descend to the operators $T_{uv}$ in the previous subsection \eqref{Tuv} under the above embedding.

\subsection*{Acknowledgement} 
We would like to thank J\"org Teschner for contributions at early stages of the project. This
research was supported in part by a grant from the Krembil Foundation. DG is supported by
the NSERC Discovery Grant program and by the Perimeter Institute for Theoretical Physics. 
Research at Perimeter Institute is supported in part by the Government of Canada through the Department of Innovation, Science and Economic
Development Canada and by the Province of Ontario through the Ministry of Colleges and
Universities. The work of FA  is funded by the Deutsche Forschungsgemeinschaft (DFG, German Research Foundation) under Germany's Excellence Strategy - EXC 2121 ``Quantum Universe'' - 390833306 and acknowledges support by the Collaborative Research Center - SFB 1624 ``Higher structures, moduli spaces and integrability'' - 506632645.

\appendix

\section{Schur pairing and the Schur index} \label{sec:phys}
In this section we will briefly review the physical origin of the fusion algebra of loop operators and of the Schur pairing on the algebra. Recall the four-dimensional ${\cal N}=2$ super-algebra:
\begin{align}
    \{Q^A_\alpha, Q^B_\beta\}&=0 \cr
    \{Q^A_\alpha, \bar Q^B_{\dot \beta}\}&=\epsilon^{AB}P_{\alpha \dot \beta} \cr
    \{\bar Q^A_{\dot \alpha}, \bar Q^B_{\dot \beta}\}&=0
\end{align}
Here $\alpha$, $\beta$ and $\dot \alpha$, $\dot \beta$ are spinor indices of the two chiralities, $A$, $B$ are doublet indices for the $SU(2)_R$ R-symmetry. 

We will always assume the existence of $SU(2)_R$. All 4d ${\cal N}=2$ gauge theories have that property, with the exception of Abelian gauge theories deformed by FI parameters. 
Theories of class $\CS$ also have $SU(2)_R$ symmetry. Furthermore, we will essentially define a Coulomb vacuum or RG flow as being $SU(2)_R$-preserving. 

SCFTs are also equipped with an $U(1)_r$ symmetry giving charge $\pm 1$ to super-charges 
depending on their chirality. The theories, vacua and RG flows we consider typically break $U(1)_r$. 

In a Coulomb vacuum, with the exception of the conformal vacuum of an SCFT, the SUSY algebra is typically deformed by central charges: 
\begin{align}
    \{Q^A_\alpha, Q^B_\beta\}&=\epsilon^{AB} \epsilon_{\alpha \beta} Z \cr
    \{Q^A_\alpha, \bar Q^B_{\dot \beta}\}&=\epsilon^{AB}P_{\alpha \dot \beta} \cr
    \{\bar Q^A_{\dot \alpha}, \bar Q^B_{\dot \beta}\}&=\epsilon^{AB} \epsilon_{\dot \alpha \dot \beta} \bar Z
\end{align}
The central charge is a complex linear combination of flavour symmetries of the theory and in particular it vanishes in the vacuum itself, but is carried by massive excitations. Indeed, it sets a lower ``BPS'' bound on the mass of particles which carry it: $M\geq |Z|$. 

Particles which saturate this bound are called BPS. Their spectrum (in the sense of an appropriate supersymmetric index) is robust and only jumps across certain walls of marginal stability in parameter space, which we will discuss below. The quantum spectrum generator $S$ captures the spectrum in a particularly convenient way and is invariant across walls of marginal stability for BPS particles. 

In superconformal or free field theories one can neatly organize local operators into supersymmetry multiplets and define superconformal indices which count special ``short'' supersymmetry multiplets in a robust manner. The Schur index we employ through this paper was originally defined in a similar manner, but it is believed to make sense for non-conformal theories as well. This is done by selecting a specific nilpotent supercharge 
\begin{equation}
    Q \equiv Q^-_- + Q^{-}_{\dot -}
\end{equation}
and expressing the Schur index as a graded Euler character/Witten index of the cohomology of that supercharge in the space of local operators. 

The super-charge $Q$ is called Holomorphic-Topological \cite{Kapustin:2006hi} because most translations are $Q$-exact except $P_{+ \dot +} = \partial_z$, where $z$ is an holomorphic coordinate in a plane $\bC \subset \bR^4$ in spacetime. Accordingly, if an operator $O$ contributes to the cohomology only the $\partial_z^n O$ derivatives will also contribute. The supercharge $Q$ commutes with the combination $\hat J = J_3-I_3$ of the rotation generator $J_3$ of $\bC$ and R-symmetry generator $I_3$. The $\partial_z$ derivatives 
increase $\hat J$ by $2$. We will assume that the cohomology $H(Q)$ of local operators has finite-dimensional $\hat J$ eigenspaces and thus the Schur index 
\begin{equation}
    I(\fq) \equiv \Tr_{H(Q)} (-1)^{2 I_3} \fq^{2J_3-2I_3}
\end{equation}
is well-defined. We can also insert fugacities $\mu$ for flavor symmetries. 

The Schur index is expected to be invariant under continuous deformations of the SQFT. In gauge theories with an explicit Lagrangian description, a standard strategy to compute the Schur index is to make the couplings arbitrarily weak and count gauge-invariant polynomials in cohomology representatives of the space of fields. Concretely, each half-hypermultiplet of matter fields 
contributes a single bosonic field and its $\partial_z$ derivatives, and each vectormultiplet of gauge fields contributes two fermionic fields and their $\partial_z$ derivatives. The outcome are the UV formulae for the Schur index we employ in this paper. 

The behavior of the Schur index under RG flow has important subtleties. Superconformal indices are often RG flow invariant, but not always. A typical proof strategy is to define a generalization of the indices which can be defined away from conformality and which remains constant along the flow. As RG flows may involve the decoupling of very massive degrees of freedom, this only works if such degrees of freedom do not contribute to the final answer. The Schur index is already defined away from conformality, but it can manifestly receive contributions from massive degrees of freedom: an hypermultiplet can be given a mass which preserves $Q$ and $\hat J$ and has non-trivial Schur index. 
Accordingly, the RG flow from a massive free hypermultiplet to a trivial theory cannot preserve the Schur index. 

The IR formulae for the Schur index were initially proposed for SCFTs, where the state-operator map 
can be used to replace local operators with states on a three-sphere $S^3$. Assuming that the $S^3 \times \bR$ setup could be deformed to allow the theory to move to a non-trivial Coulomb vacuum, it was conjectured that only massive BPS particles would contribute to the $Q$-cohomology, and only if placed at specific central charge-dependent locations on the $S^3$. The spectrum generator $S$ would then collect the contribution of half of the possible locations, leading to the $I(\fq) = \langle S|S\rangle$ formula. 

The extension of this perspective to a general SQFT is challenging but possible in the language of Holomorphic-Topological twists, which we review below. Alternatively, one could take a perspective analogue to \cite{Gaiotto:2015aoa}, where local operators are characterized in terms of the ``fans'' of BPS particles they can create out of the vacuum. 

For SCFTs, convergence of the Schur index for $|\fq|<1$ is roughly guaranteed by the state-operator map to a partition function on a certain $S^1 \times S^3$ geometry. For a general SQFT the statement is less obvious. The Holomorphic-Topological twist interpretation below provides a possible argument.

The main actors in our story are line defects which preserve 
\begin{itemize}
    \item Half of the supercharges, of the general form 
    \begin{equation}
        q^A_\alpha \equiv Q^A_\alpha + e^{i \vartheta} n_\alpha^{\dot \beta} \bar Q^A_{\dot \beta}  
    \end{equation}
    where $n$ is the (time-like) unit vector pointing in the direction of the line defect in space-time and $\vartheta$ an angle we will specify momentarily. 
    These anti-commute to 
    \begin{equation}
        \{q^A_\alpha, q^B_\beta \} =2 \epsilon^{AB} e^{i \vartheta} n\cdot P 
    \end{equation}
    i.e. to the translations along the defect. We set our conventions so that a standard line defect extending along the time direction has $n_\alpha^{\dot \beta} = \delta_\alpha^{\dot \beta}$ and thus $q^-_-=Q$ if $\vartheta=0$. 
    \item The $SO(3)$ rotation symmetry in the directions perpendicular to the defect and the full $SU(2)_R$ R-symmetry.
    \item The flavour symmetries of the theory.
\end{itemize}
These constraints are not as restrictive as it may appear. In particular, they hold for natural 't Hooft-Wilson line defects defined in gauge theory as well as in theories of class $S$. They are also preserved by the RG flows we are interested in. Furthermore, they can be defined uniformly for any choice of $\vartheta$, though one finds that a continuous evolution $\vartheta \to \vartheta + 2 \pi$ may not bring a defect back to the original. For example, in non-conformal gauge theories it will shift the electric charge of a 't Hooft-Wilson line
by a multiple of the magnetic charge, via an analogue of Witten's effect. Ultimately, this will give rise to the $\rho^2$ map on line defects.

Any such line defect $L$ will be associated to some space of local operators supported on the defect. This space is acted upon by the symmetries preserved by the line defect. Assuming that we have tuned the choice of $n$ and $\vartheta$ so that $Q=q^-_-$, e.g. as described above, we can define a Schur index $I_{L,L}(\fq)$ counting such local operators. If we have two line defects $L_1$ and $L_2$ with the same orientation in space-time, we can count defect-changing local operators between them by a Schur index $I_{L_1;L_2}(\fq)$ \cite{Dimofte:2011py,Cordova:2015nma,Cordova:2016uwk}. 

One can arrange $Q=q^-_-$ for line defects with a generic orientation in the plane transverse to the $z$ direction by adjusting $\vartheta$ to equal the angle off the (euclidean time) vertical in the plane. A collection of line defects $L_i$ with a cyclic sequence of slopes in the plane can then end at the origin and support some space of junctions. The corresponding Schur index will be denoted as 
\begin{equation}
    I_{L_1, \cdots L_k}(\fq)
\end{equation}
We do not need to specify the slopes in detail, as the index is a deformation invariant. 

Here we defined the setup with all line defects ``outgoing'' from the junction point and labeled the line defects as if they had been brought back to the vertical. Accordingly, if we bring a line all around the cyclic order we need to pay a $\rho^2$ map:
\begin{equation}
    I_{L_1; \cdots L_k}(\fq) = I_{\rho^2(L_k);L_1; \cdots L_{k-1}}(\fq) \, ,
\end{equation}
etc.

We can compare this definition with the original $I_{L_1,L_2}(\fq)$ by defining a map $\rho$ which rotates the slope of a line  defect by $\pi$ and then re-interprets it as defect along the original direction. Then 
\begin{equation}
    I_{L_1,L_2}(\fq) = I_{\rho(L_1);L_2}(\fq)=I_{L_2; \rho^{-1}(L_1)}(\fq)
\end{equation}

As the Schur indices are deformation-invariant, they can be thought of as depending on some kind of deformation classes of line defects. A general expectation is that there will be some collection of ``elementary'' line defects $L_i$ such that all other ones line defects with the same amount of symmetry sit in deformation classes which include a direct sum of $L_i$ defects dressed by auxiliary ``Chan-Paton'' vector spaces with the same action of symmetries. The Schur indices would then be expressed in terms of the Schur indices for elementary line defects and of the the characters $\Tr (-1)^{2 I_3} \fq^{2J_3-2I_3}$ of these auxiliary vector spaces, which are linear combinations of $\fq$-integers coming from $SU(2)$ irreps.  

The notion of fusion of half-BPS line defects involves the same kind of assumption: two parallel line defects, say both placed at the origin of the $z$ plane and preserving the same SUSY, can be interpreted as a single defect breaking $SO(3)$  rotation symmetry to the $SO(2)$ subgroup generated by $\hat J$. The general expectation is that such a defect would also belong to the same deformation class as some linear combination of $L_i$'s, though the CP factors will only carry an action of $SO(2)$ rotations. This leads to the notion of fusion algebra $\CA$ of line defects we employ in the main text, with coefficients which are integral Laurent polynomials in $\fq$ \cite{Kapustin:2006hi,Kapustin:2007wm,Gaiotto:2010be,Braverman:2016wma,Gaiotto:2023ezy,Gaiotto:2024fso}. 

With a bit of help from the HT twist setup described below, one can then argue that 
\begin{equation}
    I_{L_1; \cdots L_k}(\fq) = I_{L_1 \cdot \cdots L_k}(\fq)
\end{equation}
i.e. consecutive line defects can be fused without changing the Schur indices.

There is an important subtlety concerning the $I_{L_1,L_2}(\fq)$ indices and the map $\rho$.
If we dress $L_2$ by an auxiliary CP space $V$, the Schur index will be multiplied by the character of $V$. If we dress $L_1$, though, the index will be multiplied by the character of $V^\vee$. 
The same discrepancy occurs for $\rho$: if $L$ is dressed by $V$ then $\rho(L)$ should be dressed by $V^\vee$. 

The character of $V^\vee$ is obtained from the character of $V$ by flipping all fugacities. We already accounted for the effect on flavor fugacities when we set $\rho(\mu) = \mu^{-1}$. But we also need to invert the $\fq$ fugacity. At the same time, it is easy to see graphically that $\rho$ acting on the fusion of two lines should give the fusion of their $\rho$ images in the opposite order. The correct definition of $\rho$ is thus that of a map 
\begin{equation}
    \rho: \CA_\fq \to \CA^{\rm op}_{\fq^{-1}}
\end{equation}
Because of the $SO(3)$ rotation symmetry of the physical line defects we employ, we get for free an identification $A^{\rm op}_{\fq^{-1}} \simeq A_\fq$ mapping the $L_i$ to themselves and mapping $\fq \to \fq^{-1}$. This map is implicitly included in the definitions we use in the main text, 
where $\rho$ is defined as an algebra map $\CA \to \CA$ and the pairing $I_{a,b}$ as a pairing on $\CA$. In other word, we start from $\rho(L_i)$ and $I_{L_i,L_j}$ and extend them linearly to the whole $\CA$.

We do not have a solid physical argument for $I_{a,b}$ being positive-definite. In a sense, the IR formulae for the Schur index are the best evidence available for that statement. 

\subsection{RG flow of line defects}
If we consider RG flows triggered by vacua or mass deformations which preserve $SU(2)_R$ symmetry, 
a line defect which preserves the symmetries we discussed above will be described in a low-energy effective field theory as a line detect preserving the same symmetries. At the level of deformation classes, we assume that each $L_i^\UV$ line defect will be described as a direct sum of 
$L_i^\IR$ reference defects for the effective field theory, with Chan-Paton coefficients which can carry representations of the symmetries. This is the origin of the $\RG$ map from $\CA^\UV$ to $\CA^\IR$. It is an algebra map if we assume that the RG flow and fusion operations commute within a given deformation class. 

A surprising observation is that the same argument will fail when comparing RG flow and rotations. Namely, if we denote as $\RG_\vartheta$ the RG flow map for line defects of slope $\vartheta$ we will find that the map jumps at special values of $\vartheta$. The reason for that is that a massive BPS particle of central charge $Z$ can preserve the same SUSY as a line defect whose slope agrees with the argument of $Z$. Accordingly, the low energy effective description of a line defect of that slope becomes ambiguous/dependent on the scheme we use to distinguish a line defect from a line defect superimposed to a BPS particle. The definition of a deformation class of line defects as we vary their slope will break down. In particular, $\rho$ and $\RG$ do not commute. 

This problem is intimately connected to the corrections to the Schur index due to BPS particles. 
We can already see it in the presence of a single line defect of slope $\vartheta$, say $0<\vartheta<\pi$. Then the IR formula for the Schur index would involve the schematic combination 
\begin{equation}
    S_{0,\vartheta}\RG_\vartheta(L) S_{\vartheta,\pi}
\end{equation}
where $S = S_{0,\vartheta}S_{\vartheta,\pi}$ has been factored into contributions of BPS particles of central charge phase supported in the two intervals. As we change $\vartheta$, $\RG_\vartheta(L)$ passes across the phase of BPS particles and jumps in such a manner that the overall contribution remains constant and equal to $\RG(a) S = S \rho_\IR^{-1}(\RG(\rho_\UV(a)))$.

\subsection{The HT twist perspective}    
The Holomorphic Topological twist of a physical theory is defined by treating $Q$ as a BRST charge. 
The result is a new QFT which is holomorphic in the $z$ plane and topological in the two transverse directions. Essentially by definition, the Schur index counts local operators in the new theory. 

The line defects we studied become topological line defects in the twisted theory. 
The twisted theory may have many other topological line defects, thought, which do not arise in this manner. Topological line defects in the HT theory, say placed at the origin of the $\bC$ $z$ plane, 
automatically form a category $\mathrm{Lines}$ with specific properties. Namely, 
\begin{itemize}
    \item The spaces of morphism $\Hom(L_1,L_2)$ coincide with the space of defect-changing local operators and have an action of $\bC^*$ rotations of $z$ and of flavor symmetries. Their graded character is $I_{L_1,L_2}(\fq)$.
    \item Fusion of line defects equip $\mathrm{Lines}$ with a monoidal structure.
    \item Rotations of the topological plane equip $\mathrm{Lines}$ with the notion of left- and right- duals, leading to the categorical version of $\rho$. In particular, 
    \begin{equation}
        \Hom(L_1,L_2)\simeq \Hom(1,\rho(L_1)\otimes L_2)\simeq \Hom(1,L_2 \otimes \rho^{-1}(L_1)
    \end{equation}
    are the categorical versions of the relation between indices and $\rho$. 
\end{itemize}
The RG flow of line defects should map to a functor between corresponding categories $\mathrm{Lines}_\UV$ and $\mathrm{Lines}_\IR$, which we do not expect to preserve $\Hom$'s or $\rho$. 
Fixing this issue is an important open problem, see \cite{Gaiotto:2024fso} for a recent proposal.

The HT twist of gauge theories is reasonably well understood, and so is the corresponding category $\mathrm{Lines}$, see \cite{cautis2023canonicalbasescoulombbranches,Niu:2021jet}. It would be interesting to 
study the categorical analogue of our RG flow to the pure gauge theory. 

\subsection{Vacua, central charges and walls of marginal stability.}
A typical 4d ${\cal N}=2$ SQFT $T$ is endowed with an $SU(2)_R$ R-symmetry, as well as space of vacua $\cB$ with the property that $SU(2)_R$ is not spontaneously broken. 
The vacua are parameterized by the expectation values of certain ``Coulomb branch'' BPS local operators and have a complex structure such that these expectation values are holomorphic. We refer to ``$u$'' as the collection of such expectation values. 

For generic values of $u$, outside a co-dimension $1$ discriminant locus $\cD$, 
the low-energy effective description of $T$ is expected to be a pure Abelian gauge theory. The lattice $\Gamma$ is the lattice of charges for that effective theory. 
This is only defined locally on $\cB_0 \equiv \cB/\cD$ and undergoes monodromies 
around the discriminant locus, which fix the sub-lattice of flavour charges $\Gamma_f$. 

The low-energy effective action is built from a certain central charge function $Z$, a locally holomorphic function $\cB_0 \times \Gamma \to \bC$ linear in the second factor. We write it as $Z_\gamma(u)$ for a charge $\gamma$ and vacuum $u$. The central charge restricted to $\Gamma_f$ is $u$-independent and coincides with the ``mass parameters'' $m$ of $T$. An important property of the central charge is that it maps $\cB_0$ into a Lagrangian submanifold of $\bC^{\mathrm{rk}(\Gamma_g)}$, 
i.e. 
\begin{equation}
    \langle dZ, dZ \rangle =0
\end{equation}
where we used the fact that $dZ$ vanishes on $\Gamma_f$ and inverted the inner product on $\Gamma_g$. 

The central charge function controls the BPS bound 
\begin{equation}
    M\geq |Z_\gamma|
\end{equation}
on particles in the low-energy theory and insures that massive BPS particles can enter/exit the spectrum only at walls of marginal stability, where the phases of central charges of two particles align. 

It also controls the BPS bound 
\begin{equation}
    E \geq -\mathrm{Re} \, e^{- i \vartheta} Z_\gamma
\end{equation}
for BPS states in the presence of an half-BPS line defect, aka ``framed BPS states'', which insures that framed BPS states enter/exit the spectrum only at ``$\cK^\vartheta_\gamma$-walls'' of framed marginal stability, where 
\begin{align}
    &\mathrm{Im} \, e^{- i \vartheta} Z_\gamma =0 \cr
    &\mathrm{Re} \, e^{- i \vartheta} Z_\gamma <0 \, .
\end{align}
The phase $\vartheta$ controls which supersymmetries are preserved by the line defect and BPS states. We will set $\vartheta=0$ unless otherwise specified, and refer to $\cK_\gamma$-walls accordingly. 

An important idea is that the jumps across $\cK_\gamma$-walls in the spectrum of framed BPS states are controlled by the spectrum of BPS particles of charge multiple of $\gamma$ via a framed wall-crossing formula. In turn, self-consistency of the framed wall-crossing formula implies a wall-crossing formula for the BPS particles themselves. 

\subsection{Framed degeneracies and the quantum torus algebra}
As for the definition of $T$, the definition and classification of supersymmetric line defects in $T$ is also a difficult problem. For our purposes, the only information we are interested in are the graded Witten indices $F(L)_\gamma$ of the space of framed BPS states of charge $\gamma$ in the presence of the line defect $L$.
The indices are graded by a combination of spin and $SU(2)_R$ charge, with fugacity $\fq$ and we collect them in the generating function 
\begin{equation}
    F(L) \equiv \sum_{\gamma \in \Gamma} F(L)_\gamma X_\gamma \, .
\end{equation}
This generating function is expected to have a finite number of non-zero terms. As a Witten index, it is expected to be invariant under deformations unless we cross a wall of marginal stability. 

It is convenient to take the $X_\gamma$ formal variables to be generators of the quantum torus algebra. Then the product $F(L) F(L')$ counts the BPS states in the presence of both $L$ and $L'$, placed in a specific order along the axis fixed by the rotation generator used to grade the Witten index. With some work, it is possible to define the notion of fusion $L \circ L'$ of parallel BPS line defects, so that 
\begin{equation}
    F(L) F(L') = F(L \circ L')
\end{equation}
The fusion operation is an UV notion independent of a choice of vacuum. 

We can formalize this idea by defining an ``algebra of line defects'' $A$ over $\bZ[\fq, \fq^{-1}]$ \footnote{Really, the equivariant K-theory of a monoidal category.} which is a property of $T$ only, such that $L \to F(L)$ is an algebra morphism $A \to Q_\Gamma$ for every choice of $u$ and $\vartheta$. 

The definition of Witten indices $\Omega_\gamma$ which count BPS particles of charge $\gamma$ is a bit more subtle due to the position zeromodes and their super-partners. In our story, they will exclusively enter through a multi-particle index
\begin{equation}
    \cK_\gamma \equiv \prod_{n=1}^\infty \prod_s E_\fq((-\fq)^s X_{n \gamma})^{a_{n\gamma, s}}
\end{equation}
where 
\begin{equation}
    \Omega_{n \gamma} = \sum_s a_{n \gamma,s} (-\fq)^s
\end{equation}
The framed wall-crossing formula is the statement that 
\begin{equation}
    F^+(L) \cK_\gamma = \cK_\gamma F^-(L)
\end{equation}
where $F^\pm(L)$ are the framed BPS degeneracies on the two sides of a $\cK_\gamma$ wall, with $\pm$ being the sign of $\mathrm{Im} \, Z_\gamma$.

Away from a wall of framed marginal stability, define $\Gamma_+$ as the cone of charges such that $\mathrm{Im}, Z_\gamma >0$. Then the spectrum generator is defined as 
\begin{equation}
    S \equiv \prod^{\rightarrow}_{\gamma \in \Gamma_+} \cK_\gamma
\end{equation}
where the product is taken in order of $\mathrm{arg} Z_\gamma$, with larger phases to the left. This quantity is invariant across walls of marginal stability, away from walls of framed marginal stability. 

\section{Spectral problem on the annulus}
\label{spectralproblem}
In this appendix we discuss the spectral problem associated to the Discrete Liouville difference operator  that we report here for convenience:
\begin{equation}
\begin{split}
    F[w_1] \,g_{n}(\zeta)  &= g_{n+1}(\fq\,\zeta) +  g_{n-1}(\fq^{-1}\,\zeta) + \fq^{-n} \zeta \, g_n(\zeta)
\end{split}
\end{equation}
corresponding to the diagonalization of the Wilson line operator $F[w_1]$ on the annulus. 
This problem was solved in \cite{faddeev2014} in the case $\abs{\fq}^2  = 1$; here we extend their discussion to the Schur quantization regime. 
Let us go in the Fourier space representation over $\bL^2(\bZ \times \bZ)$:
\begin{equation}\label{inthamZ}
\wt F[w_1] \, g_{n,m} = \fq^m g_{n+1,m} + \fq^{-m} g_{n-1,m} + \fq^{-n}g_{n,m+1} \period
\end{equation}
Now, this operator is non-diagonal in both copies of $\bZ$. We parametrize the eingevalues as $\lambda + \lambda^{-1}$  and study the eigenproblem: 
\begin{equation}
\label{spectralprobl}
\fq^m g_{n+1,m} + \fq^{-m} g_{n-1,m} + \fq^{-n}g_{n,m+1} - (\lambda + \lambda^{-1}) g_{n,m} = 0 \comma\quad  |\lambda|^2 \in \fq^\bZ\period
\end{equation} 
It is convenient to parametrize $$\lambda = \fq^k \kappa$$ with  $k\in \bZ\comma \kappa\in S^1$. 
Let us further consider a second Fourier transformation $ \bL^2(\bZ \times S^1)   \xrightarrow[]{\mathcal{F}_1\circ \mathcal{F}_2} \bL^2(S^1 \times \bZ)$:
\begin{equation}
\label{recurrence}
(\wh F[w_1] - \lambda -\lambda^{-1}) \wh g_m(\theta) = \left(\fq^m \zeta + \fq^{-m} \zeta^{-1} - \fq^k \kappa + \fq^{-k} \kappa^{-1}  \right) \wh g_m(\zeta) + \wh g_{m+1}(\fq\, \zeta)  = 0 
\end{equation}
This is again the analogous of the recursion relation in \cite{faddeev2014}.
\paragraph{Solutions of the difference equation}
Now, let us introduce the $0<\fq^2<1$ version of the Faddeev Quantum dilogarithm introduced by Woronowicz \cite{Woronowicz1992}:
\begin{equation}
\label{quantumdil}
F_\fq(z) = \prod_{k=0}^{\infty} \frac{1 - \fq^{2k} z^\dagger}{1 - \fq^{2k} z }\comma
\end{equation}
that is continuous for \begin{equation}\zeta \in \bC^\fq =\{\zeta \in \bC : \log_\fq(|\zeta|) \in \bZ \}\period\end{equation}

This special function has a natural representation on the Schur quantization Hilbert space $\bL^2(\bZ \times S^1)$:
\begin{equation}
F_\fq(n,\zeta) = \prod_{k=0}^{\infty} \frac{1 - \fq^{2k +n}\, \zeta}{1 - \fq^{2k +n}\, \zeta^{-1}}\comma \qquad \zeta\in S^1\period
\end{equation}
$F_\fq(n,\zeta)$ admits a meromorphic continuation to the complex plane with poles and zeros respectively at:
$ \zeta  = \fq^{-2k -n}$ and 
 $\zeta  = \fq^{2k +n}$ for $k \in \mathbb{Z}$
It enjoys the reality condition:
\begin{equation}
\overline{F_{\fq}(n,\zeta)} F_{\fq}(n,\zeta) = F_\fq(n,\zeta^\dagger) F_{\fq}(n,\zeta) = 1\comma 
\end{equation}
and the inversion formula:
\begin{equation}
\label{inversion}
F_{\fq}(n+2,\zeta) F_{\fq}(-n,\zeta^{-1}) = (-\zeta)^{-n-1}\period
\end{equation}
This special function satisfies the important shift-property:
\begin{equation}\label{shiftprop}
     F_\fq(n+1,\fq^{-1}\,\zeta) = \left(1 - \fq^{n} \zeta^{-1}  \right)  F_\fq(n,\zeta)\comma 
\end{equation}
as illustrated by the direct computation:
\begin{equation}
\frac{F_\fq(n+1, \fq^{-1}\,\zeta)}{F_\fq(n,\zeta)} =  \prod_{k=0}^{\infty}\frac{1 - \fq^{2k +n} \zeta^{-1} }{1 - \fq^{2k +n+2} \zeta^{-1}} = 1 - \fq^n \zeta^{-1}
\end{equation}
The property \eqref{shiftprop} is crucial to show that:
\begin{equation}
\begin{split}
\psi_m(\zeta;\lambda) &= (-\zeta)^{m-1}   F_\fq( m + k + 1,\fq \zeta \kappa^{-1})  \, F_\fq( m - k+1,\fq\zeta \kappa ) \\
 &=(-\zeta)^{m-1} \frac{(\fq^{2+m} \lambda^\dagger\zeta;\fq^2)_\infty} {(\fq^m \lambda \zeta^{-1};\fq^2)_\infty } \frac{(\fq^{2+m} (\lambda^\dagger)^{-1} \zeta;\fq^2)_\infty}{(\fq^m \lambda^{-1} \zeta^{-1};\fq^2)_\infty}
\end{split}
\end{equation}
provides a general solution of the recurrence relation in \eqref{recurrence}, an anticipated in \eqref{eigenfunction}. One can recognize the two factors be one one of pair of chiral multiplets to the 3d superconformal index.
%Of course by $\theta+ \kappa$ we intend addition as coordinates on $S^1$.  %Denote \begin{equation} \bs k = i \kappa + \hbar k\comma \qquad  \bs p = i \theta + \hbar m \end{equation} coordinates over $\kappa \in S^1\times \bZ\ni k$ for the eigenvalue. We set \begin{equation}
%c(\bs{k}) = (-\fq)^{\Re \bs k/\hbar} \, e^{i \IM (\bs k \cdot \bs k)/2}
%\end{equation} 
%and denote the corresponding solution as:
%\begin{equation}
%\wh \varphi_{\bs k}(\bs p) = c(\bs{k})\,c(\bs{p}) \,  F_\fq(\bs{p} +\bs {k} + i \bs{\w})\,  F_\fq(\bs{p}-\bs{k}+ i \bs{\w})\comma \qquad \bs\w =  \hbar -i  \period
%\end{equation}
%Where, explicitly one has:
%\begin{equation}
%F_\fq(\bs{p} + i \bs{\w}) = \frac{\left(e^{i\phi} \fq^{2+n}; \fq^2 \right)_\infty}{\left(e^{-i\phi} %\fq^{n}; \fq^2 \right)_\infty}
%\end{equation}
%where 
The shift by a semi-period guarantees that when $m\pm k=0$, $\kappa  = \zeta^{\mp 1} $ is a pole,  that will be crucial for  the following.  Also, note that this solution satisfies the reality condition:
\begin{equation}\label{reality}
\psi_{m}(\zeta^{-1}; \lambda^\dagger) = \overline{\psi_{m}(\zeta;\lambda)}
\end{equation}
\paragraph{Scattering solution} 
From $\psi_m(\zeta;\lambda)$, one constructs the scattering solution by anti-Fourier transforming it:
\begin{equation}\label{scatteringsol}
\begin{split}
\varphi_{n}(\zeta;\lambda) &= \frac{1}{2\pi}\sum_{m\in \bZ} \int_{S^1}\mathrm{d}{\xi} \, \xi^n\,  \zeta^m \, \psi_{m}(\xi;\lambda)\comma \qquad n \in \bZ\comma \quad \zeta \in S^1\\
  \widetilde{\varphi}_{n,m}(\lambda) &= \frac{1}{2\pi} \int_{S^1}\mathrm{d}{\xi} \,\xi^n \,  \psi_{m}(\xi;\lambda)\comma \qquad n,m \in \bZ
\end{split}
\end{equation}
$\varphi_{n}(\zeta;\lambda)$ admits analytical continuation in $\bC^\fq$  by contour deformation of the Fourier transform.  By standard contour manipulation, it is straightforward to show that indeed \eqref{scatteringsol} provide solutions to \eqref{spectralprobl}.

Since $\forall m \in \bZ$\comma $\psi_{m}(\zeta;\lambda)$ is an analytical function on $\zeta \in S^1$, then its Fourier transform decays exponentially at infinity:
\begin{equation}
\lim_{n\to \pm \infty} \widetilde{\varphi}_{n,m}(\lambda)= \lim_{n\to \pm \infty} \varphi_{n}(\zeta;\lambda) = 0
\end{equation}
Explicitly, let us parametrize $\xi = e^{-i\theta}$, $\kappa = e^{-i \chi}$, so that the Fourier transform is an integral $\frac{1}{2\pi}\sum_{m\in \bZ} \int_0^{2\pi}\mathrm{d}{\theta} \, e^{-in\theta}\,  \zeta^m \, \psi_{m}(e^{-in\theta};\lambda)$ . The closest poles to the real axis of $\psi_{m}(\xi;\lambda)$ are located at:
\begin{equation}
\label{poles}
\theta  = \mp  \chi +  i (m\pm k) \hbar  \qquad m\pm k =0,\cdots, \infty\comma \quad\hbar = -\log\fq
\end{equation}
\begin{figure}
\centering
\begin{tikzpicture}[scale=2]
    % Axes
    \draw[->] (-0.5,0) -- (1.5,0) node[right] {${\rm Re
    }\theta$};
    \draw[->] (0,-0.5) -- (0,2) node[above] {${\rm Im}\theta$};

    % Vertices of the rectangle
    \coordinate (A) at (0,0);
    \coordinate (B) at (1,0) ;
    \coordinate (C) at (1,1.5) ;
    \coordinate (D) at (0,1.5);

\coordinate (P1) at (0.3, 1.5);
    \coordinate (P2) at (0.7, 1.5);
    \fill[red] (P1) circle (1pt) node[above right] {  };
    \fill[red] (P2) circle (1pt) node[above right] {};

    % Deformed rectangle contour with semicircles around poles
    \draw[thick] (A) -- (B) node[midway, above] {\small$\rightarrow$};
    \draw[thick] (B) -- (1,1.5) node[midway, left] {\small$\uparrow$};
    \draw[thick] (1,1.5) -- (0.8,1.5) ;
    \draw[thick] (0.6,1.5) -- (0.4,1.5) node[midway, below] {\small$\leftarrow$};
    \draw[thick] (0,1.5) -- (0.2,1.5);
    \draw[thick] (D) -- (A) node[midway, right] {\small$\downarrow$};
    \draw[thick] (0.8,1.493) arc[start angle=0, end angle=180, radius=0.1] -- (0.6,1.5);
    \draw[thick] (0.4,1.493) arc[start angle=0, end angle=180, radius=0.1] -- (0.2,1.5);

    % Points
    \fill (A) circle (1pt) node[below left] {$0$};
    \fill (B) circle (1pt) node[below right] {$2\pi$};
    \fill (D) circle (1pt) node[above left] {$i (m\pm k) \hbar $};
    \fill (C) circle (1pt);

\end{tikzpicture}
\caption{Integration contour in the complex $\theta$-plane}
\label{figcontour}
\end{figure}
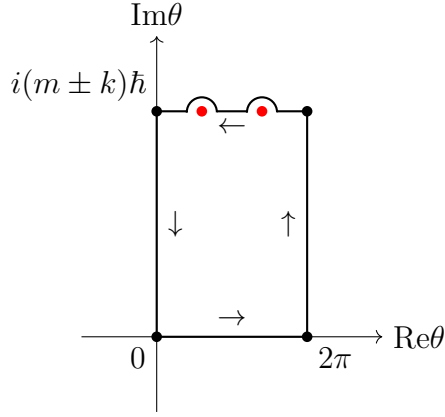
That are located in the upper or lower half plane depending on the sign of $(m\pm k)$. Let us assume for now that $m\pm k > 0$, as the situation is analogous for the opposite case.  
To determine the behavior at $n \to -\infty$, on can consider a box contour in the upper half plane as in Figure \ref{figcontour}. The contributions due to the contour along the imaginary axis is suppressed exponentially at $n\to -\infty$: $\pm \int_{0}^{\hbar (m\pm k) } e^{ n \theta } (\cdots)\mathrm{d}{\theta} \to 0$ as well as the contribution from the top straight part of the contour$ \int_{2\pi}^0 \mathrm{d}{\theta} e^{ n\hbar (m\pm k ) }e^{-i n \theta} (\cdots) \xrightarrow{n\to -\infty} 0$. The dominant contribution comes from the two residue terms located at the points \eqref{poles}. The residue of $F_\fq(n, \zeta)$ there, can be computed recursively:
\begin{equation}
i {\rm Res}_{\zeta = \fq^{-n} } \left(F_{\fq}(n, \zeta) \right) = \frac{1}{(\fq^2; \fq^2)_{n}}\comma \qquad n=0,\infty
\end{equation}
%{This is why we shifted above by $\bs \w$, otherwise at $n=0$ the residue would have been $0$}
So that, at $n\to -\infty$, one obtains the following asymptotic expansion:
\begin{equation}
\wt{\varphi}_{n,m} (\lambda) \asymp M_m(\lambda^{-1}) \fq^{-n(m + k) }\kappa^{-n} + M_m(\lambda) \fq^{ n (k - m) }\kappa^n \comma
\end{equation}
that, at $n\to-\infty$, is non zero only for:
\begin{equation}\begin{split}
\wt{\varphi}_{n,m} (\lambda) &\asymp \delta_{m,-k}M_{m}(\lambda^{-1}) \kappa^{-n} +\delta_{m,k} M_m(\lambda) \kappa^n \\
&:=  M_+(\lambda) \, \kappa^n \delta_{m,k} + M_-(\lambda^{-1})\,  \kappa^{-n}\delta_{m,-k} \period
\end{split}\end{equation}
This is analogous to what found in \cite{faddeev2014}. One can compute explicitly the functions $M_\pm(\lambda)$: \begin{equation}
M_{\pm}(\lambda) =  \frac{(-\fq)^{\pm k}}{(\fq^2; \fq^2)_0}\,   F_\fq( \mp 2k + 1,\fq \kappa^{\pm 2}) =  (-\fq)^{\pm k} \frac{(\fq^{2\mp 2k} \kappa^{\pm 2};\fq^2)_\infty} {(\fq^{\mp 2k} \kappa^{\mp 2};\fq^2)_\infty }
\end{equation}
that satisfy:
\begin{equation}
\label{Mpmmp}
M_{\mp}(\lambda) = M_{\pm} (\lambda^{-1})
\end{equation}
% Furthermore, using an adaptation of the inversion formula:
% \begin{equation}
% \frac{\left(\kappa^{2} \fq^{2-2k};\fq^2\right)_{\infty } \left(\kappa^{-2} \fq^{2k+2};\fq^2\right)_{\infty }}{\left(\kappa^{-2}\fq^{-2k};\fq^2\right)_{\infty } \left(\kappa^{2}  \fq^{2k};\fq^2\right)_{\infty }}=\frac{\fq^{2k} e^{-4i k \kappa }}{\left(1-e^{2i \kappa } \fq^{2k}\right) \left(1-e^{-2i \kappa } \fq^{2k}\right)}
% \end{equation}
Furthermore, one has that:
\begin{equation}
\frac{1}{M_+(\lambda) M_-(\lambda)} =  4 \fq^{-2k} (\lambda - \lambda^{-1})(\lambda - \lambda^{-1})^\dagger
\end{equation}
That is the equivalent of the result of \cite{faddeev2014}.

\paragraph{Jost solutions}

As $n\to -\infty$ the Fourier tranform of operator \eqref{spectralprobl} reduces to the ``free'' operator:
\begin{equation}
\fq^m g_{n+1,m} + \fq^{-m} g_{n-1,m} - 2 (\lambda + \lambda^{-1})g_{n,m} \period
\end{equation}
The free operator, admits the Jost solutions:
\begin{equation}
f^{\rm free}_\pm(n,m;\lambda) = \kappa^{\mp n} \delta_{\pm m,k} \period 
\end{equation}
It is then natural to look for Jost solutions $f_\pm$ of the ``interacting" operator, having the following  asymptotic behavior:
\begin{equation}
f_\pm(n,m; \lambda) \xrightarrow{n\to -\infty} \kappa^{\pm n} \delta_{\mp m,k} + o(1) \comma
\end{equation}
Consider the functions (remember  $\kappa = e^{i\chi}$):
\begin{equation}\begin{split}\label{jost}
f_\pm(n,m; \lambda) &= \frac{1}{4\pi\sinh( \mp \frac{2\pi\chi}{\hbar} ) M_{\pm}(\lambda)}  \int_0^{2\pi}\mathrm{d}{\theta} e^{- i n\theta}{\varphi}_{m}(e^{i \theta}; \lambda)  \left( \sinh(\frac{2\pi\theta}{\hbar} ) + \sinh(\mp \frac{ 2\pi  \chi}{\hbar}  )   \right)\period
\end{split}\end{equation}

To show that $\lim_{n\to -\infty}f_{\pm}(n,m; \lambda) = e^{\pm i n \kappa} \delta_{m,\mp k}+ o(1)$, one uses the same contour as in Figure \ref{figcontour}. Yet, due to the  term further quasi-constant piece 
$ \left( \sinh(\frac{2\pi\theta}{\hbar} ) + \sinh(\mp \frac{ 2\pi\kappa}{\hbar}  )   \right) $ , only one between the two poles of $\wh\varphi$ at $\theta = \pm \kappa$ for $ i \hbar (m\pm k) = 0$ is a singularity of each of the integrands in \eqref{jost}. 

The functions \eqref{jost} solve \eqref{spectralprobl} in its representation over $L^2(\bZ \times \bZ)$ \eqref{inthamZ}. This can be shown by noting that:
\begin{equation}
f_\pm(n,m; \lambda) = \frac{1}{4\pi\sinh(\mp \frac{2\pi \chi}{\hbar} ) M_\pm(\lambda)}  \left(\wt\varphi_{\scriptsize n+ \frac{ 2\pi i}{\hbar} ,m}(\lambda) - \wt\varphi_{\scriptsize -\frac{ 2\pi i}{\hbar} ,m}(\lambda) + \sinh(\mp \frac{2\pi\chi}{\hbar}  ) \wt\varphi_{\scriptsize n,m}(\lambda)\right)
\end{equation}
And, since $\fq^{-n \pm \frac{ 2\pi i}{\hbar}} = \fq^{-n}$, then also $\wt\varphi_{\scriptsize n \pm \frac{ 2\pi i}{\hbar} ,m}$ are  solutions of \eqref{inthamZ}.

Then, it follows that:
\begin{equation}\label{phiMf}
\wt\varphi_{n,m}(\lambda) = M_+(\lambda) f_+(n,m;\lambda) + M_-(\lambda) f_-(n,m;\lambda)
\end{equation}

\paragraph{Casorati determinant}
It is important for the following to compute the discrete version of the Wronskian, known as \textit{Casorati} determinant. Given two solutions $f,g$, the Casorati determinant is defined by:
\begin{equation}
C(f,g) (n,m;\lambda) =   \fq^m \, f(n, -m ;\lambda)\, g(n+1,m;\lambda)  - \fq^m \, f(n+1,m ;\lambda) \, g(n,-m;\lambda)
\end{equation}
Let us take the Casorati determinant between $f_-$ and $f_+$.
We have that:
\begin{equation}
\lim_{n\to -\infty} C(f_+,f_-) (n,m;\lambda) = \lambda - \lambda^{-1} + o(1)
\end{equation}
One has \begin{equation}C(f_+,f_-) (n,m;\lambda) = \lambda - \lambda^{-1} \end{equation} 
%\FA{I fill the details later, one just has to provide an estimate of growth at $x\to \infty$ to show that is bounded and entire in the strip, so it's constant.}. 
Note that: 
\begin{equation}C(f_+,f_-) (n,-m;\lambda) =  \left(\lambda - \lambda^{-1}\right)^\dagger\end{equation}
Clearly, one has:
\begin{equation}
C(f_+,\wt\varphi)(n,m;\lambda) = 2 M_-(\lambda) \left(\lambda - \lambda^{-1} \right)
\end{equation}
\paragraph{Resolvent}
The resolvent of $\wt F[w_1]$ is, by definition,  $$R(\lambda) = (\wt F[w_1] -\lambda)^{-1}$$, i.e.\ the operator such that:
\begin{equation}
\sum_{n,m\in \bZ} R(n,m; n',m'|\lambda) g_{n',m'} =  h_{n,m} \in \cD(\wt F[w_1])\comma \qquad \left(\wt F[w_1] - (\lambda + \lambda^{-1}) \right) h_{n,m} = g_{n,m} \period
\end{equation}
This, for $g\in \cD(\wt\cH)$, is equivalent to verify that:
\begin{equation}\left(\wt F[w_1]- (\lambda + \lambda^{-1}) \mathds{1}_{n,m} \right)R(n,m; n',m'|\lambda) = \delta_{n,n'}\delta_{m,m'}
\end{equation}
or, explicitly:
\begin{equation}\begin{split}\label{reseq}
\fq^m \, R(n+1,m; n',m'|\lambda)+ \fq^{-m} R(n-1,m; n',m'|\lambda)  &+ \fq^{-n} R(n,m+1; n',m'|\lambda)  + \\
- (\lambda + \lambda^{-1}) R(n,m; n',m'|\lambda)  &= \delta_{n,n'} \delta_{m,m'}
\end{split}\end{equation}
Define the operator:
\begin{equation}\label{resdefinition}
R(n,m; n',m'|\lambda) = \frac{\pi}{(\lambda - \lambda^{-1}) M_-(\bs k) } \left[ \frac{ f_+(n,m; \lambda) \wt\varphi(n',-m')(\lambda)}{1- e^{2\pi i(n-n')} e^{-2\pi i (m-m')}} + \frac{ f_+(n', -m'; \lambda) \wt\varphi_{n,m}(\lambda)}{1- e^{-2\pi i(n-n')} e^{2\pi i(m-m')}}\right]
\end{equation}
Given that both $f_-$ and $\wt\varphi$ are solutions of $(\wt F[w_1] - (\lambda + \lambda^{-1})) = 0$ and the denominators are invariant under shifts of $n,m$ by integers, it is clear that \eqref{resdefinition} satisfies \eqref{reseq} whenever $R(n,m; n',m'|\lambda)$ is not singular, so that \eqref{reseq} is trivially satisfied whenever $n\neq n'$ and $m \neq m'$. At $n = n' \wedge m = m'$, where the distribution $R$ is singular, one receives the following divergent contribution:
\begin{equation}\begin{split}
\lim_{\epsilon \to 0^+ } \frac{\pi}{(\lambda - \lambda^{-1}) M_-(\lambda) } \frac{1}{2\pi} \Bigg[ &\fq^m \frac{ f_+(n+1,m; \lambda) \wt\varphi_{n',-m'}(\lambda) - f_+(n',-m'; \lambda) \wt\varphi_{n+1,m}(\lambda)}{(n-n') - (m-m') + i \epsilon} \\ 
&+ \fq^{-m} \frac{ f_+(n-1,m; \lambda) \wt\varphi_{n',-m'}({\lambda}) - f_+(n',-m'; \lambda) \wt\varphi_{n-1,m}({\lambda})}{(n-n') - (m-m') - i \epsilon} \Bigg]\\
= \delta_{n,n'} \delta_{m,m'} &\frac{C(f_+,\wt\varphi)(n,n;\lambda)}{2(\lambda - \lambda^{-1}) M_-(\lambda)}  =  \delta_{n,n'} \delta_{m,m'} 
\end{split}
\end{equation}
that coincides with the r.h.s.\ of \eqref{reseq}. 
%\FA{Missing: show it is bounded. (very easy, just need to type)}

\paragraph{Eigenfunctions expansion theorem}
Let us denote the spectrum domain as: 
\begin{equation}
\mathcal{I} = \Big\{ \IM\,(\lambda + \lambda^{-1}) \comma \,\abs{\lambda} \in \fq^\bZ \Big\}
\end{equation}
For  $\psi\in \bL^2(\bZ\times \bZ)$, define the operator $\mathcal{U}$ by:
\begin{equation}\begin{split}
\mathcal{U}: \bL^2(\bZ \times \bZ) \to \bL^2(\mathcal{I})\comma
(\mathcal{U} \psi)(\bs k) = \sum_{n,m\in \bZ} \psi(n,m) \wt\varphi_{n,m}(\lambda) = \langle \psi, \wt\varphi(\lambda)\rangle \comma 
\end{split}\end{equation}

then, $\mathcal{U}$ maps isometrically $\bL^2(\bZ\times \bZ)$ onto $\bL^2(\mathcal{I})$ with spectral measure:
\begin{equation}
\rho(\lambda ) =\frac{\pi \fq^{-2k}}{2 M_+(\lambda) M_-(\lambda)}  = \frac{\pi}{2} (\lambda - \lambda^{-1})
(\lambda - \lambda^{-1})^\dagger \end{equation}
To show that, we claim that the following identity holds:
\begin{equation}\label{idimpo}
\psi(n,m) = \int_{\mathcal{I}} \mathrm{d}{\lambda}  \sum_{n',m'\in \bZ} \psi(n',m') \wt\varphi_{n',m'}(\lambda)\overline{\wt\varphi_{n,m}(\lambda)}\rho(\lambda) = \int_{\mathcal{I}} \mathrm{d}{\lambda}  (\mathcal{U} \psi)(\lambda) \overline{\wt\varphi_{n,m}(\lambda)}\rho(\lambda)
\end{equation}
The starting point to prove this, is the \textit{Stone's theorem} relating a resolution of the identity to the jump of the resolvent on the absolutely continuum spectrum:
\begin{equation}
    \mathds{1}_{n,n'}\mathds{1}_{n,m'} = \lim_{\epsilon\to 0^+}\int_{\mathcal{I}} \dd{\lambda} (\lambda - \lambda^{-1})\left(R(n,n',m,m'| \lambda e^{i\epsilon}) - (R(n,n',m,m'| \lambda e^{- i\epsilon})  \right)
\end{equation}
% We have:
% \begin{equation}
% \Lambda = (\lambda + \lambda^{-1})\comma \qquad  \mathrm{d}{\bs\lambda} = 2\sinh(\bs k) \mathrm{d}{\bs k}
% \end{equation}
    taking $\lambda e^{- i\epsilon}$ corresponds to sending $\lambda \to \lambda^{-1}$. 
We can compute explicitly the discontinuity of the resolvent across the continuous spectrum. 
To do that, let first write the resolvent in the form:
\small
\begin{equation}
{\textstyle R( n,m;n',m'|\lambda)} = \pi \frac{ f_+(n,m; \lambda ) \wt\varphi_{n', -m'}(\lambda) e^{-i\pi(n-n' + m-m')} - f_+(n',-m';\lambda) \wt\varphi_{n,m}(\lambda) e^{i\pi(n-n' + m-m')}}{2(\lambda - \lambda^{-1}) M_-(\lambda) \sinh(i \pi(n-n' + m-m'))}\period
\end{equation}
Using \eqref{Mpmmp}, \eqref{phiMf} and \eqref{reality} and the definition of $\wt \varphi$ in \eqref{scatteringsol}, one computes:
\normalsize
\begin{equation}\begin{split}
R(n,n',m,m'| \lambda) - R(n,n',m,m'|{ \textstyle \lambda^{-1}}) &=  \frac{\pi}{2(\lambda - \lambda^{-1})} \frac{\fq^{-2k} \wt\varphi_{n',m'}(\lambda) \, \overline{\wt\varphi_{n,m}(\lambda)}}{M_+(\lambda)M_-(\lambda) } \comma
\end{split}
\end{equation}
and therefore:
\begin{equation}
\mathds{1}_{n,n'}\mathds{1}_{n,m'} = \int_{\mathcal{I}} \mathrm{d}{\lambda} \rho(\lambda) \wt\varphi_{n',m'}(\lambda) \, \overline{\wt\varphi_{n,m}(\lambda)}\period
\end{equation}
Then, we obtain \eqref{idimpo} by simply inserting this resolution of the identity:
\begin{equation}\begin{split}
    \psi(n,m) &= \sum_{n',m' \in \bZ}  \mathds{1}_{n,n'}\mathds{1}_{n,m'}\, \psi(n',m') = \sum_{n',m' \in \bZ}  \int_{\mathcal{I}} \mathrm{d}{\lambda} \rho(\lambda)  \psi(n',m' \wt\varphi_{\lambda}(n',m') \, \overline{\wt\varphi_{n,m}(\lambda)}\\
    &= \int_{\mathcal{I}} \mathrm{d}{\lambda} \rho(\lambda) \, (\mathcal{U} \psi)(\lambda)\,  \overline{\wt\varphi_{n,m}(\lambda)}
\end{split}
\end{equation}
From this identity, by taking the inner product with $\overline{\psi(n,m)}$ follows the Plancharel theorem:
\begin{equation}
\norm{\mathcal{U} \psi}^2_{\bL^2(\mathcal{I})} = \norm{\psi}^2_{\bL^2(\bZ \times \bZ)}
\end{equation}
Where the Hilbert space $\bL^2(\mathcal{I})$ is equipped with spectral measure $\rho(\lambda) \mathrm{d}{\lambda}$. From this isometric property of the operator, follows that $\mathcal{U}^{-1}\mathcal{U} = \mathds{1}_{\bL^2(\bZ \times \bZ)}$. To show that $\mathcal{U}\mathcal{U}^{-1} = \mathds{1}_{\bL^2(\mathcal{I})}$, it is enough to show that the image of $\mathcal{U}$ coincides with $\bL^2(\mathcal{I})$. This follows, as customary by the observation that since:
\begin{equation}
\mathcal{U} R(\lambda) \mathcal{U}^{-1} = \wh{R}(\lambda) \Rightarrow \mathcal{U} (F[w_1] - \lambda -\lambda^{-1} ) \mathcal{U}^{-1} = \wh F[w_1] -\lambda -\lambda^{-1} \comma   \end{equation}
then $\wh F[w_1]$ coincides on $\bL^2(\mathcal{I})$ with the multiplication operator $(\lambda + \lambda^{-1})$.

\pdfbookmark[1]{\refname}{mono}
\bibliographystyle{JHEP}
\bibliography{Ref}

	\end{document}